\documentclass[a4paper,fleqn,usenatbib]{mnras}

\usepackage[T1]{fontenc}
\usepackage{ae,aecompl}

\usepackage{graphicx}	
\usepackage{amsmath}	
\usepackage{amssymb}	
\usepackage{array}
\usepackage{booktabs}
\usepackage{pdflscape}
\newcommand{\msun}{\mathrm{M}_{\sun}}
\newcommand{\mzams}{M_{\mathrm{ZAMS}}}
\newcommand{\mni}{M_{^{56}\mathrm{Ni}}}
\newcommand{\bvrijhk}{BV\!RI\!J\!H\!K}
\newcommand{\bvrriijhk}{BV\!(r)R(i)I\!J\!H\!K}
\newcommand{\fqbol}{F^{\bvrijhk}}
\newcommand{\phs}{\phantom{-}} 
\newcommand{\phn}{\phantom{0}} 
\newcommand{\bv}{{B\!-\!V}}
\newcommand{\vi}{{V\!-\!I}}
\newcommand{\vk}{{V\!-\!K}}
\newcommand{\ssd}{\hat{\sigma}}
\newcommand{\leff}{\bar{\lambda}}
\newcommand{\feffx}{\bar{f}_x}
\newcommand{\EGBV}{E_{\bv}^\mathrm{G}}
\newcommand{\EhBV}{E_{\bv}^\mathrm{h}}
\newcommand{\zsnhel}{z^\mathrm{SN}_\mathrm{helio}}
\newcommand{\qbcv}{\mathrm{qBC}_V}
\newcommand{\bc}{\mathrm{BC}}
\newcommand{\zp}{\mathrm{ZP}}
\newcommand{\zpsn}{\mathrm{ZP}^{\mathrm{SN}}}
\newcommand{\yfe}{\bar{y}_{\mathrm{Fe}}}
\newcommand{\ztf}{\mathrm{ZTF}}

\everycr={\noalign{\global\advance\stepno by 1}}%

\defcitealias{2013MNRAS.433.1745D}{D13}
\defcitealias{2014MNRAS.439.3694J}{J14}
\defcitealias{2017MNRAS.466...34L}{L17}
\defcitealias{2006ApJ...653.1145K}{K06}
\defcitealias{2016ApJ...821...38S}{S16}

\title[Iron yield of normal SNe II]{The Iron Yield of Normal Type II Supernovae}
\author[\'{O}. Rodr\'{i}guez et al.]{\'{O}. Rodr\'{i}guez,$^{1}$\thanks{E-mail: olrodrig@gmail.com}
N. Meza,$^{2}$
J. Pineda-Garc\'{i}a$^{3}$
and M. Ramirez$^{3}$
\\
$^{1}$School of Physics and Astronomy, Tel Aviv University, Tel Aviv 69978, Israel\\
$^{2}$Department of Physics, University of California, 1 Shields Avenue, Davis, CA 95616, USA\\
$^{3}$Departamento de Ciencias Fisicas, Universidad Andres Bello, Avda. Republica 252, Santiago, Chile
}

\date{Accepted XXX. Received YYY; in original form ZZZ}

\pubyear{2020}

\begin{document}
\label{firstpage}
\pagerange{\pageref{firstpage}--\pageref{lastpage}}
\maketitle

\begin{abstract}
We present $^{56}$Ni mass estimates for 110 normal Type II supernovae (SNe~II), computed here from their luminosity in the radioactive tail. This sample consists of SNe from the literature, with at least three photometric measurements in a single optical band within 95--320\,d since explosion. To convert apparent magnitudes to bolometric ones, we compute bolometric corrections (BCs) using 15~SNe in our sample having optical and near-IR photometry, along with three sets of SN~II atmosphere models to account for the unobserved flux. We find that the $I$- and $i$-band are best suited to estimate luminosities through the BC technique. The $^{56}$Ni mass distribution of our SN sample has a minimum and maximum of 0.005 and 0.177\,M$_{\sun}$, respectively, and a selection-bias-corrected average of $0.037\pm0.005$\,M$_{\sun}$. Using the latter value together with iron isotope ratios of two sets of core-collapse (CC) nucleosynthesis models, we calculate a mean iron yield of $0.040\pm0.005$\,M$_{\sun}$ for normal SNe~II. Combining this result with recent mean $^{56}$Ni mass measurements for other CC~SN subtypes, we estimate a mean iron yield $<$0.068\,M$_{\sun}$ for CC~SNe, where the contribution of normal SNe~II is $>$36~per~cent. We also find that the empirical relation between $^{56}$Ni mass and steepness parameter ($S$) is poorly suited to measure the $^{56}$Ni mass of normal SNe~II. Instead, we present a correlation between $^{56}$Ni mass, $S$, and absolute magnitude at 50\,d since explosion. The latter allows to measure $^{56}$Ni masses of normal SNe~II with a precision around 30~per~cent.
\end{abstract} 




\section{Introduction}\label{sec:introduction}

Supernovae (SNe) explosions are important astrophysical objects for a wide range of research fields. Among them we mention their use to measure distances and cosmological parameters, their connection with stellar evolution, and their contribution to the energetics and chemical enrichment of the interstellar medium. Indeed, regarding the latter, SNe synthesize the bulk of all the mass in the Universe residing in elements from oxygen to the iron group. Therefore, to understand chemical evolution, it is critical to determine how much elements are produced by every kind of SN: Type Ia and core-collapse (CC) SNe (see \citealt{2017hsn..book.1151H} and \citealt{2021Natur.589...29B} for current reviews of their explosion mechanisms). In the case of CC~SNe, almost all $\alpha$-elements have been produced in those explosions, while their contribution to the cosmic iron budget is comparable to that of SNe~Ia \citep[e.g.][]{2017ApJ...848...25M}.

CC~SNe are grouped into two classes: H-rich envelope SNe, historically known as Type II SNe \citep[SNe~II,][]{1941PASP...53..224M}, and stripped-envelope (SE) SNe. The latter group includes H-poor Type IIb and H-free Type Ib, Ic, and broad-line Ic (Ic-BL) SNe (see \citealt{2017hsn..book..195G} for a current review of the SN classification). Among SNe~II, some events are grouped into subtypes based on spectral and photometric characteristics: those showing narrow H emission lines in the spectra, indicative of ejecta-circumstellar material interaction \citep[SNe~IIn;][]{1990MNRAS.244..269S}\footnote{In this group we also include SNe~IIn/II and LLEV SNe~II, described in \citet{2020MNRAS.494.5882R}.}, and those having long-rising light curves similar to SN~1987A \citep{1988AJ.....95...63H,2016AA...588A...5T}; while a few SNe~II are recognized as having peculiar characteristics (e.g. OGLE14-073, \citealt{2017NatAs...1..713T}; iPTF14hls, \citealt{2017Natur.551..210A}; ASASSN-15nx, \citealt{2018ApJ...862..107B}; DES16C3cje, \citealt{2020MNRAS.496...95G}; SN~2018ivc, \citealt{2020ApJ...895...31B}). For the rest of SNe~II (about 90~per~cent, e.g. \citealt{2017PASP..129e4201S}), in order not to use the same name of the class that contains other SN~II subtypes and peculiar events, we will refer as normal SNe~II. The latter are found to form a continuum group\footnote{Some authors, however, suggest a separation into distinct groups \citep[e.g.][]{2012ApJ...756L..30A,2014MNRAS.445..554F}.} \citep[e.g.][]{2014ApJ...786...67A,2015ApJ...799..208S,2016MNRAS.459.3939V,2017ApJ...850...90G,2019MNRAS.490.2799D}, where the photometric and spectroscopic diversity depends mainly on the amount of H in the envelope at the moment of the explosion, the synthesized $^{56}$Ni mass ($\mni$), and the explosion energy \citep[e.g.][]{2017ApJ...850...90G}.

Progenitors of normal SNe~II have been directly detected on pre-explosion images. They correspond to red supergiant (RSG) stars with zero-age main-sequence (ZAMS) mass, $\mzams$, in the range $8\text{\textendash}18\,\msun$ \citep[e.g.][]{2009ARAA..47...63S,2015PASA...32...16S}. In the case of SE~SNe, evidence points toward progenitors with $\mzams$ similar to those of normal SNe~II but evolving in binary systems, and some cases with a more massive and isolated progenitor (e.g. \citealt{2019AA...628A...7A} and references therein). The $^{56}$Ni mass produced by CC~SNe depends on the explosion properties and the core structure of the progenitor \citep[e.g.][]{2019MNRAS.483.3607S}. Therefore, $\mni$ estimates are important to contrast the progenitor scenarios and explosion mechanisms of different CC~SNe. The mean $^{56}$Ni mass ($\langle\mni\rangle$) of normal SNe~II is lower than that of SE~SNe \citep[e.g.][]{2019AA...628A...7A,2020AA...641A.177M}. On the other hand, normal SNe~II account for around 60~per~cent of all CC~SNe in a volume-limited sample \citep[e.g.][]{2017PASP..129e4201S}. The latter makes normal SNe~II significant contributors to the $^{56}$Ni and iron budget of CC~SNe (e.g. see Section~\ref{sec:mean_iron_yield}). 

Normal SNe~II are characterized by having an optically thick photosphere during the first $60\text{\textendash}120$\,d after the explosion \citep[e.g.][]{2014ApJ...786...67A,2014MNRAS.442..844F,2015ApJ...799..208S,2019MNRAS.490.2799D}. In this so-called photospheric phase, the $V$-band absolute magnitudes ($M_V$) range between around $-14.5$ and $-18.5$\,mag \citep[e.g.][]{2014ApJ...786...67A,2019MNRAS.490.2799D}. In particular, normal SNe~II having ${M_V\gtrsim-15.5}$\,mag are referred as sub-luminous SNe~II (e.g. \citealt{2004MNRAS.347...74P,2014MNRAS.439.2873S}), while those having ${M_V\lesssim-17}$\,mag are referred as moderately-luminous SNe~II (e.g. \citealt{2013AA...555A.142I}). The aforementioned phase is also characterized by a period of $30\text{\textendash}70$\,d where the $V$-band magnitude remains nearly constant or declines linearly with time \citep[e.g.][]{2014ApJ...786...67A}. During this period, called plateau phase, light curves are powered by H recombination. Then, the brightness decreases by around $1.0\text{\textendash}3.5$\,mag in a lapse of about $15\text{\textendash}30$\,d, indicating that all H has recombined. After this transition phase, the luminosity starts to decrease exponentially with time. In this phase, the energy sources are the $\gamma$-rays and positrons produced by the radioactive decay of the unstable cobalt isotope $^{56}$Co (daughter of the unstable nickel isotope $^{56}$Ni) into the stable iron isotope $^{56}$Fe \citep[e.g.][]{1980NYASA.336..335W}. Therefore the luminosity in this phase, called radioactive tail, is a good estimate of the $^{56}$Ni mass ejected in the explosion.

The luminosity is given by the inverse-square law of light and the bolometric flux. The latter can be computed through the direct integration technique \citep[e.g.][]{2017PASP..129d4202L}. In this method, the available $x$-band magnitudes are converted to monochromatic fluxes ($\feffx$) and associated to $x$-band effective wavelengths ($\leff_x$). The set of (${\leff_x,\feffx}$) points defines the photometric spectral energy distribution (pSED) which, integrated over wavelength, provides the quasi-bolometric flux. For normal SNe~II in the radioactive tail, the quasi-bolometric flux in the wavelength range 0.46{\textendash}2.16\,\micron\ typically accounts for 90 per~cent of the bolometric flux (e.g. see Section~\ref{sec:fbol_and_Lbol}). To estimate the unobserved flux, some authors extrapolate fluxes assuming a Planck function \citep[e.g.][]{2001PhDT.......173H,2015ApJ...799..215P}, while others do not include the unobserved flux in the bolometric flux \citep[e.g.][]{2009ApJ...701..200B,2010MNRAS.404..981M}. In practice, the application of the direct integration technique is limited because the low number of normal SNe~II with IR photometry during the radioactive tail.

An alternative method to compute bolometric fluxes for SNe without IR photometry is the bolometric correction (BC) technique \citep[e.g.][]{2017PASP..129d4202L}. In this method, the  magnitude in a given band ($m_x$) is related to the bolometric magnitude ($m_{\mathrm{bol}}$) through ${m_{\mathrm{bol}}=m_x+\bc_x}$. Here, $\bc_x$ is calibrated using SNe with $m_{\mathrm{bol}}$ computed with the direct integration technique. Based on the $BV\!I\!J\!H\!K$ photometry of the normal SN~II~1999em and the long-rising SN~1987A, \citet{2001PhDT.......173H} reported a constant $\bc_V$ for SNe~II in the radioactive tail. Similar constant $\bc_V$ values were later reported by other authors \citep[e.g.][]{2009ApJ...701..200B,2010MNRAS.404..981M,2015ApJ...799..215P}.

The two main weaknesses in the $\bc_V$ values reported in previous works are: (1) not accounting for the unobserved flux or assuming a Planck function to estimate it, and (2) the low number of SNe used to compute $\bc_V$. At present, thanks to the development of non-local thermodynamic equilibrium (non-LTE) radiative transfer codes \citep[e.g.][]{2011MNRAS.410.1739D,2011AA...530A..45J}, it is possible to estimate the flux outside the optical/near-IR range for normal SNe~II through theoretical spectral models. From the observational side, the number of normal SNe~II observed during the radioactive tail with optical and near-IR filters had increased over time. Therefore, it is possible to improve the BC determination with the current available data. 

The goal of this work is to estimate the mean iron yield ($\yfe$) of normal SNe~II. For this, we use all normal SNe~II in the literature with useful photometry during the radioactive tail. To improve the determination of the radioactive tail luminosity through the BC technique, we also aim to enhance the BC calibration. The paper is organized as follows. In Section~\ref{sec:data_set} we outline the relevant information on the data we use. The methodology to compute BCs, $^{56}$Ni and iron masses is described in Section~\ref{sec:methodology}. In Section~\ref{sec:analysis} we present new BC calibrations, the $\mni$ distribution of our SN sample, the $\langle\mni\rangle$ and $\yfe$ values for normal SNe~II, and a new method to estimate $\mni$.
Comparisons with previous works, discussions about systematics, and future improvements are in Section~\ref{sec:discussion}. Finally, our conclusions are summarized in Section~\ref{sec:conclusions}.

\section{Data set}\label{sec:data_set}

\subsection{Supernova sample}\label{sec:SN_sample}

For this work we select normal SNe~II from the literature, having photometry in the radioactive tail (1) in at least one of the following bands: Johnson--Kron--Cousins $V\!RI$ or Sloan $ri$; (2) in the range 95{\textendash}320\,d since the explosion, corresponding to the time range where our BC calibrations are valid (Section~\ref{sec:BC_cal_others}); and (3) with at least 3 photometric epochs in order to detect possible $\gamma$-ray leakage from the ejecta (Section~\ref{sec:Ni_mass}). The final sample of 110 normal SNe~II fulfilling our selection criteria is listed in Table~\ref{table:SN_sample}. This includes the SN name (Column~1), the host galaxy name (Column~2), the Galactic colour excess $\EGBV$ (Column~3), the heliocentric SN redshift $\zsnhel$ (Column~4), the distance modulus (Column~5, see Section~\ref{sec:mu}), the host galaxy colour excess $\EhBV$ (Column~6, see Section~\ref{sec:EhBV}), the explosion epoch $t_0$ (Column~7, see Section~\ref{sec:t0}) and the references for the photometry (Column~8). Among the SNe in our set, 15 have $\bvrriijhk$ photometry in the radioactive tail. We use the latter sample to compute bolometric fluxes and BCs. Galactic colour excesses are taken from \citet{2011ApJ...737..103S} (except for SN~2002hh, see Section~\ref{sec:EhBV}), which have associated a random error of 16~per~cent \citep{1998ApJ...500..525S}. Throughout this work, for our Galaxy and host galaxies, we assume the extinction curve given by \citet{1999PASP..111...63F} with ${R_V=3.1}$ (except for SN~2002hh, see Section~\ref{sec:EhBV}).

\subsection{Theoretical models}

We use SN~II atmosphere models to compute the contribution of the flux at $\lambda<\leff_B$ and $\lambda>\leff_K$ to the bolometric one. We also employ CC nucleosynthesis models to estimate the contribution of iron stable isotopes other than $^{56}$Fe to the total ejected iron mass ($M_\mathrm{Fe}$). Among the available models from the literature, we select those using progenitors of $\mzams\leq18\,\msun$ to be consistent with observations of normal SNe~II \citep[e.g.][]{2009ARAA..47...63S,2015PASA...32...16S}. Those models were exploded with energies $\sim 10^{50}\text{\textendash}10^{51}$\,erg, corresponding to the typical range for normal SNe~II \citep[e.g.][]{2018NatAs...2..808F,2018ApJ...858...15M}. The $^{56}$Ni masses of the selected models range between 0.003 and $0.12\,\msun$ which is consistent with observations of normal SNe~II \citep[e.g.][]{2017ApJ...841..127M}.

\subsubsection{Atmosphere models}

We use the SN~II atmosphere model sets given by \citet[][hereafter \citetalias{2013MNRAS.433.1745D}]{2013MNRAS.433.1745D}, \citet[][hereafter \citetalias{2014MNRAS.439.3694J}]{2014MNRAS.439.3694J}, and \citet[][hereafter \citetalias{2017MNRAS.466...34L}]{2017MNRAS.466...34L}. Those models were generated evolving ZAMS stars with ${\mzams\geq12\,\msun}$ until the RSG stage before the CC. The explosions were simulated using a piston, while the spectra were computed through 1D non-LTE radiative transfer codes. In particular, \citetalias{2017MNRAS.466...34L} used the same methodology as \citetalias{2013MNRAS.433.1745D} but with different progenitors and explosion energies.

The selected \citetalias{2013MNRAS.433.1745D} set consists on six models ($8\text{\textendash}11$ spectra each) with ${\mzams=15\,\msun}$ and ${\mni=0.036\text{\textendash}0.121\,\msun}$. The selected \citetalias{2014MNRAS.439.3694J} set consists on three models ($3\text{\textendash}5$ spectra each) with $\mzams$ of 12, 15, and $19\,\msun$, and ${\mni=0.062\,\msun}$. The selected \citetalias{2017MNRAS.466...34L} set consists on seven models ($9\text{\textendash}10$ spectra each) with ${\mzams=12\,\msun}$, and ${\mni=0.007\text{\textendash}0.010\,\msun}$.

\subsubsection{Nucleosynthesis yields}

We use the 1D CC nucleosynthesis yields presented in \citet[][hereafter \citetalias{2006ApJ...653.1145K}]{2006ApJ...653.1145K} and \citet[][hereafter \citetalias{2016ApJ...821...38S}]{2016ApJ...821...38S}, which are updated versions of the works of \citet{1997NuPhA.616...79N} and \citet{1995ApJS..101..181W}, respectively. The selected \citetalias{2006ApJ...653.1145K} models were calibrated to be consistent with observations of normal SNe~II, where the yields are estimated using a mass cut such that ${M_\mathrm{Fe}\sim0.07\,\msun}$. The \citetalias{2016ApJ...821...38S} models were calibrated to be consistent with with the Crab SN (for $\mzams\leq12\,\msun$) and SN~1987A (for $\mzams>12\,\msun$), while the yields were estimated using the ``special trajectory'' to represent the mass cut (see \citetalias{2016ApJ...821...38S} for more details).

The selected \citetalias{2006ApJ...653.1145K} set consists on 12 models with $\mzams$ of 13, 15, and $18\,\msun$, and $\mni=0.07\text{\textendash}0.09\,\msun$. The selected \citetalias{2016ApJ...821...38S} set consist on 13 models with ${\mzams=9\text{\textendash}12\,\msun}$ and $\mni=0.003\text{\textendash}0.03\,\msun$, and 89 models with $\mzams$ between 12 and 18\,$\msun$ and ${\mni=0.05\text{\textendash}0.09\,\msun}$.

\section{Methodology}\label{sec:methodology}

\subsection{Light-curve interpolation}

\begin{figure*}
\includegraphics[width=1.0\textwidth]{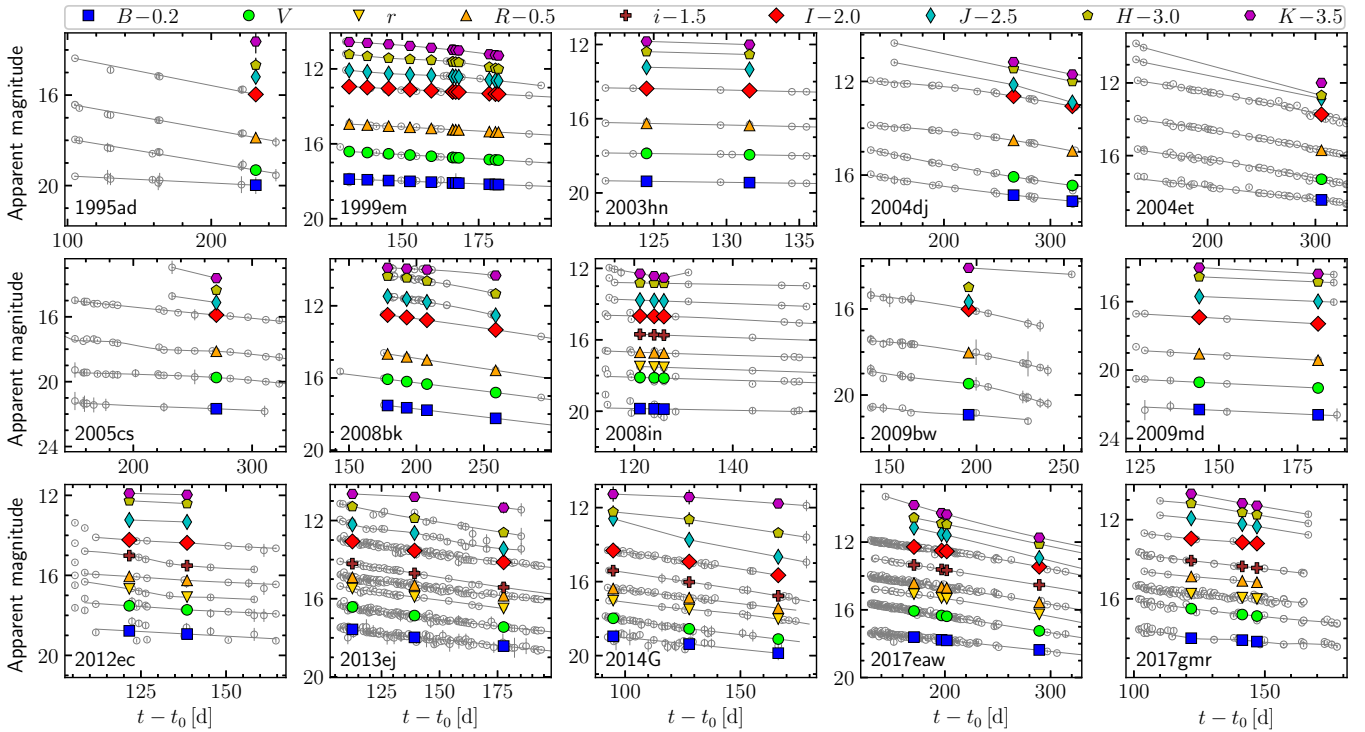}
\caption{Light curves of the 15 SNe in the BC calibration sample. Colour-filled symbols correspond to the interpolated photometry, where bands and magnitude shifts with respect to the original values are indicated in the legend. Grey empty circles correspond to the photometry from the literature, and grey lines are the light-curve fits.} 
\label{fig:BC_sample}
\end{figure*}

To compute bolometric fluxes and BCs, we first need to evaluate the $\bvrriijhk$ photometry of the 15~SNe in the calibration sample at the same set of epochs $\{t_i\}$. To determine these epochs, for each band we select the epochs covered simultaneously by the photometry in the other bands. Then, we adopt the epochs of the band with less observations as $\{t_i\}$. If the rest of bands do not have photometry at the $t_i$ epochs, then we interpolate them using the \texttt{ALR} code\footnote{\url{https://github.com/olrodrig/ALR}} \citep{2019MNRAS.483.5459R}. The \texttt{ALR} performs \texttt{loess} non-parametric regressions \citep{Cleveland_etal1992} to the input photometry, taking into account observed and intrinsic errors, along with the presence of possible outliers. If the \texttt{ALR} is not able to perform a \texttt{loess} fit (e.g. only few data points are available), then the \texttt{ALR} just performs a linear interpolation between points. In the case of SN~1995ad, we extrapolate the $BI$ photometry to the epoch of the $J\!H\!K$ photometry using a straight-line fit. Fig.~\ref{fig:BC_sample} shows the result of this process.

\subsection{Quasi-bolometric correction}

In the BC calibration sample, only 6 out of 15~SNe have $ri$ photometry. Therefore, we only use the $\bvrijhk$ photometry in order to compute quasi-bolometric fluxes with an homogeneous data set. We construct pSEDs and compute quasi-bolometric fluxes $\fqbol_i$ (in erg\,s$^{-1}$\,cm$^{-2}$) using the prescription provided in Appendix~\ref{sec:pSED}. We define the $x$-band quasi BC (qBC) as
\begin{equation}\label{eq:qBC}
\mathrm{qBC}_{x,i}=-2.5\log(\fqbol_i)-m_{x,i}^{\mathrm{cor}},
\end{equation}
where
\begin{equation}\label{eq:mcor}
m_{x,i}^{\mathrm{cor}}=m_{x,i}-R_{\leff_x}(\EGBV+\EhBV),
\end{equation}
being $m_{x,i}$ the $x$-band magnitude at epoch $t_i$, and $R_{\leff_x}$ the total-to-selective extinction ratio for $\leff_x$, listed in Table~\ref{table:zp_and_leff}.

\subsection{Bolometric correction and luminosity}\label{sec:fbol_and_Lbol}

The flux from a pSED defined in a wavelength range $\lambda_1\text{\textendash}\lambda_2$ is only an approximation of the real flux computed integrating the SED in the same wavelength range, $F_i^{\lambda_1-\lambda_2}$. In our case, to quantify the relative difference between $\fqbol_i$ and $F_i^{\leff_B-\leff_K}$, we compute $\alpha_i$ such that
\begin{equation}
F_i^{\leff_B-\leff_K} = (1+\alpha_i)\fqbol_i.
\end{equation}
For this task we use the \citetalias{2013MNRAS.433.1745D}, \citetalias{2014MNRAS.439.3694J}, and \citetalias{2017MNRAS.466...34L} spectral models. To obtain $\fqbol_i$ from models, we first compute their synthetic magnitudes (see Appendix~\ref{sec:syn_mag}) and then we compute $\fqbol_i$ with the recipe given in Appendix~\ref{sec:pSED}.

Fig.~\ref{fig:C_B-K--BVRIJHK} shows the $\alpha_i$ values computed with the three model sets. Since we do not find any correlation with colour indices, in the figure we plot $\alpha_i$ as a function of the time since explosion. For the \citetalias{2013MNRAS.433.1745D}, \citetalias{2014MNRAS.439.3694J}, and \citetalias{2017MNRAS.466...34L} models, the mean $\alpha$ values and their sample standard deviation ($\ssd$) errors are ${-3.2\pm1.8}$, ${4.6\pm3.0}$, and ${1.1\pm1.4}$~per~cent, respectively. There is a difference of at least $2.6\,\ssd$ between the mean $\alpha$ values from \citetalias{2013MNRAS.433.1745D} and \citetalias{2014MNRAS.439.3694J} models. Based on late-time optical spectra of 38 normal SNe~II, \citet{2017MNRAS.467..369S} found that the \citetalias{2014MNRAS.439.3694J} models fit better to the observations than the \citetalias{2013MNRAS.433.1745D} ones. This evidence favour the scenario where $F_i^{\leff_B-\leff_K}$ is $\sim$5~per~cent greater than $\fqbol_i$ instead of $\sim$3~per~cent lower. To be conservative, we adopt the average of the mean $\alpha$ values of the three models, i.e., ${\alpha=0.8\pm3.9}$~per~cent ($1\,\ssd$ error). This value is consistent within $\pm1\,\ssd$ with the results obtained for the three model sets.

\begin{figure}
\includegraphics[width=0.95\columnwidth]{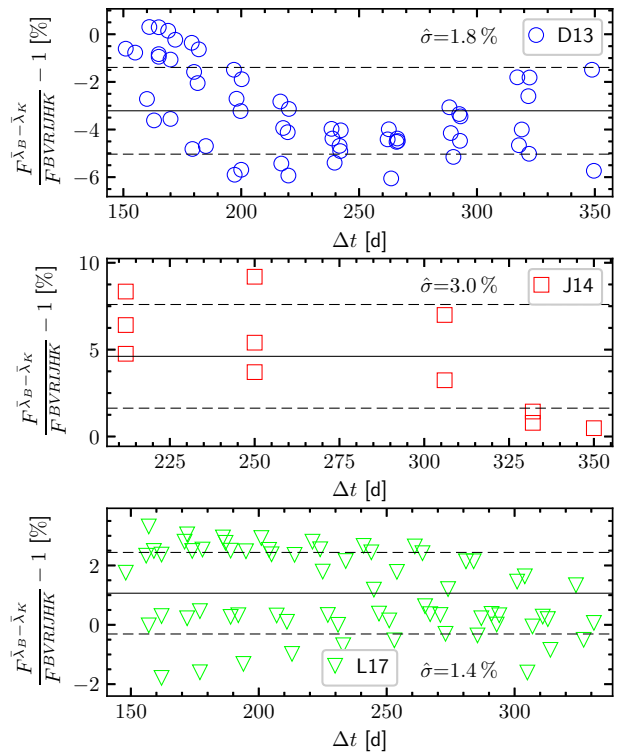}
\caption{Relative differences between the flux computed with the pSED and the SED using the \citetalias{2013MNRAS.433.1745D} (top panel), \citetalias{2014MNRAS.439.3694J} (middle panel), and \citetalias{2017MNRAS.466...34L} (bottom panel) models. Solid and dashed lines correspond to mean values and $\pm1\,\ssd$ limits, respectively.} 
\label{fig:C_B-K--BVRIJHK}
\end{figure}

To compute the bolometric flux $F^{\mathrm{bol}}$, we have to correct $F^{\leff_B-\leff_K}$ for the unobserved flux. In our case,
\begin{equation}
F_i^{\mathrm{bol}}=F_i^{\leff_B-\leff_K}+F_i^{\lambda<\leff_B}+F_i^{\lambda>\leff_K},
\end{equation}
where $F_i^{\lambda<\leff_B}$ and $F_i^{\lambda>\leff_K}$ are the unobserved fluxes at wavelengths below $\leff_B$ and beyond $\leff_K$, respectively. 

For the unobserved flux below $\leff_B$, we write
\begin{equation}
F_i^{\lambda<\leff_B} = F_i^{\leff_B-\leff_K} c_i^{\lambda<\leff_B},
\end{equation}
where
\begin{equation}
c_i^{\lambda<\leff_B}=F_i^{0.1\mu\mathrm{m}-\leff_B}/ F_i^{\leff_B-\leff_K}
\end{equation}
is the flux correction relative to $F_i^{\leff_B-\leff_K}$. We choose $0.1\,\micron$ as lower $\lambda$ value because it is the minimum wavelength in common for the three model sets. This value is also low enough to consider the flux negligible at shorter wavelengths.

\begin{figure}
\includegraphics[width=0.95\columnwidth]{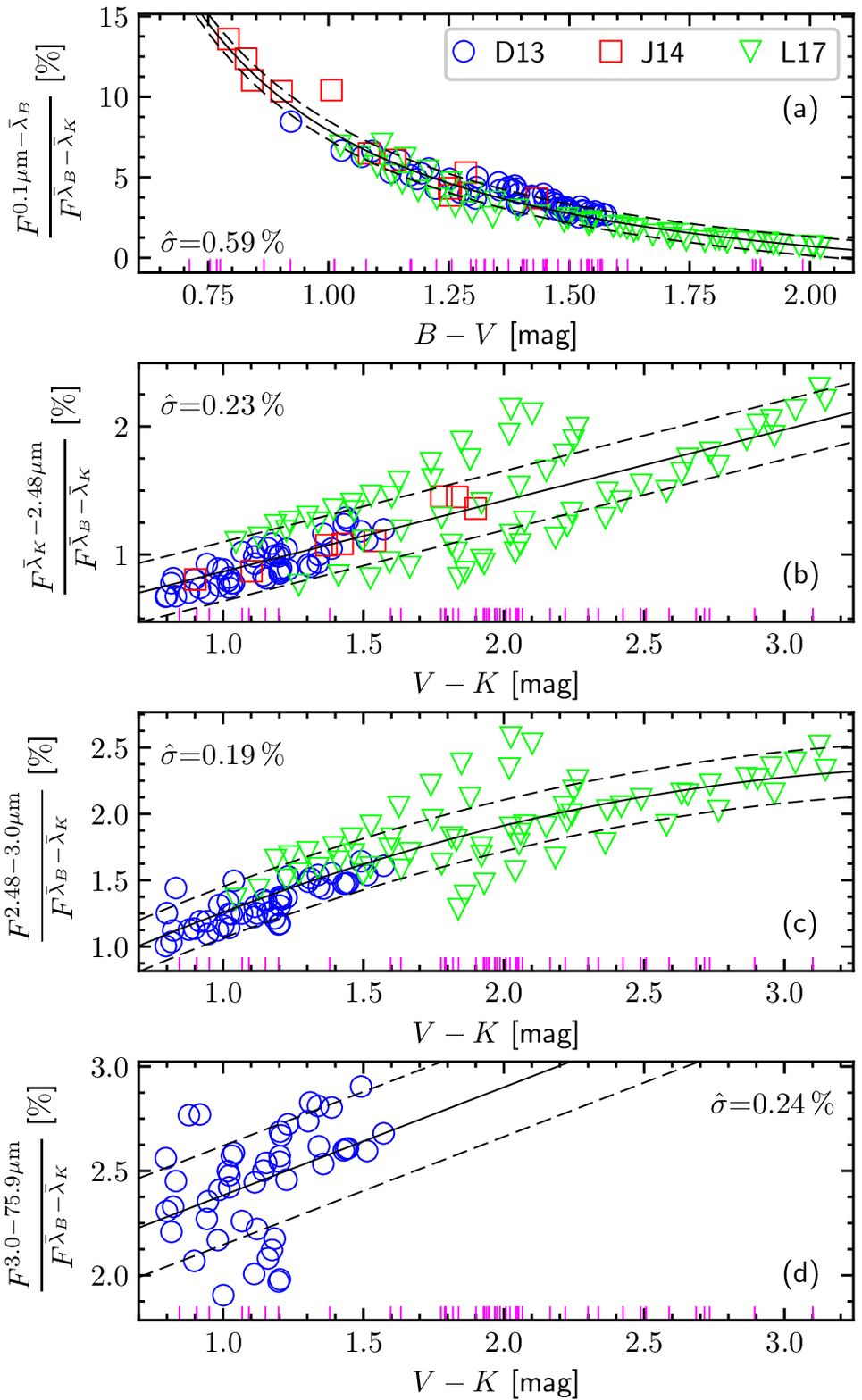}
\caption{Corrections to account for the unobserved flux in the ranges $0.1\,\micron\text{\textendash}\leff_B$ (a), $\leff_K\text{\textendash}2.48\,\micron$ (b), $2.48\text{\textendash}3.0\,\micron$ (c), and $3.0\text{\textendash}75.9\,\micron$ (d), relative to the flux in the range $\leff_B\text{\textendash}\leff_K$, using the \citetalias{2013MNRAS.433.1745D}, \citetalias{2014MNRAS.439.3694J}, and \citetalias{2017MNRAS.466...34L} models. Magenta ticks mark the colours of the SNe in the BC calibration set. Solid lines indicate the fits to the models, while dashed lines are the $\pm1\,\ssd$ limits around the fit.} 
\label{fig:C_unobserved_UV+IR}
\end{figure}

Fig.~\ref{fig:C_unobserved_UV+IR}(a) shows the $c_i^{\lambda<\leff_B}$ values as a function of $\bv$. As visible in the figure, there is a correlation between both quantities. The flux correction is lower than 2~per~cent for ${\bv>1.7}$ and can be greater than 10~per~cent for ${\bv<0.9}$. Within the colour range in common between the \citetalias{2013MNRAS.433.1745D}, \citetalias{2014MNRAS.439.3694J}, and \citetalias{2017MNRAS.466...34L} models (${1.0\leq\bv\leq 1.45}$), we see that values for the three model sets are in good agreement. We parametrize the dependence of $c_i^{\lambda<\leff_B}$ on $\bv$ as
\begin{equation}
c_i^{\lambda<\leff_B} = \sum_{j=0}^2\frac{a_j}{(\bv)_i^j},
\end{equation}
where the quadratic order was determined using the model selection described in Appendix~\ref{sec:model_selection}. Table~\ref{table:c_parameters} lists the fit parameters along with the $\ssd$ around the fit, which covers the colour range ${0.79\leq\bv\leq 2.02}$. Among the $\bv$ colours of the SNe in our BC calibration set (marked as magenta ticks in the figure), four are below the lower limit (two of SN~2014G, one of SN~1995ad, and one of SN~2004dj). In order to prevent a misestimation on $c_i^{\lambda<\leff_B}$ due to extrapolations, 
for $\bv$ colours bluer than 0.79\,mag we adopt the $c^{\lambda<\leff_B}$ correction for ${\bv=0.79}$.

\begin{table}
\caption{Parameters for the flux corrections.}
\label{table:c_parameters}
\begin{tabular}{lccccc}
\hline
Correction                      & Colour  & $a_0$ (\%) & $a_1$ (\%) & $a_2$ (\%) & $\ssd$ (\%)\\
\hline
$\alpha$                        & --      & $\phs0.80$ & --     &     --     & $3.90$ \\
$c^{\lambda<\leff_B}$           & $\bv$   &    $-3.11$ & $4.30$ & $\phs6.72$ & $0.59$ \\
$c^{\leff_K-2.48\mu\mathrm{m}}$ & $\vk$   & $\phs0.32$ & $0.55$ &     --     & $0.23$ \\
$c^{2.48-3.0\mu\mathrm{m}}$     & $\vk$   & $\phs0.31$ & $1.09$ &    $-0.14$ & $0.19$ \\
$c^{\lambda>3\mu\mathrm{m}}$    & $\vk$   & $\phs1.87$ & $0.52$ &     --     & $0.24$ \\
$c^{\lambda>\leff_K}$           & $\vk$   & $\phs2.50$ & $2.16$ &    $-0.14$ & $0.38$ \\
\hline
\multicolumn{6}{l}{\textit{Note:} $c^{\lambda>\leff_K}=c^{\leff_K-2.48\mu\mathrm{m}}+c^{2.48-3.0\mu\mathrm{m}}+c^{\lambda>3\mu\mathrm{m}}$.}
\end{tabular}
\end{table}

For the unobserved flux at $\lambda>\leff_K$ we use
\begin{equation}
F_i^{\lambda>\leff_K} = F_i^{\leff_B-\leff_K}c_i^{\lambda>\leff_K},
\end{equation}
where, since the model sets do not cover the same wavelength range, we write
\begin{equation}\label{eq:c_fIR}
c_i^{\lambda>\leff_K}=c_i^{\leff_K-2.48\mu\mathrm{m}}+c_i^{2.48-3.0\mu\mathrm{m}}+c_i^{\lambda>3\mu\mathrm{m}},
\end{equation}
being
\begin{equation}
c_i^{\leff_K-2.48\mu\mathrm{m}}=F_i^{\leff_K-2.48\mu\mathrm{m}}/ F_i^{\leff_B-\leff_K},
\end{equation}
\begin{equation}
c_i^{2.48-3.0\mu\mathrm{m}}=F_i^{2.48-3.0\mu\mathrm{m}}/ F_i^{\leff_B-\leff_K},
\end{equation}
and
\begin{equation}
c_i^{\lambda>3\mu\mathrm{m}}=F_i^{3.0-75.9\mu\mathrm{m}}/ F_i^{\leff_B-\leff_K}.
\end{equation}
Here, the values $2.48$ and $3.0\,\micron$ correspond to the maximum $\lambda$ in common for the \{\citetalias{2013MNRAS.433.1745D}, \citetalias{2014MNRAS.439.3694J}, \citetalias{2017MNRAS.466...34L}\} and \{\citetalias{2013MNRAS.433.1745D}, \citetalias{2017MNRAS.466...34L}\} model sets, respectively, while $75.9\,\micron$ is the maximum $\lambda$ for the \citetalias{2013MNRAS.433.1745D} models. 

Figs.~\ref{fig:C_unobserved_UV+IR}(b) and \ref{fig:C_unobserved_UV+IR}(c) show $c_i^{\leff_K-2.48\mu\mathrm{m}}$ and $c_i^{2.48-3.0\mu\mathrm{m}}$, respectively, as a function of $\vk$. Those flux corrections, as expected, are greater for red colours than for blue ones. We express the dependence of $c_i^{\leff_K-2.48\mu\mathrm{m}}$ and $c_i^{2.48-3.0\mu\mathrm{m}}$ on $\vk$ through polynomials, i.e.,
\begin{equation}\label{eq:c_IR_parametrization}
c_i^{\lambda_1-\lambda_2} = \sum_{j=0}^{\mathcal{O}_{\lambda_1,\lambda_2}}a_j(\vk)_i^j,
\end{equation}
being the orders $\mathcal{O}_{\lambda_1,\lambda_2}$ determined with the model selection described in Appendix~\ref{sec:model_selection}. The fit parameter values are summarized in Table~\ref{table:c_parameters}. As in the case of $c_i^{\lambda<\leff_B}$, we find a good agreement between models within the ranges in common. 

Fig.~\ref{fig:C_unobserved_UV+IR}(d) shows the $c_i^{\lambda>3\mu\mathrm{m}}$ values as a function of $\vk$, where only the \citetalias{2013MNRAS.433.1745D} models provide spectral information for ${\lambda>3\,\mu\mathrm{m}}$. In this case the best fit is a straight-line, whose parameters are reported in Table~\ref{table:c_parameters}. Since the \citetalias{2013MNRAS.433.1745D} models do not cover all the $\vk$ colours of the SNe in the BC calibration set, the straight-line fit could introduce errors due to extrapolation. However, we find that $c_i^{\lambda>3\mu\mathrm{m}}$ at $\vk=3.1$ (the reddest colour in the BC calibration set) is only 1~per~cent (in value) greater than the correction for the reddest colour in the \citetalias{2013MNRAS.433.1745D} models ($\vk=1.57$). Therefore, we adopt the linear parametrization of $c_i^{\lambda>3\mu\mathrm{m}}$ for all the $\vk$ colour range (${0.80\leq\vk\leq 3.15}$). 

Based on equation~(\ref{eq:c_fIR}), $c_i^{\lambda>\leff_K}$ is a polynomial given by equation~(\ref{eq:c_IR_parametrization}), where the coefficients $a_j$ are the sum of those of $c_i^{\leff_K-2.48\mu\mathrm{m}}$, $c_i^{2.48-3.0\mu\mathrm{m}}$, and $c_i^{\lambda>3\mu\mathrm{m}}$. Parameters for $c_i^{\lambda>\leff_K}$ are given in Table~\ref{table:c_parameters}.


Once $\alpha$, $c_i^{\lambda<\leff_B}$, and $c_i^{\lambda>\leff_K}$ are determined, we define the apparent bolometric magnitude and the $x$-band BC as
\begin{equation}
m_{\mathrm{bol},i}=-2.5\log(\fqbol_i)-\kappa_i
\end{equation}
and
\begin{equation}\label{eq:BC}
\bc_{x,i}=m_{\mathrm{bol},i}-m_{x,i}^{\mathrm{cor}}=\mathrm{qBC}_{x,i}-\kappa_i,
\end{equation}
respectively. The model-based correction, $\kappa_i$, is given by
\begin{equation}
\kappa_i= 2.5\log\left(1+\alpha\right)+2.5\log\left(1+c_i^{\lambda<\leff_B}+c_i^{\lambda>\leff_K}\right),
\end{equation}
where the error in $\kappa_i$ is dominated by the error in $\alpha$ (see Table~\ref{table:c_parameters}).
For the SNe in the BC calibration set, the model-based correction ranges between 0.09\,mag (SN~2005cs) and 0.22\,mag (SN~2014G), with a median of 0.11\,mag. Therefore, during the radioactive tail, the observed $\fqbol$ typically corresponds to 90~per~cent of the bolometric flux.

Once $\bc_x$ values are calibrated with observations (section~\ref{sec:BC_cal}), luminosities ($L$, in $10^{43}\,\mathrm{erg\,s}^{-1}$) can be estimated through the BC technique, given by
\begin{equation}\label{eq:Lbol}
\log{L_i}=(\mu-\bc_{x,i}-m_{x,i}^{\mathrm{cor}})/2.5-2.922,
\end{equation}
where the constant provides the conversion from magnitude to cgs units.

\subsection{$^{56}$Ni Mass}\label{sec:Ni_mass}

During the radioactive tail, the energy sources powering the ejecta are the $\gamma$-rays and positrons produced in the radioactive decay of $^{56}$Co into $^{56}$Fe. The latter deposit energy in the ejecta at a rate $Q_{\mathrm{dep}}$. Using equations~(10){\textendash}(12) of \citet{2019MNRAS.484.3941W}, and assuming that the deposited energy is immediately emitted, we can write the relation between $Q_{\mathrm{dep}}$ (in $10^{43}\,\mathrm{erg\,s}^{-1}$) and 
$\mni$ (in $\msun$) as
\begin{equation}\label{eq:Qbol_vs_MNi}
\log Q_{\mathrm{dep},i}=\log\mni-0.39\left[\frac{\Delta t_i}{100\,\mathrm{d}}\right]+0.154+D_i.
\end{equation}
Here, $\Delta t_i = (t_i-t_0)/(1+\zsnhel)$ is the time since explosion in the SN rest frame, and
\begin{equation}\label{eq:Dfdep}
D_i=\log(0.97f_{\mathrm{dep},i}+0.03),
\end{equation}
where $f_{\mathrm{dep}}$ is the $\gamma$-ray deposition function, which describes the fraction of the generated $\gamma$-ray energy deposited in the ejecta. Knowing the deposition function, $\log\mni$ can be inferred by equating equations~(\ref{eq:Lbol}) and (\ref{eq:Qbol_vs_MNi}).

If all the $\gamma$-ray energy is deposited in the ejecta, then ${f_{\mathrm{dep},i}=1}$ (${D_i=0}$), otherwise ${f_{\mathrm{dep},i}<1}$ (${D_i<0}$). In the first case, we expect the $\log\mni$ estimates (one for each $\log{L_i}$ measurement) to be consistent with a constant value. In the second case, the $\log\mni$ estimates decrease with time as the ejecta becomes less able to thermalize $\gamma$-rays (Section~\ref{sec:MNi_estimates}), which makes it necessary to correct for $f_{\mathrm{dep}}$. For the latter we adopt the model of \citet{1999astro.ph..7015J}:
\begin{equation}\label{eq:fdep}
f_{\mathrm{dep},i}=1-\exp{[-(T_0/\Delta t_i)^2]},
\end{equation}
where $T_0$ is a characteristic time-scale (in d) that represents the $\gamma$-ray escape time. This parameter is estimated such that the $\log\mni$ estimates are consistent with a constant value.

To detect possible $\gamma$-ray leakage from the ejecta, we need at least three photometric points as we have to infer $T_0$ and $\log\mni$. The detailed recipe to compute $\log\mni$ and check if it is necessary to correct for $f_{\mathrm{dep}}$ is provided in Appendix~\ref{sec:MNi_equation}\footnote{The code implementing this algorithm (\texttt{SNII\_nickel}) is available at \url{https://github.com/olrodrig/SNII_nickel}.}. In order to properly convert $\log\mni$ into $\mni$, we use the formalism provided in Appendix~\ref{sec:lognormal_mean}.

\subsection{Iron mass}\label{sec:iron_mass}
 
The inferred $\mni$ provides a good estimate of the ejected $^{56}$Fe mass ($M_{^{56}\mathrm{Fe}}$). The $M_\mathrm{Fe}$ value, however, is greater than $M_{^{56}\mathrm{Fe}}$ since it is also composed of the stable isotopes $^{54,57,58}$Fe \citep[e.g.][]{2019ApJ...870....2C}. Using the relation
\begin{equation}\label{eq:eta}
\eta = M_\mathrm{Fe}/\mni=1+M_{^{54,57,58}\mathrm{Fe}}/M_{^{56}\mathrm{Fe}}
\end{equation}
and the CC nucleosynthesis yield models of \citet{1999ApJS..125..439I}, \citet{2008NewA...13..606B} computed ${\eta=1.08}$.

\begin{figure}
\includegraphics[width=1.0\columnwidth]{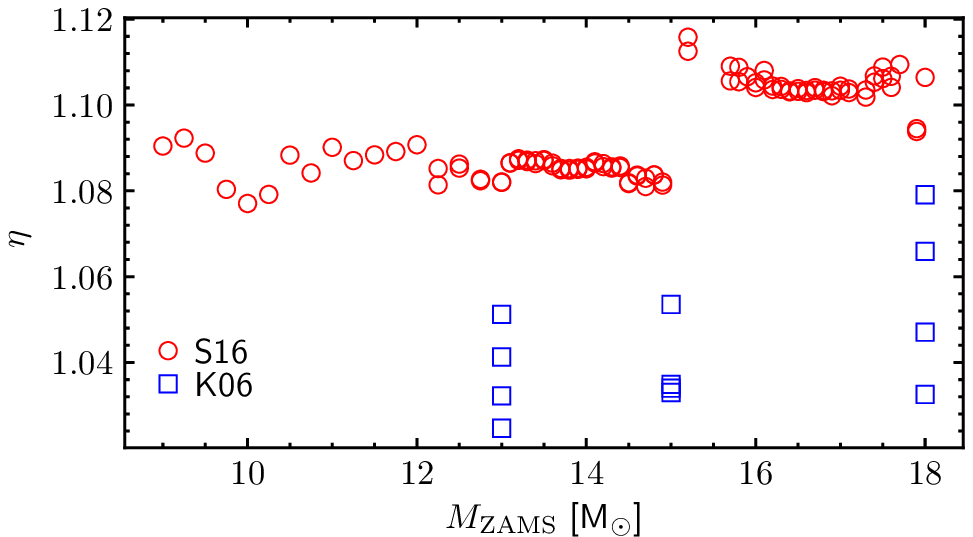}
\caption{$M_\mathrm{Fe}/M_{^{56}\mathrm{Fe}}$ against $\mzams$. Blue squares and red circles correspond to \citetalias{2006ApJ...653.1145K} and \citetalias{2016ApJ...821...38S} models, respectively.}
\label{fig:MFe_M56Fe}
\end{figure}

Fig.~\ref{fig:MFe_M56Fe} shows the $\eta$ values obtained with the \citetalias{2006ApJ...653.1145K} and \citetalias{2016ApJ...821...38S} models. Since we do not find any correlation with model parameters, in the figure we plot $\eta$ against $\mzams$. The $\eta$ values range between 1.03 and 1.11, where the \citetalias{2016ApJ...821...38S} values for ${\mzams\leq15\,\msun}$ are around 0.02 lower than those for ${\mzams>15\,\msun}$. The $\eta$ values computed with the \citetalias{2006ApJ...653.1145K} models are up to 0.06 lower than those from \citetalias{2016ApJ...821...38S} models. Since neither model is preferred, to be conservative we adopt ${\eta=1.07\pm0.04}$ (the mid-point between 1.03 and 1.11, with the error being half the range). This value, consistent with that reported in \citet{2008NewA...13..606B}, indicates that the measured $\mni$ accounts to about 93~per~cent of 
$M_\mathrm{Fe}$.

\subsection{Host galaxy distance moduli and colour excesses}\label{sec:mu_and_EhBV}

The error on $\log\mni$ is mainly dominated by uncertainties on $\mu$ and $\EhBV$ \citep[e.g.][]{2015ApJ...806..225P}. To reduce their errors, we measure $\mu$ and $\EhBV$ with various methods. 

\subsubsection{Distance moduli}\label{sec:mu}

To estimate $\mu$ for the SN host galaxies, we use distance moduli obtained with the Cepheids period-luminosity relation ($\mu_\mathrm{CPL}$), the Tip of the Red Giant Branch method ($\mu_\mathrm{TRGB}$), and the Tully-Fisher relation ($\mu_\mathrm{TF}$). We compile $\mu_\mathrm{CPL}$ and $\mu_\mathrm{TRGB}$ values from the literature, and $\mu_\mathrm{TF}$ from the Extragalactic Distance Database\footnote{\url{http://edd.ifa.hawaii.edu/}\label{edd_web}} \citep[EDD,][]{2009AJ....138..323T}. If a host galaxy does not have $\mu_\mathrm{CPL}$ nor $\mu_\mathrm{TRGB}$, then we include distances (1) computed with the Hubble-Lema\^{i}tre law ($D_\mathrm{HLL}$) using a local Hubble-Lema\^{i}tre constant ($H_0$) of $74.03\pm1.42$\,km\,s$^{-1}$\,Mpc$^{-1}$ \citep{2019ApJ...876...85R} and including a velocity dispersion of 382\,km\,s$^{-1}$\ to account for the effect of peculiar velocities over $D_\mathrm{HLL}$; and (2) from distance-velocity calculators based on smoothed velocity fields ($D_\mathrm{SVF}$) given by \citet{2017ApJ...850..207S} for ${D_\mathrm{SVF}<38}$\,Mpc and \citet{2019MNRAS.488.5438G} for ${D_\mathrm{SVF}<200}$\,Mpc. These calculators are available on the EDD website\textsuperscript{\ref{edd_web}} and described in \citet{2020AJ....159...67K}. Since the latter do not provide distance uncertainties, we adopt the typical distance error of the neighbouring galaxies as a conservative estimate, or a 15~per~cent error if the host galaxy is isolated (Ehsan Kourkchi, private communication). We convert $D_\mathrm{HLL}$ ($D_\mathrm{SVF}$) into $\mu_\mathrm{HLL}$ ($\mu_\mathrm{SVF}$) using the recipe provided in Appendix~\ref{sec:lognormal_mean}.

Table~\ref{table:mu_values} summarizes the aforementioned distance moduli. From this compilation, we adopt as $\mu$ the weighted average of $\mu_\mathrm{CPL}$, $\mu_\mathrm{TRGB}$, and $\mu_\mathrm{TF}$, if the first ones are available, otherwise we adopt the weighted average of $\mu_\mathrm{TF}$, $\mu_\mathrm{HLL}$, and $\mu_\mathrm{SVF}$. In the case of SN~2006my, whose host galaxy is within the Virgo Cluster, we include the distance modulus reported in \citet{2014MNRAS.442.3544F} based on the planetary nebula luminosity function. The $\mu$ values are in Column~5 of Table~\ref{table:SN_sample}. The typical $\mu$ error is of 0.18\,mag.

\subsubsection{Colour excesses}\label{sec:EhBV}

To calculate $\EhBV$ we use the following methods:
\begin{itemize}

\item[1.] The colour-colour curve (C3) method \citep{2014AJ....148..107R,2019MNRAS.483.5459R}. This technique assumes that, during the plateau phase, all normal SNe~II have similar linear $\vi$ versus $\bv$ C3s. Under this assumption, the $\EhBV$ value of an SN can be inferred from the vertical displacement of its observed C3 with respect to a reddening-free C3 (for a graphical representation, see Fig.~3 of \citealt{2014AJ....148..107R}). Using the C3 method (Appendix~\ref{sec:C3}), implemented in the \texttt{C3M} code\footnote{\url{https://github.com/olrodrig/C3M}}, we measure the colour excesses ($E_{\bv}^\mathrm{h,C3}$) of 71~SNe in our set. Those values are reported in Column~2 of Table~\ref{table:Eh_values}. The typical $E_{\bv}^\mathrm{h,C3}$ uncertainty is of 0.085\,mag.

\item[2.] The colour method \citep[e.g.][]{2010ApJ...715..833O}. This technique assumes that all normal SNe~II have the same intrinsic $\vi$ colour at the end of the plateau phase. \citet{2010ApJ...715..833O} defined this epoch as 30\,d before the middle of the $V$-band transition phase ($t_{\mathrm{PT},V}$, see Section~\ref{sec:MNi_vs_S}). The prescription provided by \citet{2010ApJ...715..833O} to compute colour excesses can be written as
\begin{equation}
E_{\bv}^{\mathrm{h},\vi}=0.812[(\vi)_{-30}-0.656],
\end{equation}
\begin{equation}
\sigma_{E_{\bv}^{\mathrm{h},\vi}}=0.812\sqrt{\sigma_{(\vi)_{-30}}^2+0.079^2}.
\end{equation} 
Here, $(\vi)_{-30}$ is the $\vi$ colour measured at $t_{\mathrm{PT},V}-30$\,d corrected for $\EGBV$ and $K$-correction (e.g. \citealt{2019MNRAS.483.5459R}). We compute $E_{\bv}^{\mathrm{h},\vi}$ values for 59~SNe in our sample. For SNe~2012aw and 2013am we adopt the values provided in the literature. Column~3 of Table~\ref{table:Eh_values} lists the $E_{\bv}^{\mathrm{h},\vi}$ values, which have a typical uncertainty of 0.074\,mag.

\item[3.] Spectrum-fitting method \citep[e.g.][]{2008ApJ...675..644D,2010ApJ...715..833O}. This technique consists on inferring the colour excess ($E_{\bv}^\mathrm{h,spec}$) of an SN from the comparison between its spectra and those of reddening-corrected SNe or spectral models. We compile $E_{\bv}^\mathrm{h,spec}$ values from the literature for 22~SNe in our set. We also compute $E_{\bv}^\mathrm{h,spec}$ for 36~SNe in our sample, using the prescription given in Appendix~\ref{sec:spec_EhBV}. The $E_{\bv}^\mathrm{h,spec}$ are collected in Column~4 of Table~\ref{table:Eh_values}. The typical $E_{\bv}^\mathrm{h,spec}$ uncertainty is of 0.091\,mag.

\end{itemize}

\begin{figure}
\includegraphics[width=1.0\columnwidth]{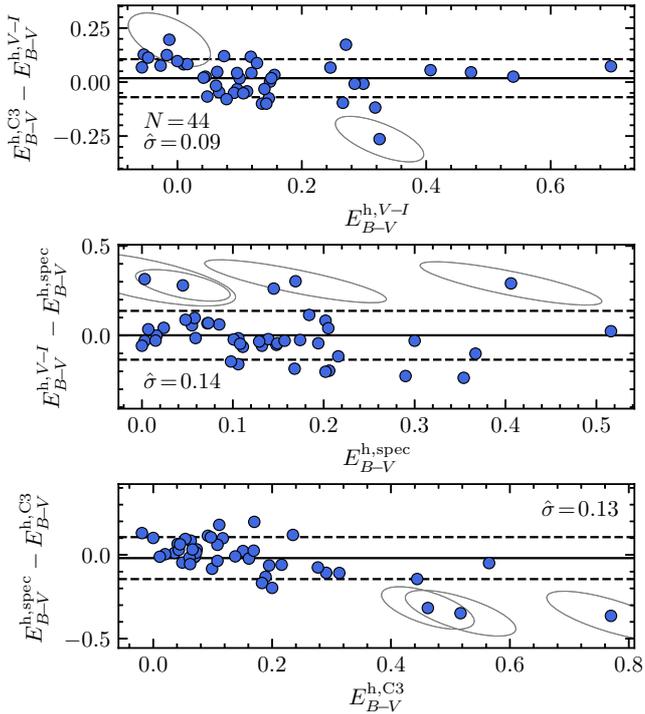}
\caption{Residuals about the one-to-one relation between $\EhBV$ values computed with the C3 method and the colour method (top panel), the colour method and the spectrum-fitting technique (middle panel), and the spectrum-fitting technique and the C3 method (bottom panel). Solid lines are mean values and dashed lines are $\pm1\,\ssd$ limits. Ellipses are $1\,\sigma$ confidence regions, which for clarity are drawn only for SNe outside $\pm2\,\ssd$ from the mean.}
\label{fig:Eh_comparison}
\end{figure}

Fig~\ref{fig:Eh_comparison} shows residuals about the one-to-one relation between $E_{\bv}^\mathrm{h,C3}$ and $E_{\bv}^{\mathrm{h},\vi}$ ($r^\mathrm{C3}_\vi$, top panel), $E_{\bv}^{\mathrm{h},\vi}$ and $E_{\bv}^\mathrm{h,spec}$ ($r^\vi_\mathrm{spec}$, middle panel), and between $E_{\bv}^\mathrm{h,spec}$ and $E_{\bv}^\mathrm{h,C3}$ ($r^\mathrm{spec}_\mathrm{C3}$, bottom panel) for the 44~SNe in our set having $E_{\bv}^\mathrm{h,C3}$, $E_{\bv}^{\mathrm{h},\vi}$, and $E_{\bv}^\mathrm{h,spec}$.
For $r^\mathrm{C3}_\vi$ we obtain a mean, $\ssd$, and typical error of 0.02, 0.09, and 0.10\,mag, respectively.
For $r^\vi_\mathrm{spec}$ we compute a mean, $\ssd$, and typical error of 0.00, 0.14, and 0.12\,mag, respectively.
For $r^\mathrm{spec}_\mathrm{C3}$ we calculate a mean, $\ssd$, and typical error of $-0.02$, 0.13, and 0.12\,mag, respectively.
Since the $\ssd$ values are quite similar to the typical residual errors, the observed dispersion is mainly due to colour excess errors. The mean offsets are statistically consistent with zero within $\pm1.3\ssd/\sqrt{N}$. Therefore, we do not detect systematic differences between the colour excesses inferred with the three aforementioned methods. 
Based on the latter, for the 77~SNe in our set having $E_{\bv}^\mathrm{h,C3}$, $E_{\bv}^{\mathrm{h},\vi}$, and/or $E_{\bv}^\mathrm{h,spec}$ estimates, we adopt the weighted mean of those values as $\EhBV$. For SNe~1980K, 2006my, 2008gz, 2014cx, and 2017it we obtain negative $\EhBV$ values (see Column~6 of Table~\ref{table:SN_sample}). The values of the first four objects are consistent with zero within $1.3\,\sigma$, while for SN~2017it the offset is of $-2.1\,\sigma$. Although negative $\EhBV$ values have no physical meaning, we keep those values as we do not have evidence to discard them.

For SNe in out sample without $\EhBV$ estimates, we evaluate to use colour excesses inferred from the pseudo-equivalent width of the host galaxy \ion{Na}{I}\,D absorption line ($\mathrm{pEW_{NaID}}$). We compile $\mathrm{pEW_{NaID}}$ values from the literature\footnote{If $\mathrm{pEW_{NaID}}$ is not reported but $\EhBV$ is, then we recover $\mathrm{pEW_{NaID}}$ using the corresponding $E_\bv(\mathrm{pEW_{NaID}})$ calibration.} for 89~SNe in our sample. With those values, we compute colour excesses ($E_{\bv}^\mathrm{h,NaID}$) using the relation of \citet{2012MNRAS.426.1465P}, and adopting a relative $E_{\bv}$ error of 68~per~cent \citep{2013ApJ...779...38P}. The $\mathrm{pEW_{NaID}}$ becomes insensitive to estimate the colour excess for $\mathrm{pEW_{NaID}}>0.1$\,nm \citep[e.g.][]{2013ApJ...779...38P}, equivalent to ${E_\bv>0.21\pm0.14}$\,mag in the \citet{2012MNRAS.426.1465P} relation. Therefore, we assume the latter lower limit for all SNe with $\mathrm{pEW_{NaID}}$ greater than 0.1\,nm. The $E_{\bv}^\mathrm{h,NaID}$ values are listed in Column~4 of Table~\ref{table:Eh_values}.

\begin{figure}
\includegraphics[width=1.0\columnwidth]{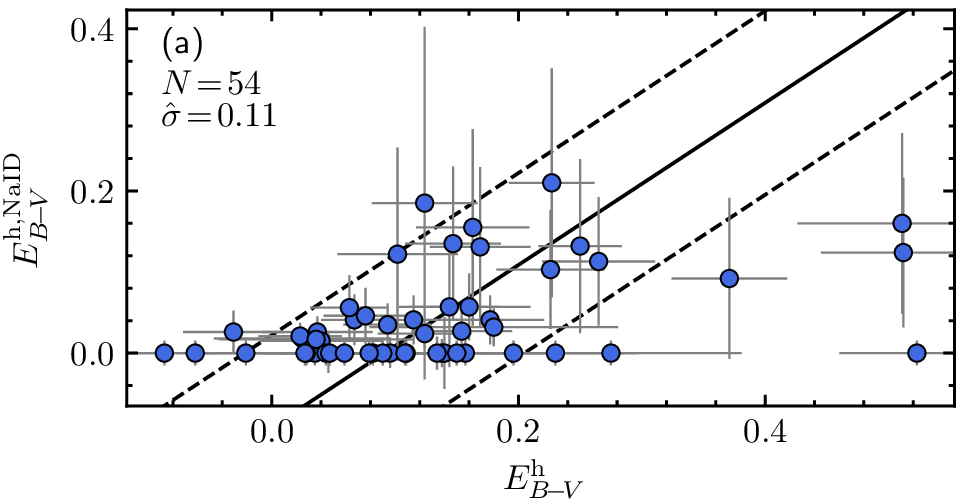}
\includegraphics[width=1.0\columnwidth]{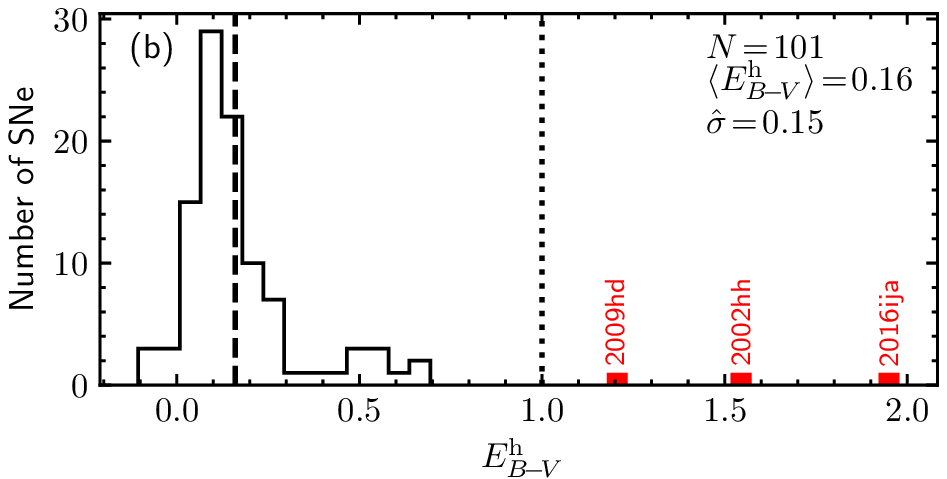}
\caption{Panel (a): colour excesses computed from $\mathrm{pEW_{NaID}}$ versus the $\EhBV$ values adopted in this work. The solid line is a straight line fit with a slope of unity, and dashed lines are the $\pm1\,\ssd$ limits around the fit. Error bars are $1\,\sigma$ errors. Panel (b): histogram of $\EhBV$. The dotted line is the Chauvenet upper rejection limit ($\EhBV=1$\,mag). The dashed line indicates the mean $\EhBV$ of the 101~SNe with $\EhBV<1$\,mag. Red bins are outliers.}
\label{fig:Eh_NaID}
\end{figure}

Fig~\ref{fig:Eh_NaID}(a) shows $E_{\bv}^\mathrm{h,NaID}$ against $\EhBV$ (54~SNe). Fitting a straight line with a slope of unity, we measure an offset and $\ssd$ value of $-0.09$ and 0.11\,mag, respectively. The offset is equivalent to $-6.0\,\ssd/\sqrt{N}$, which means that the $E_{\bv}^\mathrm{h,NaID}$ values are systematically lower than $\EhBV$. Therefore, for the 24~SNe in our sample without $\EhBV$ estimates but with $E_{\bv}^\mathrm{h,NaID}$ values we adopt ${\EhBV=E_{\bv}^\mathrm{h,NaID}+0.09}$\,mag, including in quadrature an error of 0.11\,mag.

For the highly reddened SNe~2002hh and 2016ija (both without $\EhBV$ and with ${E_{\bv}^\mathrm{h,NaID}>0.21}$\,mag) we adopt the $\EhBV$ values reported by \citet{2006MNRAS.368.1169P} and \citet{2018ApJ...853...62T}, respectively. In the case of SN~2002hh, the colour excess has two components: ${E_{\bv}=1.065}$\,mag (which includes $\EGBV$) with ${R_V=3.1}$, and ${E_{\bv}=1.545\pm0.182}$\,mag with ${R_V=1.1}$. For simplicity in the forthcoming analyses, we consider the first component as $\EGBV$ and the second one as $\EhBV$. 

Fig.~\ref{fig:Eh_NaID}(b) shows the histogram of $\EhBV$. To identify extreme values in the $\EhBV$ distribution, we use the \citet{1863mspa.book.....C} criterion. We find that SNe~2002hh, 2009hd, and 2016ija have $\EhBV$ values greater than the Chauvenet upper rejection limit ($\EhBV=1$\,mag), so we consider them as outliers. The $\EhBV$ distribution (removing the extreme values) has a mean and $\ssd$ of 0.16 and 0.15\,mag, respectively. For the six SNe in our set without $\EhBV$ (SNe~2004eg, PTF11go, PTF11htj, PTF11izt, PTF12grj, and LSQ13dpa) we adopt the mean and $\ssd$ of the latter distribution as $\EhBV$ and its error, respectively.

The adopted $\EhBV$ values are in Column~6 of Table~\ref{table:SN_sample}. The typical $\EhBV$ error is of 0.08\,mag.

\subsection{Explosion epochs}\label{sec:t0}

The SN explosion epoch is typically estimated as the midpoint between the last non-detection $t_\mathrm{ln}$ and the first SN detection $t_\mathrm{fd}$. In order to improve the $t_0$ estimates for the SNe in our set, we use the \texttt{SNII\_ETOS} code\footnote{\url{https://github.com/olrodrig/SNII_ETOS}} \citep{2019MNRAS.483.5459R}. The latter computes $t_0$ given a set of optical spectra as input and a uniform prior on $t_0$ provided by $t_\mathrm{ln}$ and $t_\mathrm{fd}$ (for more details, see \citealt{2019MNRAS.483.5459R}). If \texttt{SNII\_ETOS} is not able to compute $t_0$ for an SN, then we adopt the midpoint between $t_\mathrm{ln}$ and $t_\mathrm{fd}$ if ${t_\mathrm{fd}-t_\mathrm{ln}<20}$\,d, otherwise we use the $t_0$ value reported in the literature.

Table~\ref{table:t0_values} lists the $t_\mathrm{ln}$ (Column~2), $t_\mathrm{fd}$ (Column~3), and the adopted $t_0$ (Column~4) values (also reported in Column~7 of Table~\ref{table:SN_sample}) for our SN set. The typical $t_0$ error is of 3.8\,d.

\section{Analysis}\label{sec:analysis}

\subsection{Quasi-bolometric and bolometric corrections}\label{sec:BC}

\begin{figure}
\includegraphics[width=0.95\columnwidth]{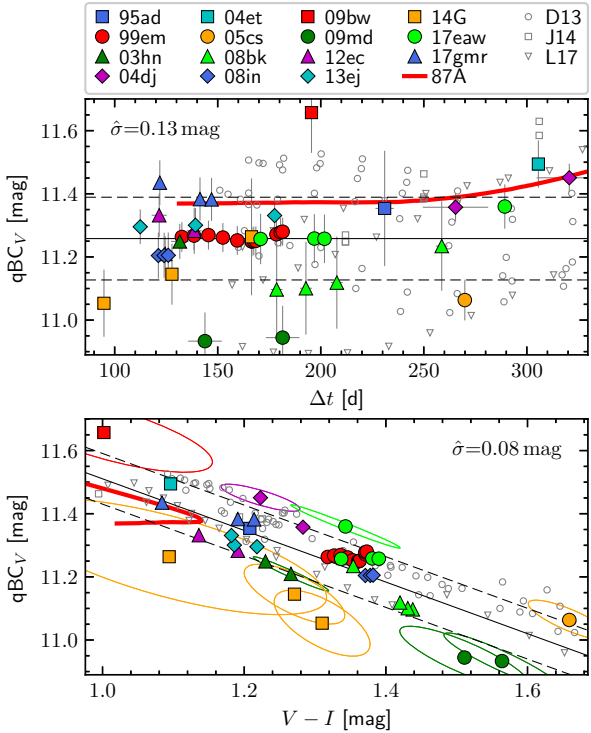}
\caption{q$\bc_V$ values for the SNe in the BC calibration set (filled symbols), the \citetalias{2013MNRAS.433.1745D}, \citetalias{2014MNRAS.439.3694J}, and \citetalias{2017MNRAS.466...34L} models (empty symbols), and the long-rising SN~1987A (red thick line), as a function of time since explosion (top panel) and $\vi$ colour (bottom panel). Solid lines are the best polynomial fit to the observed data, while dashed lines indicate the $\pm1\,\ssd$ limits around the fit. Error bars are $1\,\sigma$ errors and ellipses indicate $1\,\sigma$ confidence regions, which for clarity are drawn only for SNe outside the $\pm1\,\ssd$ limits.}
\label{fig:qBC_V}
\end{figure}

The top panel of Fig.~\ref{fig:qBC_V} shows the $\qbcv$ values for the 15~SNe in the BC calibration set against the time since explosion. For comparison we include the \citetalias{2013MNRAS.433.1745D}, \citetalias{2014MNRAS.439.3694J}, and \citetalias{2017MNRAS.466...34L} models, along with the long-rising SN~1987A, which is typically used to estimate $\bc_V$ for normal SNe~II \citep[e.g.][]{2001PhDT.......173H,2009ApJ...701..200B,2010MNRAS.404..981M}. In the figure we see that, except for SN~2014G, the four SNe with three or more points at ${\Delta t<250}$\,d and a time baseline greater than 30\,d (SNe~1999em, 2008bk, 2013ej, and 2017eaw) seems to be consistent with a constant $\qbcv$ value, as in the case of SN~1987A. For SNe~2017eaw and 2008bk we notice that the $\qbcv$ values at ${\Delta t>250}$\,d are $0.1${\textendash}$0.15$\,mag greater than the values at ${\Delta t<210}$\,d. This could be due to the effect of newly formed dust. We also notice differences in $\qbcv$ of around $0.3\text{\textendash}0.7$\,mag between the sub-luminous SNe~2005cs, 2008bk, 2009md \citep[e.g.][]{2014MNRAS.439.2873S} and the moderately-luminous SNe~2004et and 2009bw \citep[e.g.][]{2013AA...555A.142I}.

The bottom panel of Fig.~\ref{fig:qBC_V} shows $\qbcv$ versus $\vi$. We detect a correlation between both quantities, in the sense that the redder the SN the lower the $\qbcv$. This correlation is also displayed by models (empty symbols), which are consistent with the linear fit to the observations (solid line) within $\pm1\,\ssd$ (dashed lines). Since the errors in $\EhBV$, $\EGBV$, and $V$-band photometry affect the $\qbcv$ and $\vi$ values, the confidence region of each observation is an elongated ellipse. We see that the confidence regions are nearly oriented in the direction of the $\qbcv$ versus $\vi$ correlation. Therefore, the errors in $\EGBV$, $\EhBV$, and $V$-band photometry are not the main sources of the observed dispersion.

\begin{figure}
\includegraphics[width=0.95\columnwidth]{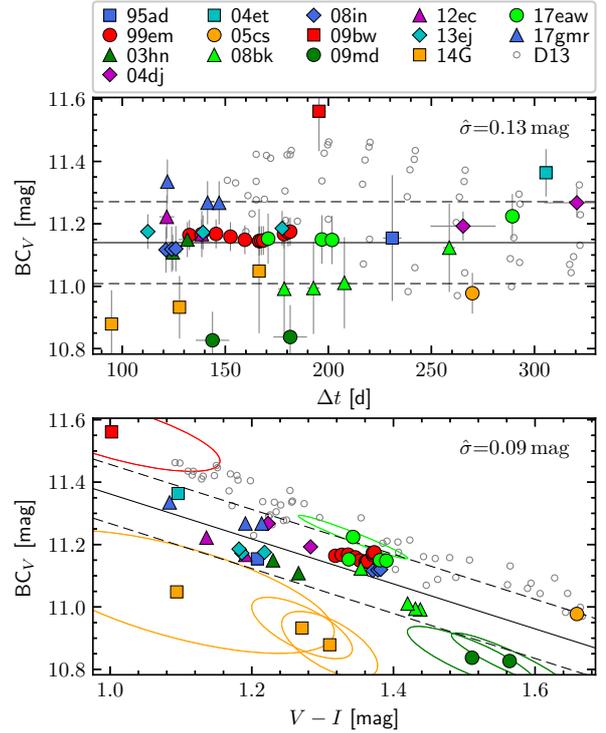}
\caption{$\bc_V$ values as a function of time since explosion (top panel) and $\vi$ (bottom panel). Symbols, lines, error bars, and ellipses have the same meaning than in Fig.~\ref{fig:qBC_V}.}
\label{fig:BC_V}
\end{figure}

Fig.~\ref{fig:BC_V} shows $\bc_V$ against the time since explosion (top panel) and $\vi$ (bottom panel). As we can see in the figure, the behaviour of $\bc_V$ is the same as that of $\qbcv$.

\subsection{BC calibration}\label{sec:BC_cal}

\subsubsection{$\bc_V$ versus $\vi$}\label{sec:BCV_vs_VI}

To calibrate the dependence of $\bc_V$ on $\vi$ displayed in the bottom panel of Fig.~\ref{fig:BC_V}, we use the expression
\begin{equation}\label{eq:BC_psi}
\bc_x=\zp_x^{\mathrm{BC}}+\Psi_x(X).
\end{equation}
Here, $\zp_x^{\mathrm{BC}}$ is the zero-point for the $x$-band BC calibration, and $\Psi_x$ is a polynomial function (without the zero-order term) representing the dependence of $\bc_x$ on the independent variable $X$ (in our case, ${X=\vi}$). To compute the polynomial parameters, we minimize
\begin{equation}
s^2=\sum_{\mathrm{SN}}\sum_i\left[\mathrm{BC}_{V,i}^{\mathrm{SN}} + \delta^{\mathrm{SN}}-\Psi_V\left((\vi)_i^{\mathrm{SN}}\right)\right]^2,
\end{equation}
where $\delta^{\mathrm{SN}}$ is an additive term to normalize the $\bc_V$ values of each SN to the same scale, and the polynomial order is determined with the model selection described in Appendix~\ref{sec:model_selection}.

\begin{figure}
\includegraphics[width=1.0\columnwidth]{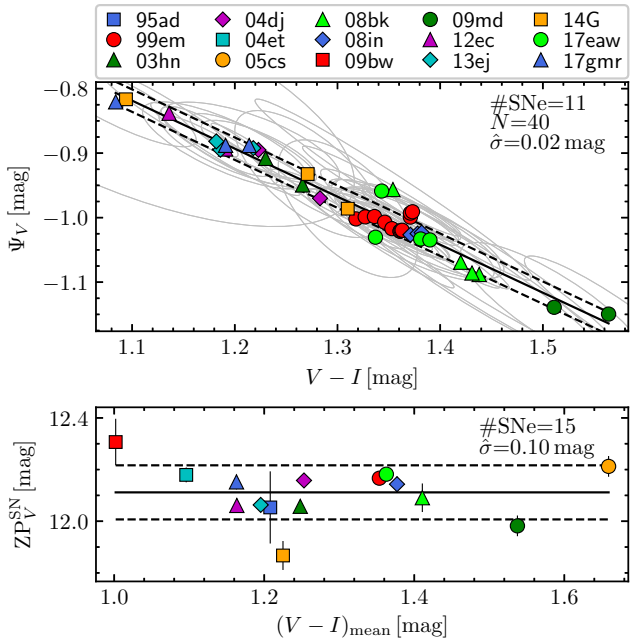}
\caption{Top panel: $\Psi_V$ against $\vi$, where the solid line is a linear fit and ellipses are the $1\,\sigma$ confidence regions. Bottom panel: $\zpsn_V$ as a function of the mean $\vi$ colour, where the solid line indicates the mean value. Dashed lines are the $\pm1\,\ssd$ limits.}
\label{fig:BC_V_cal}
\end{figure}

The top panel of Fig.~\ref{fig:BC_V_cal} shows the result of the aforementioned process, where we exclude SNe~1995ad, 2004et, 2005cs, and 2009bw because they have only one $\bc_V$ estimate. From this analysis we obtain that data are well represented by a straight line with slope ($\beta_V$) of ${-0.745\pm0.002}$.

To compute the $\zp_V^{\mathrm{BC}}$ value for each SN ($\zp_V^{\mathrm{SN}}$), we arrange equations~(\ref{eq:BC_psi}), (\ref{eq:BC}), and (\ref{eq:mcor}), obtaining
\begin{equation}\label{eq:ZP_SN}
\zp_V^{\mathrm{SN}}=\langle m_{\mathrm{bol},i}-(1+\beta_V)V_i+\beta_V I_i\rangle + (\EGBV+\EhBV)\xi,
\end{equation}
where ${\xi=(1+\beta_V)R_{\leff_V} -\beta_V R_{\leff_I}=2.01}$ (using the $R_{\leff_x}$ values provided in Table~\ref{table:zp_and_leff}). The angle brackets in equation~(\ref{eq:ZP_SN}) denote a weighted mean with weights
\begin{equation}
w_i=\left[\sigma_{m_{\mathrm{bol},i}}^2+(1+\beta_V)^2\sigma_{V_i}^2+\beta_V^2 \sigma_{I_i}^2\right]^{-1},
\end{equation}
where $\sigma_{m_{\mathrm{bol},i}}$ 
only includes the error on photometry. The random error for each $\zp_V^{\mathrm{SN}}$ value is given by
\begin{equation}\label{eq:eZP_SN}
\sigma_{\zp_V^{\mathrm{SN}}} =\left[\frac{1}{\sum_i w_i}+(\sigma_{\EGBV}^2+\sigma_{\EhBV}^2)\zeta^2\right]^{1/2},
\end{equation}
where ${\zeta=\xi-R_{\mathrm{p}}}$, being $R_{\mathrm{p}}$ the pSED total-to-selective extinction ratio (see Appendix~\ref{sec:pSED}). For our BC calibration set ${R_{\mathrm{p}}=1.68}$, so the $\EGBV$ and $\EhBV$ errors are scaled by 0.33. For example, an $\EhBV$ uncertainty of 0.08\,mag (the typical value for our sample) induces an error on $\zp_V^{\mathrm{SN}}$ of 0.03\,mag.

The bottom panel of Fig.~\ref{fig:BC_V_cal} shows the $\zpsn_V$ values for the SNe in the BC calibration set. To verify whether a residual correlation between $\zpsn_V$ and $\vi$ exists, in the figure we plot the $\zpsn_V$ values against mean $\vi$ colours. Using the model selection given in Appendix~\ref{sec:model_selection}, we find that data are consistent with $\zpsn_V$ being constant, meaning that all the dependence of $\zpsn_V$ on $\vi$ was captured by $\Psi_V$. We compute a mean and $\ssd$ value of 12.11 and 0.10\,mag, respectively. The typical $\zpsn_V$ error is about $0.03$\,mag, so the observed $\ssd$ value is mainly due to intrinsic differences between SNe. Therefore we adopt ${\zp_V^{\mathrm{BC}}=12.11\pm0.10}$\,mag.

\subsubsection{BCs for other combinations}\label{sec:BC_cal_others}

In addition to $\bc_V$ as a function of $\vi$, we also calibrate the dependence of $\bc_x$ for the $V\!r\!RiI$ bands as a function of different independent variables. In the BC calibration set there are only six SNe having $ri$ photometry in the radioactive tail (namely SNe~2008in, 2012ec, 2013ej, 2014G, 2017eaw, and 2017gmr). For the remaining nine SNe we convert the $RI$ magnitudes to $ri$ ones (see Appendix~\ref{sec:mag_transf}).

To calibrate the dependence of $\bc_x$ on a given $X$ variable, we perform the same analysis as in Section~\ref{sec:BCV_vs_VI}. As independent variables, we consider the ten colour indices that can be defined with the $V\!r\!RiI$ bands along with $\Delta t$. We find that the BC calibrations providing the lowest $\ssd$ values are $\bc_V$ as a function of $\vi$, $\bc_r$ as a function of $r\!-\!I$, $\bc_R$ as a function of $R\!-\!I$, $\bc_i$ as a function of $i\!-\!I$, and $\bc_I$ as a function of $\Delta t$. In all cases the dependence of $\bc_x$ on the $X$ variable is linear. The slopes of the linear relations ($\beta_x$) are reported in Column~3 of Table~\ref{table:BC_parameters}.

\begin{table}
\caption{BC calibrations.}
\label{table:BC_parameters}
\begin{tabular}{cccc}
\hline
$x$  & $X$                          & $\beta_x$           & $\zp_x^{\bc}$ (mag) \\ 
\hline
 $V$ & $\vi$                        & $-0.745\pm0.002$    & $12.11\pm0.10$ \\ 
 $r$ & $r\!-\!I$                    & $-0.837\pm0.004$    & $12.35\pm0.10$ \\
 $R$ & $R\!-\!I$                    & $-0.755\pm0.007$    & $12.35\pm0.10$ \\ 
 $i$ & $i\!-\!I$                    & $-0.963\pm0.005$    & $12.42\pm0.10$ \\ 
 $I$ & $\Delta t/(100\,\mathrm{d})$ & $\phs0.036\pm0.002$ & $12.37\pm0.10$ \\ 
 $V$ & --                           & --                  & $11.15\pm0.18$ \\
 $r$ & --                           & --                  & $11.89\pm0.16$ \\ 
 $R$ & --                           & --                  & $12.07\pm0.14$ \\
 $i$ & --                           & --                  & $11.91\pm0.12$ \\ 
 $I$ & --                           & --                  & $12.44\pm0.11$ \\ 
\hline
\multicolumn{4}{m{0.95\linewidth}}{\textit{Notes.} $\bc_x=\zp_x^{\bc}+\beta_x X$, valid for $\Delta t$ between 95 and 320\,d. $\zp_x^{\bc}$ errors do not include the uncertainty due to the $\alpha$ error.}
\end{tabular}
\end{table}

\begin{figure}
\includegraphics[width=1.0\columnwidth]{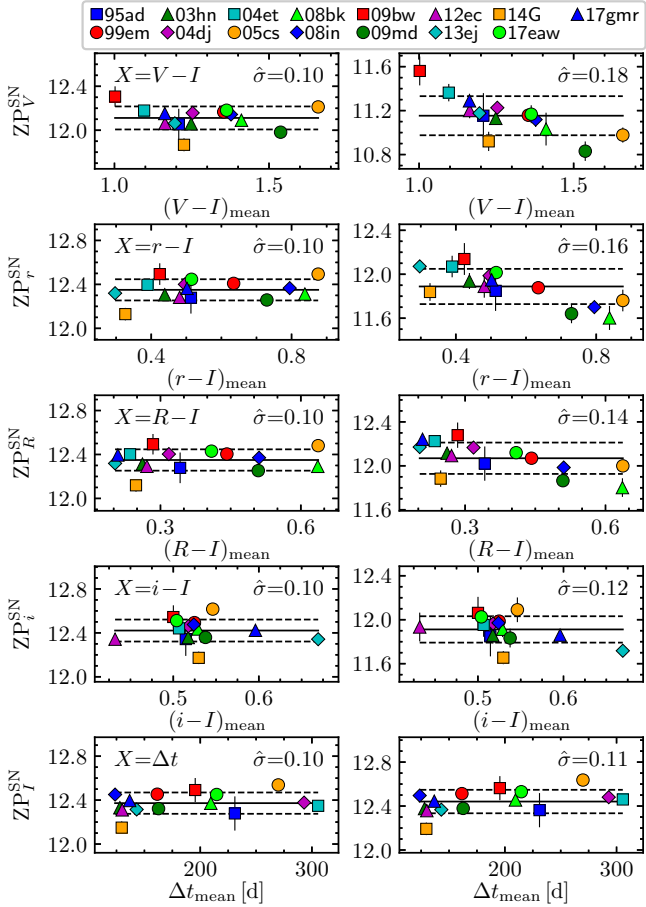}
\caption{$\zpsn_x$ versus the mean $X$ value. Left-hand panels: using the calibration with the lowest $\ssd$ value. Right-hand panels: assuming a constant BC. Solid and dashed lines indicate mean values and $\pm1\,\ssd$ limits around the mean, respectively.}
\label{fig:BCx_cal}
\end{figure}

The left-hand side of Fig.~\ref{fig:BCx_cal} shows the $\zpsn_x$ values for the aforementioned calibrations. For each one we find that the $\zpsn_x$ estimates are consistent with a constant value (using the model selection given in Appendix~\ref{sec:model_selection}). As for $\bc_V$ as a function of $\vi$, for $r\!RiI$ bands we adopt the mean and $\ssd$ value as $\zp_x^{\mathrm{BC}}$ and its error, respectively. Those values are listed in Column~4 of Table~\ref{table:BC_parameters}. Given the domain of the data (top panel of Fig.~\ref{fig:qBC_V}), our BC calibrations are valid for $\Delta t$ between 95 and 320\,d. Since the BC calibration set includes sub-luminous (SNe~2005cs, 2008bk, and 2009md) and moderately-luminous (e.g. SNe~2004et, 2009bw, and 2017gmr) SNe~II, we assume that our $\bc_x$ calibrations are valid for all normal SNe~II. 

We also compute $\bc_x$ calibrations assuming they are constant. The $\zp_x^{\mathrm{BC}}$ values are given in Column~4 of Table~\ref{table:BC_parameters}, while the $\zpsn_x$ estimates are shown in the right-hand side of Fig.~\ref{fig:BCx_cal}. Since $\bc_x$ is not actually constant but depends on a specific $X$ variable, there is a dependence of $\zpsn_x$ on the mean $X$ values. The latter, as we can see in right-hand side of Fig.~\ref{fig:BCx_cal}, is more evident for $V\!r\!R$ bands.

The BCs with the best precision are those including the $I$-band in the calibration (${\ssd=0.10}${\textendash}0.11\,mag), followed by $\bc_i$ as a constant (${\ssd=0.12}$\,mag). The latter means that, among the $V\!r\!RiI$ bands, the $I$- and $i$-band magnitudes are more correlated with the bolometric one. In order to compute luminosities through the BC technique (equation~\ref{eq:Lbol}), the $I$-band photometry along with $\bc_I$ as a function of $\Delta t$ must be preferred. If the $I$-band photometry is not available, then the $i$-, $R$-, $r$-, or $V$-band photometry along with the constant $\bc_x$ can be used to estimate luminosities. In the latter case, however, we caution that the derived luminosities of sub-luminous and moderately-luminous SNe~II will be systematically under and overestimated, respectively, specially for the $V\!r\!R$ bands.

\subsection{$^{56}$Ni mass distribution}\label{sec:MNi_distribution}

\subsubsection{$\log\mni$ estimates}\label{sec:MNi_estimates}

Armed with BCs for normal SNe~II in the radioactive tail, we compute $\log\mni$ using the recipe given in Appendix~\ref{sec:MNi_equation}. 

\begin{figure}
\includegraphics[width=1.0\columnwidth]{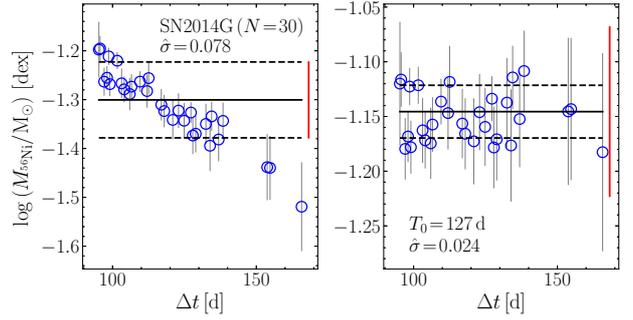}
\caption{$\log\mni$ estimates of SN~2014G versus the time since explosion, assuming the complete $\gamma$-ray trapping scenario (left-hand panel), and correcting for $f_{\mathrm{dep}}$ (right-hand panel). Solid horizontal lines correspond to the $\log\mni$ values that maximize the likelihood, and dashed lines are the $\pm1\,\ssd$ limits around them. Error bars are $1\,\sigma$ errors due to uncertainties on photometry, while red vertical bars depict the $1\,\sigma$ errors due to the uncertainty on $\mu$, colour excesses, $t_0$, and BC.}
\label{fig:2014G}
\end{figure}

For 25~SNe in our sample we find it is necessary to correct for the deposition function. As example, the left-hand side of Fig.~\ref{fig:2014G} shows the $\log\mni$ estimates of SN~2014G as a function of the time since explosion, assuming the complete $\gamma$-ray trapping scenario (i.e. $D_i=0$ in equation~\ref{eq:Qbol_vs_MNi}). As visible in the figure, the $\log\mni$ estimates are not concentrated around a constant but they decrease with time. This trend emerges because the ejecta becomes less able to thermalize $\gamma$-rays with time. In this case of $\gamma$-ray leakage, the inferred $\log\mni$ corresponds to a lower limit. The right-hand side of the figure shows the $\log\mni$ estimates corrected for $f_{\mathrm{dep}}$. We can see that, using $T_0=127$\,d, the systematic with time disappears and the $\log\mni$ estimates are consistent with a constant value.

\begin{figure}
\includegraphics[width=1.0\columnwidth]{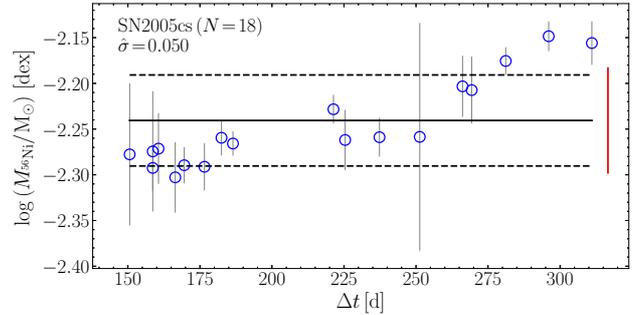}
\caption{$\log\mni$ estimates of SN~2005cs versus the time since explosion, assuming the complete $\gamma$-ray trapping scenario. Error bars and lines have the same meaning than in Fig.~\ref{fig:2014G}.}
\label{fig:2005cs}
\end{figure}

In our SN set, SNe~2004dj, 2005cs, 2006my, 2013am, and 2013bu have $\log\mni$ estimates that increase with time, which is shown in Fig.~\ref{fig:2005cs} for the case of SN~2005cs. This trend indicates that the observed luminosity increases with time relative to that expected from the radioactive decay. The latter suggests that (1) the SN ejecta during the early radioactive tail is not optically thin enough, so the energy deposited in the ejecta is not immediately emitted (i.e. ${L<Q_{\mathrm{dep}}}$); and/or (2) there is an additional source of energy (i.e. ${L>Q_{\mathrm{dep}}}$), whose relative contribution to the observed flux increases with time. For the latter scenarios, the higher and lower $\log\mni$ estimates are closer to the real value, respectively. In this work, in order to be conservative about the origin of the observed tendency, for the aforementioned five SNe we adopt the $\log\mni$ value obtained with ${D_i=0}$.

In the case of SNe~1988A, 2003iq, 2005dx, PTF10gva, 2010aj, LS13dpa, 2015cz, and 2016ija we obtain ${D_i=0}$ because the characteristics of their photometry (number of data, photometry errors, and time baseline) are not good enough to detect departures from a constant $\log\mni$ value. In our sample 30 out of 102~SNe are consistent with ${D_i\neq0}$. Therefore, we expect only two out of the aforementioned SNe to have ${D_i\neq0}$, which should not impact our results.

Table~\ref{table:Ni_masses} lists the derived $\log\mni$ and $\mni$ values (Columns~6 and 7, respectively), along with the bands and numbers of photometric points used to compute $\log\mni$ (Columns~2 and 4, respectively). The $T_0$ values of the 24~SNe corrected for the deposition function are listed in Column~5.

Table~\ref{table:error_budget} shows the error budget for the $\log\mni$ estimates computed with the $I$-band photometry (the preferred one), adopting the typical errors in our SN sample. The uncertainty on $\mu$ dominates the error budget, accounting for about 50~per~cent of the total error. The $\zp_I^{\mathrm{BC}}$ error, on the other hand, is the main source of systematic uncertainty. Errors in photometry, $t_0$, and $\EGBV$ induce only $\sim$3~per~cent of the total $\log\mni$ error. The typical error of the measured $\log\mni$ is of 0.102\,dex ($\mni$ error of 24~per~cent).

\begin{table}
\caption{Error budget for the $\log\mni$ estimates.}
\label{table:error_budget}
\begin{tabular}{llccc}
\hline
Error & Error       & Typical & Error in      & \% of total \\
type  & source      & error   & $\log\mni$ & error       \\
      &             &         & (dex)         &             \\
\hline
 Random     & $\mu$    & 0.18\,mag  & 0.072     &  50.4 \\
            & $\EhBV$  & 0.08\,mag  & 0.054     &  28.0 \\
            & $t_0$    & 3.8\,d     & 0.013     &   1.6 \\
            & $m_I$    & 0.05\,mag  & 0.012$^*$ &   1.3 \\
            & $\EGBV$  & 0.01\,mag  & 0.007     &   0.4 \\
            & All      &            & 0.092     &  81.7 \\
 Systematic & $\zp_I^{\bc}$ & 0.10\,mag  & 0.040     &  15.5 \\
            & $\alpha$      & 3.9\,\%    & 0.017     &   2.8 \\
            & All           &            & 0.043     &  18.3 \\
 Total      &               &            & 0.102     & 100.0 \\
\hline
\multicolumn{5}{l}{$^*$Considering three photometric points.}
\end{tabular}
\end{table}

Similar to the $I$-band, the $\log\mni$ error budgets for the $V\!r\!Ri$ bands are dominated by uncertainties on $\mu$ and $\zp_x^{\mathrm{BC}}$. Errors in photometry, $t_0$, and $\EGBV$ induce about 2{\textendash}3~per~cent of the total $\log\mni$ error, while the typical $\log\mni$ errors are of 0.143, 0.128, 0.121, and 0.109\,dex for the $V\!r\!Ri$ bands, respectively.

\subsubsection{Outliers}\label{sec:MNi_vs_Mv50d}

\begin{figure}
\includegraphics[width=1.0\columnwidth]{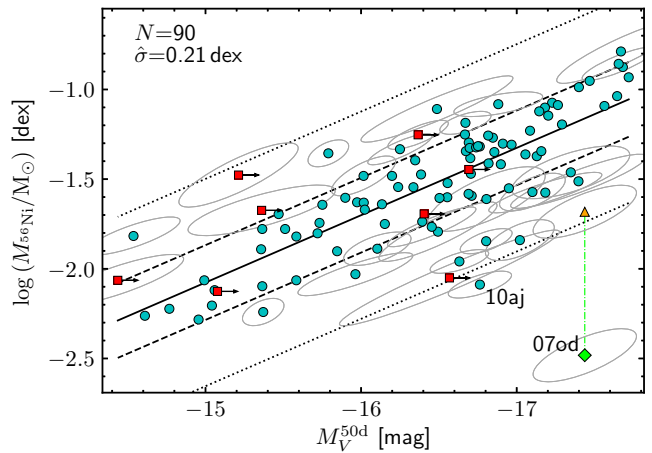}
\caption{$\log\mni$ against $M_V^{50\mathrm{d}}$. The solid line is a straight-line fit to the data (cyan circles, 90~SNe), dashed lines indicate the $\pm1\,\ssd$ limits around the fit, while dotted lines are the Chauvenet rejection limits. SNe with upper limits on $M_V^{50\mathrm{d}}$ are indicated as red squares. The green diamond corresponds to SNe~2007od, while the orange triangle is SN~2007od with a $\log\mni$ correction of 0.8\,dex. Ellipses indicate $1\,\sigma$ confidence regions, which for clarity are drawn only for SNe outside the $\pm1\,\ssd$ limits.}
\label{fig:MNi_vs_MV50d}
\end{figure}

Fig.~\ref{fig:MNi_vs_MV50d} shows $\log\mni$ versus the absolute $V$-band magnitude at 50\,d since the explosion ($M_V^{50\mathrm{d}}$, listed in Column~8 of Table~\ref{table:Ni_masses}). As reported by \citet{2003ApJ...582..905H} and other authors (e.g. \citealt{2014MNRAS.439.2873S}, \citealt{2015ApJ...799..215P,2015ApJ...806..225P}, \citealt{2016MNRAS.459.3939V}, \citealt{2017ApJ...841..127M}, \citealt{2019ApJ...882...68S}), we see a correlation between both quantities. As noted by \citet{2015ApJ...799..215P}, SN~2007od is an outlier in the $\log\mni$ versus $M_V^{50\mathrm{d}}$ distribution. The latter is a consequence of the increase in extinction due to the newly formed dust during the radioactive tail \citep{2010ApJ...715..541A,2011MNRAS.417..261I}.

Using the model selection procedure (Appendix~\ref{sec:model_selection}), we find that the correlation between $\log\mni$ and $M_V^{50\mathrm{d}}$ can be represented by the straight line
\begin{equation}\label{eq:MNi_vs_Mv50d}
\log\mni=-7.708(\pm0.445)-0.3755(\pm0.0271)M_V^{50\mathrm{d}},
\end{equation}
where the parameter errors (in parentheses) are obtained performing $10^4$ bootstrap resamplings. 

To identify outliers other than SN~2007od in the $\log\mni$ versus $M_V^{50\mathrm{d}}$ distribution, we use the Chauvenet's criterion. In Fig.~\ref{fig:MNi_vs_MV50d} we see that SN~2010aj is located below the Chauvenet lower rejection limit but consistent with it within $2\,\sigma$, which means that we cannot confirm that SN as an outlier. SN~2010aj was presented in \citet{2013AA...555A.142I}, which suggested that it may be affected by newly formed dust. In that work, however, the lack of further evidence did not allow confirmation of the above scenario. 

In the case of SNe~2007od we consider its $\log\mni$ values as lower limits. Indeed, based on the $\mni$ values reported by \citet{2011MNRAS.417..261I}, the inclusion of the IR light excess to the luminosity of SN~2007od increases its $\log\mni$ in $\sim$0.8\,dex. Applying this correction, SN~2007od moves in Fig.~\ref{fig:MNi_vs_MV50d} from $-6.3$ to $-2.5\,\ssd$ below the fit, becoming consistent with the $\log\mni$ versus $M_V^{50\mathrm{d}}$ distribution.

\subsubsection{Sample completeness}\label{sec:completeness}

The SNe in our set were selected from the literature by having at least three photometric points in the radioactive tail, so our sample is potentially affected by the selection bias. In order to correct for the latter bias and construct an SN sample as complete as possible, we use as reference the volume-limited SN sample of \citet{2017PASP..129e4201S}. Most of the SNe~II in that sample have completeness $\gtrsim$95~per~cent at the cut-off distance of 38\,Mpc ($\mu=32.9$\,mag), while sub-luminous and highly reddened SNe have completeness $\gtrsim$70~per~cent (see Fig.~4 of \citealt{2011MNRAS.412.1441L}). Therefore, we assume that the set of normal SNe~II at $\mu<32.9$ in the \citet{2017PASP..129e4201S} sample is roughly complete.

From the \citet{2017PASP..129e4201S} sample we select the 28 normal SNe~II with $\mu<32.9$\,mag (hereafter the RC set), where we use $\mu$ values computed with the procedure described in Section~\ref{sec:mu} (reported in Column~8 of Table~\ref{table:mu_values}). We recalibrate their absolute $R$ magnitudes ($M_R$) at maximum ($M_R^{\max}$, listed in \citealt{2011MNRAS.412.1441L}) using our $\mu$ values, and correcting for $\EhBV$ (estimated with the procedure described in Section~\ref{sec:EhBV}, and reported in Column~6 of Table~\ref{table:Eh_values}). We also replace the \citet{1998ApJ...500..525S} $\EGBV$ values used in \citet{2011MNRAS.412.1441L} by the new ones provided by \citet{2011ApJ...737..103S}. Table~\ref{table:MRmax} lists the $M_R^{\max}$ estimates for the RC sample.

\begin{figure}
\includegraphics[width=1.0\columnwidth]{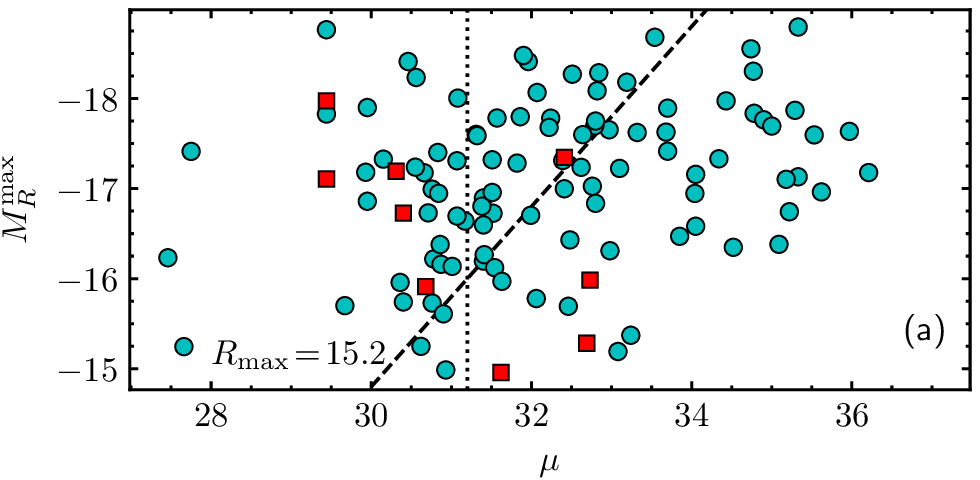}
\includegraphics[width=1.0\columnwidth]{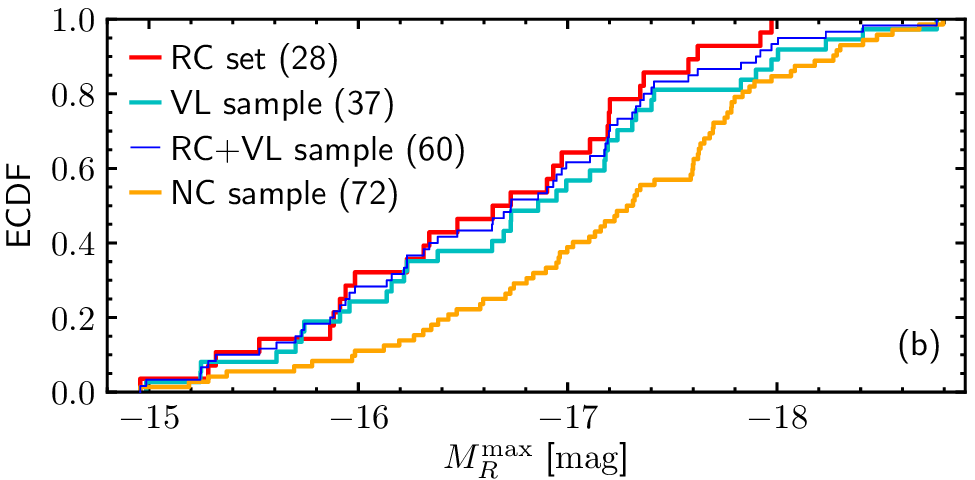}
\caption{Panel (a): $M_R^{\max}$ against $\mu$ for the SNe in our sample, where red squares are the SNe in common with the RC set. The dashed line corresponds to ${R_{\max}}=15.2$\,mag, while the dotted line (${\mu=31.2}$) is the limit for our VL sample. Panel (b): ECDF for the $M_R^{\max}$ values in the RC (red line), VL (cyan line), and RC+VL (thin blue line) samples, and for the $M_R^{\max}$ values of the SNe in our sample with ${\mu>31.2}$ (orange line).}
\label{fig:loss_edf}
\end{figure}

Among the 109~SNe in our sample, nine have $M_R^{\max}$ estimates provided in the RC set, so we adopt those values for consistency. Out of the remaining 100~SNe:
\begin{itemize}
\item[1.] Forty-seven SNe have $R$- or $r$-band light curves during the maximum light, where $r$ magnitudes are converted into $R$ ones using ${r\!-\!R=0.12}$ (see Appendix~\ref{sec:mag_transf}). We measure $M_R^{\max}$ performing an \texttt{ALR} fit to the maximum photometry or adopting the brightest $M_R$ value ($M_R^\mathrm{br}$) as $M_R^{\max}$. The average of the rise time ($t_R^\mathrm{rise}$) for the latter SNe is of $15\pm6$\,d ($1\,\ssd$ error), while the mean $V\!-\!R$ and $R\!-\!I$ colours at time $t_R^\mathrm{rise}$ are of $0.15\pm0.08$ and $0.04\pm0.07$\,mag, respectively.
\item[2.] Thirty SNe have $R/r$ light curves where the maximum light cannot be determined; six SNe have $V\!I$ photometry, which we convert into $R$ magnitudes using ${V\!-\!R=0.05+0.59(\vi)}$ (see Appendix~\ref{sec:mag_transf}); and eight SNe (one SN) only have $V$-band ($I$-band) photometry, which we convert into $R$ magnitudes using $V\!-\!R=0.15$ ($R\!-\!I=0.04$). For each of these 45~SNe we fit a straight line to the light curve, compute $M_R$ at $t_R^\mathrm{rise}$ ($M_R(t_R^\mathrm{rise})$), and adopt the minimum between $M_R^\mathrm{br}$ and $M_R(t_R^\mathrm{rise})$ as $M_R^{\max}$.
\item[3.] Eight SNe have photometry starting at ${\Delta t>80}$\,d. Two of them (SNe~1997D and 2004eg) are sub-luminous SNe~II, which tend to have flat light curves during the photospheric phase \citep[e.g.][]{2014MNRAS.439.2873S}, so we adopt $M_R^\mathrm{br}$ as $M_R^{\max}$. For the remaining six SNe we estimate $M_R^\mathrm{max}$ using their $\log\mni$ values and $M_R^{\max}=-20.01-1.86\log\mni$, obtained from the 47~SNe with well-defined maximum light.
\end{itemize}

The $M_R^{\max}$ values for the SNe in our sample are reported in Table~\ref{table:MRmax} and plotted in Fig.~\ref{fig:loss_edf}(a) against $\mu$. The mean apparent $R$-band magnitude at maximum ($R_{\max}$, corrected for reddening) is of 15.2\,mag, which is indicated as a dashed line. As we move to greater $R_{\max}$ values (right-hand side of the dashed line) we see a decrement in the number of SNe, which is due to (1) bright SNe (in apparent magnitude) are more likely to be selected for photometric monitoring in the radioactive tail than faint ones; and (2) faint SNe have in general less photometric points in the radioactive tail than bright ones, so they are more likely not to meet our selection criterion of having at least three photometric points. In order to minimize the effect of the selection bias, we construct a volume-limited (VL) sample with the 37~SNe at $\mu\leq31.2$ such that the selection bias could be relevant only in the small region between $R_{\max}>15.2$ and $\mu\leq31.2$.

Fig.~\ref{fig:loss_edf}(b) shows the empirical cumulative distribution function (ECDF) for the $M_R^{\max}$ values in the VL (cyan line) and the RC (red line) samples. To test whether both $M_R^{\max}$ samples are drawn from a common unspecified distribution (the null hypothesis), we use the two-sample Anderson-Darling (AD) test \citep[e.g.][]{Scholz_Stephen1987}. We obtain a standardized test statistic ($T_\mathrm{AD}$) of $-0.65$ with a $p$-value of 0.74, meaning that the null hypothesis cannot be rejected at a significance level $>$74~per~cent. Since the $M_R^{\max}$ values of the RC and the VL samples are likely drawn from the same $M_R^{\max}$ distribution, we can assume that the completeness of both samples is quite similar, so we combine them into a single data set (RC+VL). The $M_R^{\max}$ ECDF of the RC+VL sample (blue thin line) has a minimum, maximum, mean and $\ssd$ of $-18.8$, $-15.0$, $-16.7$, and $0.9$\,mag, respectively.

Fig.~\ref{fig:loss_edf}(b) also shows the ECDF for the $M_R^{\max}$ values of the 72~SNe at $\mu>31.2$ (orange line), which we refer as the non-complete (NC) sample. Using the two-sample AD test to test the null hypothesis for the $M_R^{\max}$ values of the RC+VL and the NC samples, we obtain a $T_\mathrm{AD}$ of $5.66$ and a $p$-value of 0.002. Thus, the null hypothesis can be rejected at a significance level of 0.2~per~cent, which is expected since the NC sample is affected by the selection bias.

In order to roughly quantify the number and magnitudes of the SNe missing from the NC sample, and therefore from our full sample, we proceed as follows. First, we divide the $M_R^{\max}$ distribution into four bins of width 1\,mag (Column~1 of Table~\ref{table:completeness}), and register the number of SNe within each bin for the NC and the RC+VL samples (Columns~2 and 3, respectively). Then, since bright SNe are less affected by the selection bias, we scale the number of SNe in the RC+VL sample by a factor of 15/8 (Column~4) in order to match the number of SNe with $M_R^{\max}<-17.8$ to that of the NC sample. In other words, the numbers in Column~4 are the SNe we would expect for a roughly complete sample with 15~SNe at $M_R^{\max}<-17.8$. The number of expected SNe minus the observed ones (i.e. the NC sample) is listed in Column~5. Thus, to correct our full SN sample for selection bias, we have to include 3, 23, and 15~SNe of magnitude ${-17.8\leq M_R^{\max}<-16.8}$, ${-16.8\leq M_R^{\max}<-15.8}$, and ${M_R^{\max}\geq-15.8}$, respectively. The latter SNe can be randomly selected from the VL or the RC+VL sample within the corresponding $M_R^{\max}$ bins.

\begin{table}
\caption{Histogram for $M_R^{\max}$.}
\label{table:completeness}
\begin{tabular}{lcccc}
\hline
$M_R^{\max}$ range & NC  &RC+VL& Expected & Missing \\
\hline
$-17.8,\,-18.8$  & 15   &  8 &  15 & 0  \\
$-16.8,\,-17.8$  & 36   & 21 &  39 & 3  \\
$-15.8,\,-16.8$  & 15   & 20 &  38 & 23 \\
$-14.8,\,-15.8$  &  6   & 11 &  21 & 15 \\
\hline
\multicolumn{5}{l}{\textit{Notes}: $\mathrm{Expected}=15/8\mathrm{(RC+VL)}$. $\mathrm{Missing}=\mathrm{Expected}\!-\!\mathrm{NC}$.}
\end{tabular}
\end{table}

\subsubsection{Mean $^{56}$Ni mass}\label{sec:mean_nickel_mass}

\begin{figure}
\includegraphics[width=1.0\columnwidth]{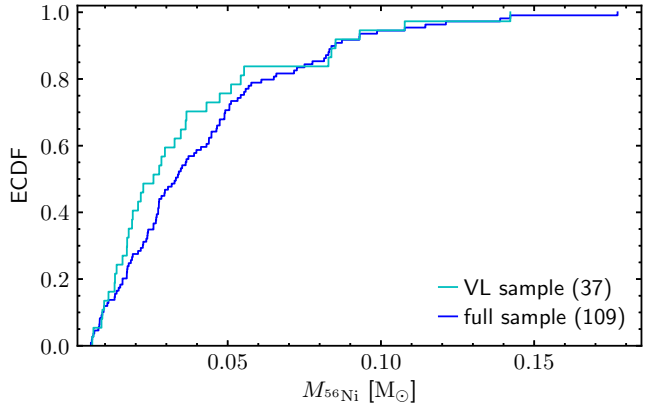}
\caption{ECDF for the $^{56}$Ni masses in the VL (cyan line) and the full sample (blue line).}
\label{fig:cdf}
\end{figure}

Fig.~\ref{fig:cdf} shows the ECDFs for the $\mni$ values in the VL (cyan line) and the full (blue line) samples. The ECDF of the VL sample has a $\langle\mni\rangle$ and $\ssd$ of 0.037 and 0.032\,$\msun$, respectively, while the random error (ran) on the mean ($\ssd/\sqrt{N}$) is of $0.005\,\msun$. On the other hand, the $\mni$ distribution of the full sample has a minimum, maximum, mean, and $\ssd$ of 0.005, 0.177, 0.042, and $0.033\,\msun$, respectively,  with $\ssd/\sqrt{N}=0.003\,\msun$. The latter mean value corresponds to $\langle\mni\rangle$ uncorrected for selection bias ($\langle\mni\rangle^{\mathrm{unc}}$) which, as expected, is greater than the $\langle\mni\rangle$ estimate for the VL sample. We note that for models based on the neutrino-heating mechanism (the generally accepted one for CC~SNe, e.g. \citealt{2021Natur.589...29B}) the upper limit for the synthesized $\mni$ is around 0.15 and 0.23\,$\msun$ \citep[e.g.][]{2012ApJ...757...69U,2019MNRAS.483.3607S}, which is consistent with the maximum $\mni$ of our full SN sample.

The random error on $\langle\mni\rangle^{\mathrm{unc}}$ is made up of the sampling error along with the uncertainties induced by errors in $\mu$ and $\EhBV$. To estimate the random error in $\langle\mni\rangle^{\mathrm{unc}}$ induced by uncertainties in $\mu$, we perform $10^5$ simulations varying randomly $\mu$ according to its errors (assumed normal). For each realization, we rescale the $\log\mni$ values of the SNe in our full sample using the simulated $\mu$ values and calculate the mean $^{56}$Ni mass. Using those $10^5$ simulated mean values we compute a $\ssd$ around $\langle\mni\rangle^{\mathrm{unc}}$ of $0.0009\,\msun$, which we adopt as the error induced by uncertainties in $\mu$. We repeat the same process for $\EhBV$, obtaining $0.0011\,\msun$. The random error on $\langle\mni\rangle^{\mathrm{unc}}$ is $0.0028\,\msun$ greater in quadrature than the error induced by $\mu$ and $\EhBV$. We adopt the latter value as the sampling error.

As mentioned in Section~\ref{sec:completeness}, to correct our full SN sample for selection bias we have to include 41~SNe. The mean $^{56}$Ni mass of the selection-bias-corrected sample can be written as
\begin{equation}
\langle\mni\rangle = \langle\mni\rangle^{\mathrm{unc}} -\mathrm{sbc}.
\end{equation}
Here,
\begin{equation}
\mathrm{sbc} = \frac{41}{109+41}(\langle\mni\rangle^{\mathrm{unc}}-\langle\mni\rangle_{41}^{\mathrm{unc}})
\end{equation}
is the selection bias correction, where $\langle\mni\rangle_{41}^{\mathrm{unc}}$ is the mean $^{56}$Ni mass computed with the 41~SNe that we have to add to our full SN sample. Performing $10^5$ simulations, where the missing SNe (Column~5 of Table~\ref{table:completeness}) are randomly selected from the VL sample within the corresponding $M_R^{\max}$ bins, we obtain a sbc of $0.005\pm0.001\,\msun$. Therefore, our best estimate of $\langle\mni\rangle$ for normal SNe~II is of $0.037\pm0.003\,\mathrm{(ran)}\,\msun$, with a systematic error due to the uncertainty on $\zp^{\bc}$ and $\alpha$ of $0.004\,\msun$. This result compares to the $\langle\mni\rangle$ value of $0.037\pm0.005\,\mathrm{(ran)}\,\msun$ obtained with the VL sample.

Table~\ref{table:Fe_yield_error_budget} summarizes the error budget for $\langle\mni\rangle$. The $\zp_I^{\mathrm{BC}}$ error, accounting for 46~per~cent of the total uncertainty, dominates the error budget. The sampling error, which is the main source of random uncertainty, accounts for 31~per~cent of the total error.

\begin{table}
\caption{Error budget for the mean $^{56}$Ni mass.}
\label{table:Fe_yield_error_budget}
\begin{tabular}{llccc}
\hline
Error type  & Error       & Typical         &Error in              & \% of total \\
            & source      & error           & $\langle\mni\rangle$ & error       \\
            &             &                 & ($\msun$)            &             \\
\hline
 Random     & Sampling    & 0.0028\,$\msun$ & 0.0028               &   31.0      \\
            & sbc         & 0.0014\,$\msun$ & 0.0014               &    7.7      \\
            & $\EhBV$     & 0.08\,mag       & 0.0011               &    4.8      \\
            & $\mu$       & 0.18\,mag       & 0.0009               &    3.2      \\
            & All         &                 & 0.00344              &   46.7      \\
 Systematic & $\zp^{\bc}$ & 0.10\,mag       & 0.0034               &   45.6      \\
            & $\alpha$    & 3.9\,\%         & 0.0014               &    7.7      \\
            & All         &                 & 0.00368              &   53.3      \\
 Total      &             &                 & 0.00504               &  100.0      \\
\hline
\end{tabular}
\end{table}

\subsection{Mean iron yield}\label{sec:mean_iron_yield}

With our $\langle\mni\rangle$ measurement along with equation~(\ref{eq:eta}) and ${\eta=1.07\pm0.04}$ (see Section~\ref{sec:iron_mass}), we obtain a $\yfe$ value of ${0.040\pm0.005\,\msun}$ for normal SNe~II. 

In addition, we evaluate $\yfe$ for CC~SNe employing recent estimations of mean $^{56}$Ni masses for other CC~SN subtypes. The mean $^{56}$Ni mass for CC~SNe, using the SN rates provided in \citet{2017PASP..129e4201S}, is given by
\begin{equation}\label{eq:yNiCC}
f_\mathrm{CC}^\mathrm{Ni}=0.696f^\mathrm{Ni}_\mathrm{II}+0.304f^\mathrm{Ni}_\mathrm{SE}.
\end{equation}
Here,
\begin{equation}\label{eq:yNiII}
f_\mathrm{II}^\mathrm{Ni}=0.891f^\mathrm{Ni}_\text{II-normal}+0.067f^\mathrm{Ni}_\mathrm{IIn}+0.042f^\mathrm{Ni}_\text{long-rising}
\end{equation}
and
\begin{equation}\label{eq:yNiSE}
f_\mathrm{SE}^\mathrm{Ni}=0.360f_\mathrm{IIb}^\mathrm{Ni}+0.356f_\mathrm{Ib}^\mathrm{Ni}+0.247f_\mathrm{Ic}^\mathrm{Ni}+0.037f_\text{Ic-BL}^\mathrm{Ni},
\end{equation}
where $f^\mathrm{Ni}$ denotes the mean $^{56}$Ni mass for the subscripted CC~SN types and subtypes.



In equation~(\ref{eq:yNiII}) we adopt ${f^\mathrm{Ni}_\text{long-rising}=0.086\,\msun}$ \citep{2019AA...628A...7A} and for SNe~IIn we assume ${f^\mathrm{Ni}_\mathrm{IIn}=f^\mathrm{Ni}_\text{II-normal}}$, thus obtaining ${f^\mathrm{Ni}_\mathrm{II}=0.039\,\msun}$.

For SNe~IIb, Ib, Ic and Ic-BL we adopt the mean $\mni$ values from the compilation of \citet{2019AA...628A...7A}: 0.124, 0.199, 0.198, and 0.507\,$\msun$, respectively. The $^{56}$Ni masses of the SE~SNe compiled by \citet{2019AA...628A...7A} were mainly computed with the \citet{1982ApJ...253..785A} rule, which overestimates the $^{56}$Ni mass of SE~SNe by $\sim$50~per~cent \citep{2015MNRAS.453.2189D,2016MNRAS.458.1618D}. Including this correction to the mean $\mni$ values of SE~SNe, with equation~(\ref{eq:yNiSE}) we get ${f^\mathrm{Ni}_\mathrm{SE}=0.122\,\msun}$. Recently, by using the radioactive tail luminosity, \citet{2020arXiv200906683A} estimated mean $\mni$ values of 0.06, 0.11, 0.20, and $0.15\,\msun$ for SNe~IIb, Ib, Ic, and Ic-BL, respectively\footnote{Mean values were computed with $\leq8$~SNe per subtype, so we caution that those values are not statistically significant.}. Inserting these values in  equation~(\ref{eq:yNiII}) we get ${f^\mathrm{Ni}_\mathrm{SE}=0.116\,\msun}$, which is similar to the previous finding. Since the SN samples of \citet{2019AA...628A...7A} and \citet{2020arXiv200906683A} are not corrected for selection bias, we adopt ${f^\mathrm{Ni}_\mathrm{SE}<0.12\,\msun}$

Replacing the $f^\mathrm{Ni}_\mathrm{SE}$ and $f^\mathrm{Ni}_\mathrm{II}$ values in equation~(\ref{eq:yNiCC}), we get ${f^\mathrm{Ni}_\mathrm{CC}<0.064\,\msun}$, where the contribution of normal SNe~II to $f^\mathrm{Ni}_\mathrm{CC}$ is $>$36~per~cent. Finally, if we assume ${\eta=1.07}$ for all CC~SN subtypes, then from equation~(\ref{eq:eta}) we obtain ${\yfe<0.068\,\msun}$ for CC~SNe.

\subsection{Steepness as $^{56}$Ni mass indicator}\label{sec:MNi_vs_S}

From the analysis of nine normal SNe~II and the long-rising SN~1987A, \citet{2003AA...404.1077E} reported a linear correlation between $\log\mni$ and the maximum value of $dV/dt$ during the transition phase, called $V$-band steepness $S_V$. This correlation was also rebuilt by \citet{2018MNRAS.480.2475S}, which included another 30~SNe to the sample of \citet{2003AA...404.1077E}. The observed correlation is proposed to be a consequence of the $^{56}$Ni heating during the transition phase \citep[e.g.][]{2011ApJ...741...41P,2013MNRAS.434.3445P}. The latter produces slower transitions of the luminosity from the end of the plateau phase to the beginning of the radioactive tail as $\mni$ increases. Since the correlation between $\log\mni$ and steepness has not been studied for bands other than $V$, in this work we will include the $gr\!RiI$ bands in the analysis.

\begin{figure}
\includegraphics[width=1.0\columnwidth]{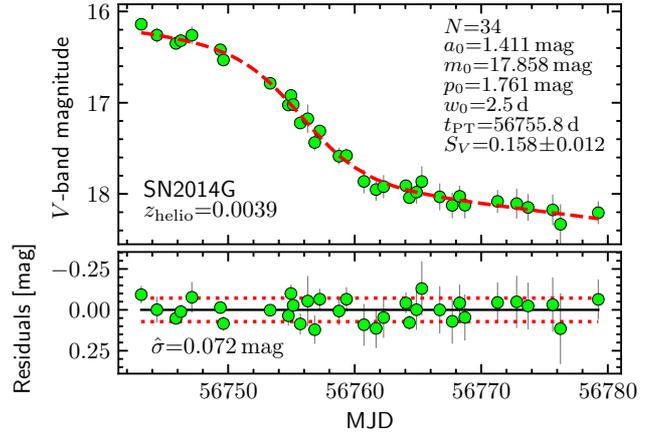}
\caption{Top panel: $V$-band light curve of SN~2014G in the transition phase, where the dashed line corresponds to the best fit. Bottom panel: best fit residuals, where dotted lines indicate the $\pm1\,\ssd$ limits. Error bars are $1\,\sigma$ errors.}
\label{fig:2017gmr_O10_fit}
\end{figure}

In order to measure the $x$-band steepness, $S_x$, we represent the light-curve transition phase by the function 
\begin{equation}\label{eq:transition_phase_fit}
m_x(t)=m_{0,x}-\frac{a_{0,x}}{1+e^{(t-t_{\mathrm{PT},x})/w_{0,x}}}+p_{0,x} \frac{t-t_{\mathrm{PT},x}}{100\,\mathrm{d}}
\end{equation}
\citep[e.g.][]{2010ApJ...715..833O,2016MNRAS.459.3939V}. The parameters $m_{0,x}$, $a_{0,x}$, $t_{\mathrm{PT},x}$ (the middle of the transition phase), $w_{0,x}$, and $p_{0,x}$ are obtained maximizing the model log-likelihood (equation~\ref{eq:lnL}). Fig.~\ref{fig:2017gmr_O10_fit} shows this analytical fit applied to the $V$-band photometry of SN~2014G. Using the aforementioned parametric function, the $x$-band steepness (in mag\,d$^{-1}$) in the SN rest frame is given by
\begin{equation}
S_x=\left(\frac{a_{0,x}}{4w_{0,x}}+\frac{p_{0,x}}{100\,\mathrm{d}}\right)/(1+\zsnhel).
\end{equation}

Columns~2--7 of Table~\ref{table:S} list the $gV\!r\!RiI$-band $S_x$ values and their bootstrap errors. As visible in the table, the $S_x$ estimates are in general not available for each of the $gV\!r\!RiI$ bands. In those cases we estimate $S_x$ indirectly from the steepnesses in bands other than $x$ (see Appendix~\ref{sec:Sx_from_Sb}). These indirect $S_x$ values ($S_x^*$) for the $gV\!r\!RiI$ bands and their errors are in Columns~8--13 of Table~\ref{table:S}. Table~\ref{table:mean_Sdiff} lists the mean and $\ssd$ values of the $S_x^*\!-\!S_x$ estimates for the $gV\!r\!RiI$ bands. The mean values are statistically consistent with zero within $1\,\ssd/\sqrt{N}$. Therefore, for SNe without an specific $S_x$ value, we can use their respective $S_x^*$ as a proxy.

\begin{table}
\caption{Mean $S_x^*\!-\!S_x$ values.}
\label{table:mean_Sdiff}
\begin{tabular}{cccccccc}
\hline
$x$   & $\langle S_x^*\!-\!S_x\rangle$ & $\ssd$ & $N$ & $x$ & $\langle S_x^*\!-\!S_x\rangle$ & $\ssd$          & $N$ \\
\hline
$g$ & $-0.002$ & 0.016 & 16 & $V$ & $-0.003$ & 0.024 & 41 \\
$r$ & $0.003$  & 0.029 & 20 & $R$ & $-0.002$ & 0.017 & 32 \\
$i$ & $-0.004$ & 0.036 & 22 & $I$ & $0.002$  & 0.018 & 31 \\
\hline
\multicolumn{8}{l}{\textit{Note.} Mean and $\ssd$ values are in mag\,d$^{-1}$ units.}\\
\end{tabular}
\end{table}

\begin{figure}
\includegraphics[width=1.0\columnwidth]{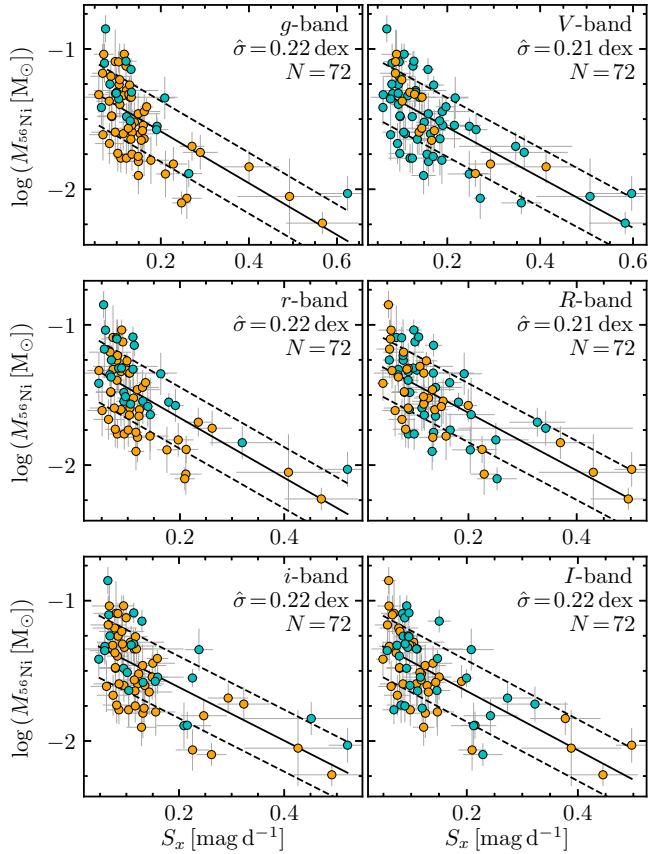}
\caption{$\log\mni$ against $S_x$ for $gV\!r\!RiI$ bands. Cyan circles are SNe with $S_x$ measured from their $x$-band light curves, while orange circles are SNe with $S_x$ estimated from the steepnesses in other bands. Solid lines are straight-line fits to the data, dashed lines are $\pm1\,\ssd$ limits around the fits, and error bars are $1\,\sigma$ errors.}
\label{fig:lMNi_vs_S}
\end{figure}

Fig.~\ref{fig:lMNi_vs_S} shows $\log\mni$ versus $S_x$ for the $gV\!r\!RiI$ bands. Through the model selection procedure (Appendix~\ref{sec:model_selection}), we find that the correlation between $\log\mni$ and $S_x$ is well represented by the straight line
\begin{equation}\label{eq:lMNi_vs_Sx}
\log\mni=c_{0,x}+c_{1,x} S_x
\end{equation}
which, for the $V$-band, corresponds to the best fit proposed by \citet{2003AA...404.1077E}. The parameters $c_{0,x}$ and $c_{1,x}$ (and their bootstrap errors) for the $gV\!r\!RiI$ bands are reported in Table~\ref{table:lMNi_vs_S_pars}. To evaluate the linear correlation between $\log\mni$ and $S_x$, we calculate the Pearson correlation coefficient $r_x$, listed in Column~6 of Table~\ref{table:lMNi_vs_S_pars}. The probability of obtaining ${|r_x|\geq0.65}$ from a random population with ${N=72}$ is $<$0.001~per~cent.

\begin{table}
\caption{$\log\mni$ versus $S_x$ calibrations.}
\label{table:lMNi_vs_S_pars}
\begin{tabular}{cccccc}
\hline
$x$ & $c_{0,x}$      & $c_{1,x}$      & $\ssd$ & $N$ & $r_x$     \\
    & (dex)          & (dex)          & (dex)  &     &           \\
\hline
$g$ & $-1.219\pm0.049$ & $-1.844\pm0.256$ & 0.217 & 72 & $-0.67$ \\
$V$ & $-1.201\pm0.043$ & $-1.789\pm0.183$ & 0.210 & 72 & $-0.69$ \\  
$r$ & $-1.238\pm0.047$ & $-2.130\pm0.292$ & 0.219 & 72 & $-0.66$ \\
$R$ & $-1.217\pm0.042$ & $-2.053\pm0.217$ & 0.210 & 72 & $-0.69$ \\
$i$ & $-1.237\pm0.044$ & $-1.901\pm0.231$ & 0.221 & 72 & $-0.65$ \\
$I$ & $-1.223\pm0.045$ & $-2.106\pm0.253$ & 0.219 & 72 & $-0.66$ \\
\hline
\multicolumn{6}{m{0.95\columnwidth}}{\textit{Notes.} $\log\mni=c_{0,x}+c_{1,x} S_x$. $\mni$ and $S_x$ are in $\msun$ and mag\,d$^{-1}$ units, respectively.}
\end{tabular}
\end{table}

The observed $\ssd$ of 0.21{\textendash}0.22\,dex for the $gV\!r\!RiI$ bands indicates that there is no preferred band for the $\log\mni$ versus $S_x$ correlation. Since the $\ssd$ values are $\sim$0.17\,dex greater in quadrature than the typical random error (0.13\,dex), the observed dispersion is mainly intrinsic. The latter was also pointed out by \citet{2013MNRAS.434.3445P}. Indeed, the shape of light curves in the transition phase not only depends on $\mni$ but also, among others, on the H mass retained before the explosion and the $^{56}$Ni mixing \citep[e.g.][]{2004ApJ...617.1233Y,2011ApJ...729...61B,2019MNRAS.483.1211K}.

\begin{figure}
\includegraphics[width=1.0\columnwidth]{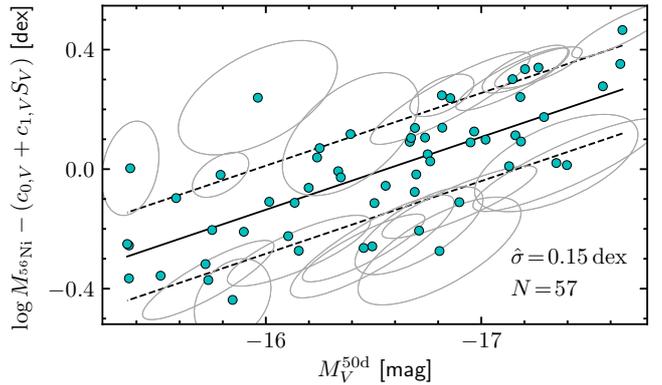}
\caption{Residuals of the $\log\mni$ versus $S_V$ correlation, as a function of $M_V^{50\mathrm{d}}$. The solid and dashed lines have the same meaning than in Fig~\ref{fig:lMNi_vs_S}. Ellipses are 1\,$\sigma$ confidence regions, which for clarity are drawn only for SNe outside the dashed lines.}
\label{fig:rlMNi_MV50d}
\end{figure}

Fig.~\ref{fig:rlMNi_MV50d} shows the residuals of the $\log\mni$ versus $S_V$ correlation (i.e. upper-right panel of Fig.~\ref{fig:lMNi_vs_S}) plotted against $M_V^{50\mathrm{d}}$, where we detect a linear dependence of the residuals on $M_V^{50\mathrm{d}}$. Therefore the \citet{2003AA...404.1077E} relation over and underestimates the $\log\mni$ of sub-luminous and moderately-luminous SNe II, respectively, by up to $\sim$0.3\,dex. This fact, along with the low statistical precision of the \citet{2003AA...404.1077E} relation to measure $\mni$ (around 50~per~cent), makes the latter method poorly suited for $^{56}$Ni mass measurements.

\subsection{Nickel-magnitude-steepness relation}\label{sec:NMS}

Our previous finding suggests a correlation of $\log\mni$ as a function not only of $S_x$ but also of the $x$-band absolute magnitude at ${\Delta t=50}$\,d ($M_x^{50\mathrm{d}}$). The $M_x^{50\mathrm{d}}$ values for $gr\!RiI$ are listed in Table~\ref{table:Mx50d}. If $M_x^{50\mathrm{d}}$ is not available for a given band, then we estimate it using photometry in other bands and magnitude transformation formulae (see Appendix~\ref{sec:mag_transf}).

\begin{figure*}
\includegraphics[width=0.98\textwidth]{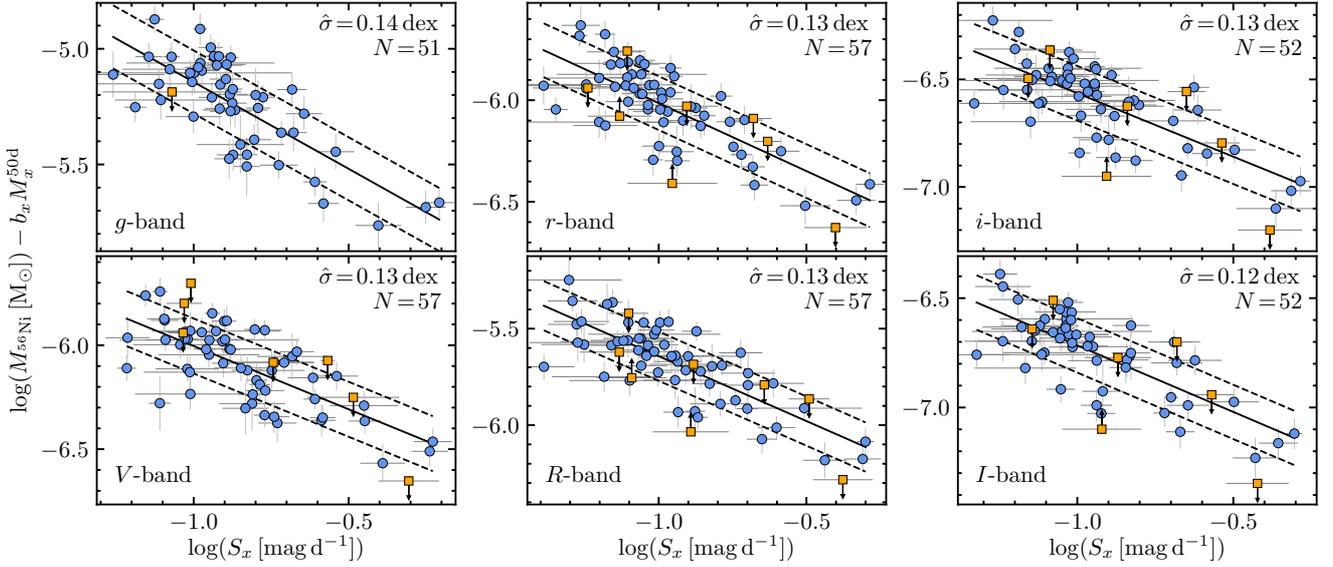}
\caption{NMS relations for the $gV\!r\!RiI$ bands. Solid lines are the best fits to the data (circles), while dashed lines indicate the $\pm1\,\ssd$ limits around the fits. Orange squares correspond to SNe with lower/upper limits on $M_x^{50\mathrm{d}}$. Error bars are $1\,\sigma$ errors.}
\label{fig:lMNi_S_M}
\end{figure*}

Using the model selection procedure (Appendix~\ref{sec:model_selection}), we find that the correlation of $\log\mni$ as a function of $S_x$ and $M_x^{50\mathrm{d}}$ can be represented by
\begin{equation}\label{eq:SM_cal}
\log\mni=a_x +b_xM_x^{50\mathrm{d}}+c_x\log S_x,
\end{equation}
where $\log S_x$ is given by equation~(\ref{eq:lognormal_mean2}). Fig.~\ref{fig:lMNi_S_M} shows the nickel-magnitude-steepness (NMS) relation for the $gV\!r\!RiI$ bands, while Table~\ref{table:lMNi_M_S_pars} lists the parameters (and their bootstrap errors) of equation~(\ref{eq:SM_cal}) along with the ranges of $M_x^{50\mathrm{d}}$ and $\log S_x$ where the relation is valid. To evaluate the linear correlation of $\log\mni$ on $M_x^{50\mathrm{d}}$ and $\log S_x$, we calculate the multiple correlation coefficient ($R_x^\mathrm{mc}$, reported in Column~10 of Table~\ref{table:lMNi_M_S_pars}). The probability of obtaining $R_x^\mathrm{mc}\geq0.89$ from a random population is $<$0.001~per~cent.

\begin{table*}
\caption{NMS relation parameters.}
\label{table:lMNi_M_S_pars}
\begin{tabular}{cccccccccc}
\hline
$x$ & $a_x$            & $b_x$              & $c_x$            & $\ssd$  & $\sigma_{0,x}$ &$N$  & $M_x^{50\mathrm{d}}$ range & $\log S_x$ range     & $R_x^\mathrm{mc}$\\ 
    & (dex)            & (dex\,mag$^{-1}$)  &                  & (dex)   & (dex)          &     &  (mag)                     & (dex) &\\
\hline
$g$ & $-5.894\pm0.461$ & $-0.2311\pm0.0306$ & $-0.750\pm0.097$ & $0.136$ & $0.112$ & 51 & $-14.9,\,-17.4$ & $-1.26,\,-0.21$ & 0.89 \\ 
$V$ & $-6.612\pm0.409$ & $-0.2778\pm0.0266$ & $-0.608\pm0.083$ & $0.132$ & $0.106$ & 57 & $-15.4,\,-17.7$ & $-1.22,\,-0.23$ & 0.89  \\ 
$r$ & $-6.683\pm0.459$ & $-0.2710\pm0.0294$ & $-0.669\pm0.087$ & $0.132$ & $0.107$ & 57 & $-15.6,\,-18.1$ & $-1.39,\,-0.29$ & 0.90  \\ 
$R$ & $-6.315\pm0.449$ & $-0.2487\pm0.0287$ & $-0.673\pm0.082$ & $0.128$ & $0.094$ & 57 & $-15.7,\,-18.2$ & $-1.39,\,-0.30$ & 0.90  \\ 
$i$ & $-7.154\pm0.442$ & $-0.3066\pm0.0281$ & $-0.590\pm0.079$ & $0.127$ & $0.098$ & 52 & $-15.4,\,-17.7$ & $-1.33,\,-0.28$ & 0.90  \\ 
$I$ & $-7.324\pm0.430$ & $-0.3083\pm0.0267$ & $-0.608\pm0.080$ & $0.124$ & $0.098$ & 52 & $-15.7,\,-18.1$ & $-1.32,\,-0.30$ & 0.91  \\ 
\hline
\multicolumn{10}{l}{\textit{Notes.} $\log\mni=a_x+b_x M_x^{50\mathrm{d}}+c_x\log S_x$. $\mni$, $M_x^{50\mathrm{d}}$, and $S_x$ are in units of $\msun$, mag, and mag\,d$^{-1}$, respectively.} 
\end{tabular}
\end{table*}

The NMS relation allows to measure $\log\mni$ with a statistical precision of 0.12{\textendash}0.14\,dex ($\mni$ error of $\sim$30~per~cent). The observed random error is about 0.08\,dex, so the intrinsic random error on the NMS relation ($\sigma_{0,x}$, listed in Column~6 of Table~\ref{table:lMNi_M_S_pars}) is around 0.10\,dex. Since $\sim$80~per~cent of the SNe used to calibrate equation~(\ref{eq:SM_cal}) have $\log\mni$ computed with $I$- or $i$-band photometry, we adopt a systematic uncertainty due to the $\zp^\bc$ errors of 0.044\,dex (the average between the errors on $\zp_I^{\mathrm{BC}}$ and $\zp_i^{\mathrm{BC}}$ in dex scale). The total systematic error ($\sigma_{\mathrm{sys}}$), including the uncertainty due to $\alpha$, is of 0.047\,dex, while the total error on $\log\mni$ provided by the NMS relation is given by
\begin{equation}
\sigma_{\log\mni}=\sqrt{(b_x\sigma_{M_x^{50\mathrm{d}}})^2+(c_x\sigma_{\log Sx})^2+\sigma_{0,x}^2+\sigma_{\mathrm{sys}}^2}.
\end{equation}
Since $|b_x|<0.4$, the $\log\mni$ values estimated with the NMS relation are less dependent on $\EhBV$ and $\mu$ than those computed with the radioactive tail photometry and the BC technique (Appendix~\ref{sec:MNi_equation}).

\begin{figure}
\includegraphics[width=1.0\columnwidth]{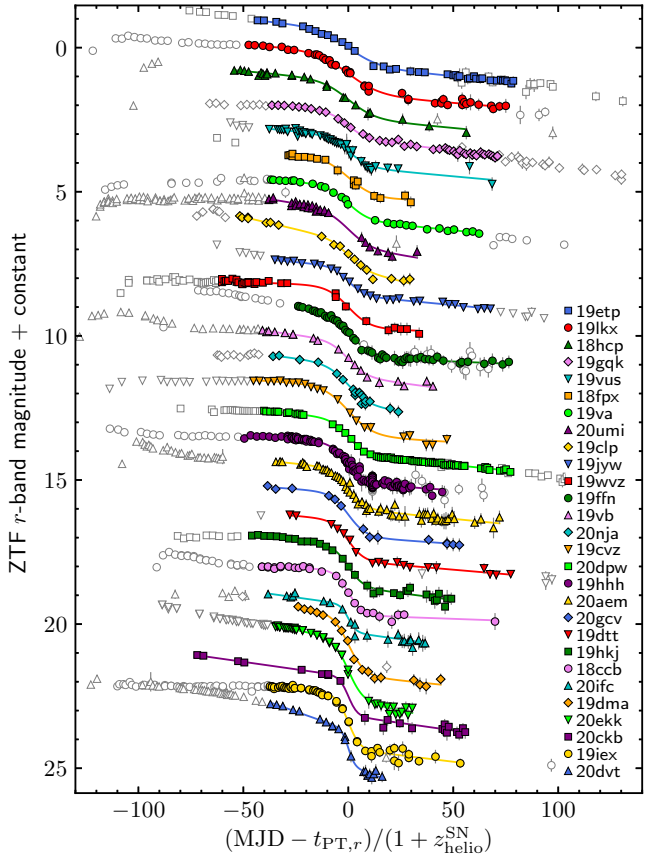}
\caption{$r_\ztf$ light curves of the 28 normal SNe~II in the SZB sample. Data used to measure the steepness are shown with colour-filled symbols, while solid lines are analytical fits.}
\label{fig:ZTF_r}
\end{figure}

As a first application of the NMS relation, we compute $\log\mni$ estimates ($\log\mni^\mathrm{NMS}$) for a sample of normal SNe~II observed by the Zwicky Transient Facility (ZTF; \citealt{2019PASP..131a8002B,2019PASP..131g8001G}). Specifically, we employ SNe from the ZTF bright transient survey\footnote{\url{https://sites.astro.caltech.edu/ztf/bts/explorer.php}} \citep[BTS;][]{2020ApJ...895...32F,2020ApJ...904...35P}, which consists on SNe brighter than 19\,mag. From this magnitude-limited survey, we select SNe spectroscopically classified as SNe~II, discarding those (1) classified as Type IIb or IIn, (2) with long-rising light curves, (3) with less than three $r_\ztf$ photometric points in the radioactive tail, (4) with $t_0$ errors greater than 10\,d, and (5) with absolute magnitudes and steepnesses outside the range where the NMS relation is valid. The selected ZTF BTS (SZB) sample of 28 normal SNe~II and their main properties are summarized in Table~\ref{table:ZTF_sample}, while Fig.~\ref{fig:ZTF_r} shows their $r_\ztf$ light curves\footnote{Photometry obtained from the ALeRCE \citep{2021AJ....161..242F} website (\url{https://alerce.online/}).}. Out of the SNe in the SZB set, 24 have $M_R^{\max}<-17$\,mag, so the sample consists mainly of moderately-luminous SNe~II.

Given that ${r_\ztf\!-\!R=0.14}$\,mag from the photospheric to the radioactive tail phase (see Appendix~\ref{sec:mag_transf}), for the $r_\ztf$ band we adopt the $R$-band NMS relation, but with ${a_{r_\ztf}=a_R-0.14b_R}$ equal to $-6.281$\,dex. Since we cannot measure $\EhBV$ for the SZB sample with the available data, we assume $\EhBV=0.16\pm0.15$\,mag (the average of the $\EhBV$ distribution shown in Fig.~\ref{fig:Eh_NaID}). Column~9 of Table~\ref{table:ZTF_sample} lists the inferred $\log\mni^\mathrm{NMS}$ values. For comparison, we also compute $\log\mni$ with the radioactive tail luminosity ($\log\mni^\mathrm{tail}$, Column~10 of Table~\ref{table:ZTF_sample}), using a constant $\bc_{r_\ztf}=\bc_R-0.14$ equivalent to $11.93\pm0.14$\,mag.

The mean offset between the $\log\mni^\mathrm{tail}$ and $\log\mni^\mathrm{NMS}$ values is of 0.038\,dex with a $\ssd$ of 0.110\,dex. The $\ssd$ value is similar to that expected for the NMS relation (0.13\,dex), while the offset is consistent with zero within $1.9\,\ssd/\sqrt{N}$. As stated in Section~\ref{sec:BC_cal_others}, the use of the constant $\bc_R$ reported in Table~\ref{table:BC_parameters} (${\bc_R=12.07}$) overestimates the radioactive tail luminosities (and therefore the $\log\mni^\mathrm{tail}$ values) of moderately-luminous SNe~II. To roughly estimate a more appropriate constant $\bc_R$ for the SZB sample, we use the $\zp_R^{\mathrm{SN}}$ values of the nine SNe in the BC calibration set with $M_R^{\max}<-17$\,mag, obtaining a $\bc_R$ of $12.13$\,mag. Increasing $\bc_{r_\ztf}$ by 0.06\,mag decreases the offset from 0.046 to 0.014\,dex, being consistent with zero within $0.7\,\ssd/\sqrt{N}$. In addition, the offset decreases from $1.9\,\ssd/\sqrt{N}$ to zero if we adopt $\EhBV=0.05$\,mag. Therefore, part of the offset could be due to an overestimation of the adopted $\EhBV$. The results we obtain with the SZB sample provides further evidence supporting the usefulness of the NMS relation for $\log\mni$ measurements.

\section{Discussion}\label{sec:discussion}

\subsection{Comparison with other works}

Table~\ref{table:BC_parameters} summarizes the BC values for SNe~II in the radioactive tail reported by \citet{2001PhDT.......173H}, \citet{2009ApJ...701..200B}, \citet{2010MNRAS.404..981M}, and \citet{2015ApJ...799..215P}. From each value we subtract the zero-point used to define the apparent bolometric magnitude scale ($-10.89$\,mag for \citealt{2001PhDT.......173H} and \citealt{2010MNRAS.404..981M}, $-11.64$\,mag for \citealt{2009ApJ...701..200B}, and $-11.48$\,mag for \citealt{2015ApJ...799..215P}). Since previous works reported constant BC values, for the comparison we use our estimates assuming a constant BC, which are also listed in the table. The BC values reported in the literature are consistent with our estimations within $\pm1.1\,\sigma$. Except for \citet{2015ApJ...799..215P}, which did not report BC errors, the uncertainties provided in previous works are lower than those reported here. The latter is due to the few SNe used in previous studies to compute BCs. Indeed, they used only SN~1999em and the long-rising SN~1987A to compute BCs. It is interesting to note the good agreement between our constant BC values for $r\!RiI$ bands and those of \citet{2015ApJ...799..215P} within $\pm0.04$\,mag.

\begin{table}
\caption{Constant BC values for the radioactive tail.}
\label{table:BC_literature}
\begin{tabular}{cccl}
\hline
$\bc_x$ (mag)  & $x$ & $N$  &         Reference           \\
\hline
$11.15\pm0.06$ & $V$ & 2$^a$  & \citet{2001PhDT.......173H} \\
$10.94\pm0.05$ & $V$ & 1$^a$  & \citet{2009ApJ...701..200B} \\
$11.22\pm0.06$ & $V$ & 2$^a$  & \citet{2010MNRAS.404..981M} \\
$11.27$        & $V$ & 26$^b$ & \citet{2015ApJ...799..215P} \\
$11.15\pm0.18$ & $V$ & 15     & This work                   \\
$11.90$        & $r$ & 26$^b$ & \citet{2015ApJ...799..215P} \\
$11.89\pm0.16$ & $r$ & 15     & This work                   \\
$12.07$        & $R$ & 26$^b$ & \citet{2015ApJ...799..215P} \\
$12.07\pm0.14$ & $R$ & 15     & This work                   \\
$11.95$        & $i$ & 26$^b$ & \citet{2015ApJ...799..215P} \\
$11.91\pm0.12$ & $i$ & 15     & This work                   \\
$12.41$        & $I$ & 26$^b$ & \citet{2015ApJ...799..215P} \\
$12.44\pm0.11$ & $I$ & 15     & This work                   \\
\hline
\multicolumn{4}{m{0.80\linewidth}}{$^a$It includes the long-rising SN~1987A.}\\
\multicolumn{4}{m{0.80\linewidth}}{$^b$Only six SNe with optical and near-IR photometry in the radioactive tail.}\\
\end{tabular}
\end{table}

Table~\ref{table:mean_MNi} collects the mean and $\ssd$ values of the normal SN~II $\mni$ distributions presented in \citet{2008NewA...13..606B} and \citet{2017ApJ...841..127M}. Despite the \citet{2008NewA...13..606B} sample includes the long-rising SN~1987A, we find that removing that SN only marginally modifies the reported mean and $\ssd$ value. In addition, we include the mean $^{56}$Ni mass computed with the 107 normal SNe~II in the sample of \citet{2019AA...628A...7A}\footnote{$\mni$ values are reported in \citet{2020AA...641A.177M}, from which we remove SN~2007od, the long-rising SNe~1987A, 1998A, 2000cb, 2006V, 2006au, and 2009E, and the LLEV SN~2008bm.}. Since previous estimates are not corrected for selection bias, for the comparison we use our $\langle\mni\rangle^\mathrm{unc}$ value. The latter value and those from the aforementioned samples are consistent within $1.6\,\ssd/\sqrt{N}$. It is worth mentioning that the collected $\langle\mni\rangle^\mathrm{unc}$ values are not independent since they were computed with SN samples having objects in common. The number of SNe in common between a given set and our full sample is indicated in Column~5. In particular, the similarity between our result and that obtained from the \citet{2019AA...628A...7A} sample is because both analyses have 70~per~cent of SNe in common.

\begin{table}
\caption{Mean $^{56}$Ni mass values for normal SNe~II.}
\label{table:mean_MNi}
\begin{tabular}{cccccl}
\hline
$\langle\mni\rangle^{\mathrm{unc}\dagger}$ & $\ssd$  & $N$  & $\ssd/\sqrt{N}$ & ${N_c}^*$ & Reference$^\ddagger$           \\
($\msun$)            &($\msun$)&      & ($\msun$)                     & &                    \\
\hline
$0.066$              & $0.082$ &  28  & 0.015 & 17 & B08 \\
$0.046$              & $0.048$ &  38  & 0.008 & 33 & M17 \\
$0.042$              & $0.044$ & 107  & 0.004 & 75 & A19 \\
$0.042$              & $0.033$ & 109  & 0.003 & -- & This work\\
\hline
\multicolumn{6}{l}{$^\dagger\langle\mni\rangle$ uncorrected for selection bias.}\\
\multicolumn{6}{l}{$^*$Number of SNe in common with our full sample (109~SNe).}\\
\multicolumn{6}{m{0.95\linewidth}}{$^\ddagger$B08: \citet{2008NewA...13..606B}; M17: \citet{2017ApJ...841..127M}; A19: \citet{2019AA...628A...7A}, selecting only normal SNe~II.}
\end{tabular}
\end{table}

We also compare the $\log\mni$ values of the SNe in common between our SN set and the samples analysed in \citet{2017ApJ...841..127M}, \citet{2016MNRAS.459.3939V}, and \citet{2020MNRAS.496.4517S}. \citet{2017ApJ...841..127M} employed the methodology of \citet{2015ApJ...799..215P}, which computes $\log\mni$ using the luminosity at ${\Delta t=200}$\,d and equation~(3) of \citet{2003ApJ...582..905H}. \citet{2016MNRAS.459.3939V} used 
\begin{equation}\label{eq:S14}
{\mni/\msun=0.075L_{\mathrm{SN}}^{\mathrm{opt}}(\Delta t)/L_{\mathrm{87A}}^{\mathrm{opt}}}(\Delta t),
\end{equation}
being $L_{\mathrm{SN}}^{\mathrm{opt}}(\Delta t)$ and $L_{\mathrm{87A}}^{\mathrm{opt}}(\Delta t)$ the optical quasi-bolometric luminosity in the radioactive tail of a specific SN and of the long-rising SN~1987A, respectively. \citet{2020MNRAS.496.4517S} used the radioactive tail luminosity and the set of equations presented in \citet{2019MNRAS.484.3941W} to derive $\log\mni$. For each sample we compute the differences between its $\log\mni$ measurements, and calculate the mean offset ($\Delta$) and its $\ssd$. Then, to track the main source of the observed dispersion, we repeat the previous process, but recomputing our $\log\mni$ values without correcting for the $\gamma$-ray leakage (except for \citealt{2020MNRAS.496.4517S}, which included this correction) and using the $\mu$ and $\EhBV$ values of the comparison work. The $\Delta$ and $\ssd$ values are listed in Table~\ref{table:MNi_comparison}.

\begin{table}
\caption{Mean $\log{\mni}$ differences between different works.}
\label{table:MNi_comparison}
\begin{tabular}{lccc}
\hline
$\log{\mni}$ differences$^\dagger$ &$N$ & $\Delta$ (dex)  & $\ssd$ (dex) \\
\hline
M17$-$here                         & 33 &    $-0.01$ & $0.29$ \\
M17$-$here(a)                      & 33 & $\phs0.01$ & $0.28$ \\
M17$-$here(a,b)                    & 33 & $\phs0.07$ & $0.13$ \\
M17$-$here(a,b,c)                  & 33 & $\phs0.05$ & $0.08$ \\
V16$-$here                         & 33 &    $-0.15$ & $0.30$ \\
V16$-$here(a,b,c)                  & 33 &    $-0.05$ & $0.14$ \\
S20$-$here                         &  7 &    $-0.09$ & $0.15$ \\
S20$-$here(b,c)                    &  7 & $\phs0.02$ & $0.05$ \\
\hline
\multicolumn{4}{m{0.95\linewidth}}{$^\dagger$M17: \citet{2017ApJ...841..127M}; S20: \citet{2020MNRAS.496.4517S}; V16: \citet{2016MNRAS.459.3939V}; here: this work, uncorrected for $\gamma$-ray leakage (a), and using the distance moduli (b) and colour excesses (c) of the comparison work.}\\
\end{tabular}
\end{table}

From the comparison with the \citet{2017ApJ...841..127M} sample we obtain ${\ssd=0.29}$\,dex. This value decreases to 0.28\,dex when we do not correct for $\gamma$-ray leakage, to 0.13\,dex when we use the distance moduli of \citet{2017ApJ...841..127M}, and to 0.08\,dex if we also use their colour excesses. This indicates that differences between our $\log\mni$ estimates and those of \citet{2017ApJ...841..127M} are mainly due to differences in the adopted distance moduli and colour excesses. Therefore, the $\ssd$ value of 0.08\,dex represents the typical $\log\mni$ error for single SNe due to differences in the methodology used to compute $\log\mni$. For the other two comparison samples we arrive at similar results.

The $\Delta$ value from the comparison with the sample of \citet{2017ApJ...841..127M} is equivalent to $3.5\,\ssd/\sqrt{N}$, which indicates the presence of a systematic offset. We note that, of the 0.05\,dex offset, 0.03\,dex is due to the numerical coefficients of the equation used in \citet{2015ApJ...799..215P} to compute $\log\mni$\footnote{Equation~(3) of \citet{2003ApJ...582..905H} is not accurate. To estimate $\log\mni$ we recommend the set of equations presented in \citet{2019MNRAS.484.3941W}.}. The remaining 0.02\,dex is statistically consistent with zero within 1.4\,$\ssd/\sqrt{N}$. The $\Delta$ values from the comparison with the samples of \citet{2016MNRAS.459.3939V} and \citet{2020MNRAS.496.4517S} are statistically consistent with zero within 2.0 and $1.0\,\ssd/\sqrt{N}$, respectively.

\subsection{Systematics}

The models we use in this work (SN~II spectra and nucleosynthesis yields) were generated by adopting 
many approximations and assumptions that help to characterize the underlying complex physical processes of SN explosions. Therefore, our $\zp_x^\bc$, $\log\mni$, $\langle\mni\rangle$, and $\yfe$ values are potentially affected by systematics on the spectral models, while $\yfe$ is also affected by systematics on nucleosynthesis yield models. An analysis and quantification of those systematics is beyond the scope of this study.

\subsubsection{Local Hubble-Lema\^{i}tre constant}

The distance moduli of 92~SNe in our sample were estimated as the weighted average of $\mu_\mathrm{TF}$, $\mu_\mathrm{HLL}$, and $\mu_\mathrm{SVF}$ (see Section~\ref{sec:mu}). The latter are anchored to Cepheid-calibrated $H_0$ values of around 75\,km\,s$^{-1}$\,Mpc$^{-1}$. If we adopt a TRGB-calibrated $H_0$ value between 69.6 and 72.4\,km\,s$^{-1}$\,Mpc$^{-1}$ \citep[e.g.][]{2020ApJ...891...57F,2019ApJ...886...61Y}, then the $\mu$ values of our SN sample increase by about 0.08{\textendash}0.16\,mag. In this case, 
we obtain $\langle\mni\rangle$ values around 0.040{\textendash}0.044\,$\msun$. If we assume that the true local $H_0$ value lies between 71 and 75\,km\,s$^{-1}$\,Mpc$^{-1}$ with a uniform probability, then the systematic offset in $\langle\mni\rangle$ induced by the $H_0$ uncertainty ranges between 0.0 and 0.005\,$\msun$, with a mean of ${0.002\pm0.001\,\msun}$. This systematic error is lower than that induced by the $\zp^\bc$ error ($0.003\,\msun$). Therefore the $H_0$ uncertainty is not so relevant for our current analysis.

\subsubsection{Early dust formation}

To compute $\log\mni$ we assumed that the extinction along the SN line of sight is constant for ${\Delta t\leq320}$\,d.  However, normal SNe~II are dust factories\footnote{The amount of newly formed dust, however, is still unclear \citep[e.g.][]{2020MNRAS.497.2227P}.}, where the onset of the dust formation is different for each SN. In some cases, the dust formation begins as early as ${\Delta t\sim100}\text{\textendash}200$\,d (e.g. SN~2007od, \citealt{2010ApJ...715..541A}, \citealt{2011MNRAS.417..261I}; SN~2011ja, \citealt{2016MNRAS.457.3241A}; SN~2017eaw, \citealt{2018ApJ...864L..20R}, \citealt{2019ApJ...873..127T}). The latter indicates that some normal SNe~II may experience a non-negligible increase of the extinction at ${\Delta t\leq320}$\,d, with a consequent decrease in their $^{56}$Ni masses inferred from optical light. 

In the case of SN~2007od, the newly formed dust decreases the $\log\mni$ inferred from optical light in $\sim$0.8\,dex (see Section~\ref{sec:MNi_vs_Mv50d}). On the other hand, for SN~2017eaw (which shows evidence of dust formation at ${\Delta t\sim120}$\,d, \citealt{2018ApJ...864L..20R}) we measure a $\log\mni$ of $-1.087$\,dex (${\mni=0.083\,\msun}$). This value is about 0.1{\textendash}0.2\,dex greater than the predicted from the NMS relation. Moreover, our $\mni$ estimate is consistent with the $^{56}$Ni mass of $0.084\,\msun$ used in the models found by \citet{2018ApJ...864L..20R} to be consistent with the near-IR spectra of SN~2017eaw. Therefore, the early dust formation does not necessarily translate into a non-negligible increase of the extinction.

In the case of strong extinction due to newly formed dust, if it only affects to SNe~2007od, and possibly to SN~2010aj in our set, then the fraction of these events is around 1{\textendash}2~per~cent. Therefore, they should not be a severe contaminant in the mean $^{56}$Ni mass of normal SNe~II.

\subsection{Future improvements}

Future works on improving the precision of $\langle\mni\rangle$ should focus on reducing the random and $\zp^\bc$ errors.

The $\zp_x^\bc$ values we present in Section~\ref{sec:BC_cal} are based on only 15~SNe, so their errors could be misestimated due to the small sample size. Indeed, the real error on $\zp_x^\bc$ for the $I$-band (the preferred one to measure $\log\mni$) can be as low as 0.07\,mag or as large as 0.19\,mag at a confidence level of 99~per~cent\footnote{Assuming that $\zp_I^\mathrm{BC}$ has a normal parent distribution with standard deviation $\sigma$, for which the quantity $(\ssd/\sigma)^2\nu$ has a chi-square distribution with $\nu$ degrees of freedom \citep[e.g.][]{Lu_1960}.}. Moreover, the small sample size of our BC calibration set does not allow us to robustly detect outliers. For example, using the Chauvenet's criterion, we find that SN~2014G is a possible outlier in the $\zp_I^\mathrm{BC}$ distribution (Fig.~\ref{fig:BCx_cal}). On the other hand, using the Chauvenet's criterion over $10^4$ bootstrap resampling, we find that SN~2014G is consistent with the $\zp_I^\mathrm{BC}$ distribution in 65~per~cent of the realizations. Increasing the number of SNe used to calibrate BCs (i.e. observed at optical and near-IR filters in the radioactive tail) is, therefore, a necessary step to improve the $\zp^\bc$ error estimation. 

One of the current surveys providing optical photometry is the ZTF, which observe about 2300 normal SNe~II brighter than 20\,mag per year \citep{2019JCAP...10..005F}. Based on our SN sample, the $R$-band magnitude of normal SNe~II during the first 50\,d of the radioactive tail is between 1.5{\textendash}3.8\,mag dimmer than at the maximum light, with an average of 2.5\,mag. This means that roughly 3~per~cent of all normal SNe~II observed by the ZTF have $r_\ztf$ photometry useful to estimate $\log\mni$ with the radioactive tail photometry. Therefore, it is possible to construct an SN set of similar size to that we used here with two years of ZTF data. Within a few years, the Rubin Observatory Legacy Survey of Space and Time (LSST) will be the main source of photometric data, which will observe $\sim$10$^5$ SNe II per year \citep{2009JCAP...01..047L}. With one year of LSST data ($\sim$2500 normal SNe~II with radioactive tail photometry), it will be feasible to reduce the random error in $\langle\mni\rangle$ from 9~per~cent (estimated in this work) to around 2~per~cent.

The NMS relation provides a method to measure $\log\mni$ virtually independent on the radioactive tail photometry. Therefore, the $\log\mni$ estimates computed with the radioactive tail luminosity and the NMS relation could be combined to further reduce the random error. Since the transition phase lasts $\lesssim$30\,d, $\log\mni$ measurements with the NMS relation require light curves sampled with a cadence of $\sim$5\,d.  As we see in Figure~\ref{fig:ZTF_r}, the cadence of the ZTF (around 3\,d) is more than enough to estimate the steepness parameter. In the case of the LSST, in order to have light curves sampled with a cadence of $\sim$5\,d, it will be necessary to combine light curves in different bands into a single one.

Finally, based on the empirical correlation between absolute magnitude and expansion velocity at $\Delta t=50$\,d ($v_{50\mathrm{d}}$) \citep[e.g.][]{2002ApJ...566L..63H,2003ApJ...582..905H}, we expect a relation between $\log\mni$, $v_{50\mathrm{d}}$, and $S_x$. If confirmed, the latter relation will provide a method to measure $\log\mni$ independent of the distance and colour excess.

\section{SUMMARY and CONCLUSIONS}\label{sec:conclusions}

In this work we computed the $^{56}$Ni masses of 110 normal SNe~II from their luminosities in the radioactive tail. To estimate those luminosities we employed the BC technique. We used 15~SNe with $\bvrriijhk$ photometry and three theoretical spectral models to calibrate the BC values. In order to convert $^{56}$Ni masses to iron masses, we used iron isotope ratios of CC nucleosynthesis models. We also analysed the correlation of the $^{56}$Ni mass on the steepness parameter and on the absolute magnitude at $\Delta t=50$\,d.

Our main conclusion are the following:

\begin{itemize}

\item[(1)] The $I$- and $i$-band are best suited to estimate radioactive tail luminosities through the BC technique. In particular, the $\bc_V$ value is not constant as reported in previous studies but it is correlated with the $\vi$ colour.

\item[(2)] We obtained $\langle\mni\rangle=0.037\pm0.005\,\msun$ for normal SNe~II, which translates into a $\yfe$ of $0.040\pm0.005\,\msun$. 
Combining this result with recent mean $^{56}$Ni mass measurements for other CC~SN subtypes, we estimated a mean CC~SN iron yield $<$0.068\,$\msun$. The contribution of normal SNe~II to this yield is $>$36~per~cent.

\item[(3)] The relation between $\log\mni$ and $S_V$ suggested by \citet{2003AA...404.1077E} is poorly suited to estimate $\mni$. Instead we proposed the NMS method, based on the correlation of $\log\mni$ on $M_x^{50\mathrm{d}}$ and $\log S_x$, which allows to measure $\log\mni$ with a precision of 0.13\,dex. Using the $r_\ztf$ photometry of 28 normal SNe~II from the ZTF BTS, we obtained further evidence supporting the usefulness of the NMS relation to measure $\log\mni$.

\end{itemize}

Future works with ZTF and LSST data during the first years of operation will allow to verify our $\langle\mni\rangle$ measurement. In particular, it will be feasible to reduce its random error by a factor of four with one year of LSST data. On the other hand, to reduce the error due to the BC ZP uncertainty, it will be necessary to carry out an observational campaign to increase the number of normal SNe~II observed with optical and near-IR filters during the radioactive tail.

\section*{Acknowledgements}

This paper is part of a project that has received funding from the European Research Council (ERC) under the European Union’s Seventh Framework Programme, Grant agreement No. 833031 (PI Dan Maoz). MR thanks the support of the National Agency for Research and Development, ANID-PFCHA/Doctorado-Nacional/2020-21202606. This research has made use of the NASA/IPAC Extragalactic Database (NED) which is operated by the Jet Propulsion Laboratory, California Institute of Technology, under contract with the National Aeronautics and Space Administration. This work has made use of the Weizmann Interactive Supernova Data Repository (\url{https://www.wiserep.org}).

\section*{Data availability}

The data underlying this article will be shared on reasonable request to the corresponding author.




\bibliographystyle{mnras}
\bibliography{references}



\appendix

\section{Synthetic magnitudes}\label{sec:syn_mag}

Given a SED $f_{\lambda}$ (in erg\,s$^{-1}$\,cm$^{-2}\,\mbox{\normalfont\AA}^{-1}$), we can compute the synthetic magnitude in the $x$-band using
\begin{equation}\label{eq:syn_mag}
m_x=-2.5\log{\int d\lambda S_{x,\lambda} \frac{\lambda f_{\lambda}}{hc}} + \zp_x^{\mathrm{mag}}
\end{equation}
\citep[e.g.][]{2001PhDT.......173H}. Here $\lambda$ is the observed wavelength (in \AA), $S_{x,\lambda}$ is the peak-normalized $x$-band transmission function (considering a photon-counting detector), ${hc=1.986\times10^{-8}}$\,erg\,\AA, and $\zp_x^{\mathrm{mag}}$ is the zero-point for the magnitude system. To compute $\zp_x^{\mathrm{mag}}$ values in the Vega system, we use equation~(\ref{eq:syn_mag}) along with the Vega SED published by \citet{2004AJ....127.3508B}\footnote{\url{https://ssb.stsci.edu/cdbs/current_calspec/alpha_lyr_stis_010.fits}} and the Vega magnitudes published by \citet{1996AJ....111.1748F}: ${B=0.03}$, ${V=0.03}$, and ${I=0.024}$\,mag, and by \citet{1999AJ....117.1864C}: ${J=-0.001}$, ${H=0.000}$, and ${K=-0.001}$\,mag. For Johnson-Kron-Cousins $BV\!RI$ bands we adopt the transmission functions given in \citet{2005PASP..117..810S}, while for 2MASS $J\!H\!K$ bands we adopt the transmission of \citet{2003AJ....126.1090C}\footnote{Since they consider an energy-counting detector, we have to divide the transmissions by $\lambda$ before to use them in equation~(\ref{eq:syn_mag}).\label{ec}}. To compute $\zp_x^{\mathrm{mag}}$ values in the AB system, we use equation~(\ref{eq:syn_mag}), ${f_{\lambda}=1/\lambda^2}$, and ${m_x=-2.408}$. For the Sloan $gri$ bands we use the transmission functions of the SDSS Data Release 7\footnote{\url{http://classic.sdss.org/dr7/instruments/imager/index.html}}. To compute the $r_\ztf$ transmission function, we use equation~(1) of \citet{2007MNRAS.376.1301P}, adopting the corresponding ZTF filter transmission\textsuperscript{\ref{ec}} and CCD quantum efficiency\footnote{\url{https://github.com/ZwickyTransientFacility/ztf_information}}, the atmospheric extinction at Palomar Observatory of \citet{1975ApJ...197..593H} (assuming an airmass of 1.2) combined with atmospheric telluric lines, a standard aluminium reflectivity curve, and a constant lens throughput. Column~3 of Table~\ref{table:zp_and_leff} lists the $\zp_x^{\mathrm{mag}}$ values.

\begin{table}
\caption{Properties of the filters used in this work.}
\label{table:zp_and_leff}
\begin{tabular}{l c c c c c c}
\hline
$x$                 & System & $\zp_x^{\mathrm{mag}}$   & $\zp_x^{\mathrm{mflux}}$& $\leff_x$ & $R_{\leff_x}$ & $R_x^{\mathrm{pt}}$\\
                    &        & (mag)  &(mag)      & (\AA)      &      & \\
\hline
 $B$                & Vega   & 15.300 & $-20.462$ & \phn$4610$ & 3.86 & 3.99 \\ 
 $g$                & AB     & 15.329 & $-20.770$ & \phn$4860$ & 3.62 & 3.68 \\
 $V$                & Vega   & 14.865 & $-21.074$ & \phn$5600$ & 2.96 & 3.04 \\
 $r$                & AB     & 14.986 & $-21.361$ & \phn$6330$ & 2.49 & 2.55 \\
 $r_\ztf$           & AB     & 15.212 & $-21.436$ & \phn$6490$ & 2.40 & 2.45 \\
 $R$                & Vega   & 15.053 & $-21.629$ & \phn$6610$ & 2.33 & 2.40 \\
 $i$                & AB     & 14.710 & $-21.780$ & \phn$7450$ & 1.93 & 1.91 \\
 $I$                & Vega   & 14.538 & $-22.354$ & \phn$8090$ & 1.68 & 1.71 \\
 $J$                & Vega   & 13.729 & $-23.787$ &    $12470$ & 0.80 & 0.80 \\ 
 $H$                & Vega   & 13.412 & $-24.886$ &    $16520$ & 0.51 & 0.51 \\
 $K$                & Vega   & 12.691 & $-25.948$ &    $21630$ & 0.35 & 0.35 \\ 
\hline
\end{tabular}
\end{table}

\section{Photometric SED}\label{sec:pSED}

To compute pSEDs, we need to convert $x$-band magnitudes to monochromatic fluxes $\feffx$ (in erg\,s$^{-1}$\,cm$^{-2}$\,$\mbox{\normalfont\AA}^{-1}$). For this, we use ${f_{\lambda}=\feffx}$ in equation~(\ref{eq:syn_mag}), thus obtaining
\begin{equation}
\feffx=10^{-0.4(m_x-\mathrm{ZP}_x^{\mathrm{mflux}})},
\end{equation}
where
\begin{equation}
\mathrm{ZP}_x^{\mathrm{mflux}}=2.5\log{\int d\lambda S_{x,\lambda} \frac{\lambda}{hc} } +\mathrm{ZP}_x^{\mathrm{mag}}.
\end{equation}
Values of $\mathrm{ZP}_x^{\mathrm{mflux}}$ are reported in Column~4 of Table~\ref{table:zp_and_leff}.

The effective wavelength of the $x$-band as a function of the SED is given by
\begin{equation}
\lambda_x=\frac{\int d\lambda S_{x,\lambda} \lambda^2 f_{\lambda}}{\int d\lambda S_{x,\lambda} \lambda f_{\lambda}}
\end{equation}
\citep[e.g.][]{2012PASP..124..140B}. To estimate this value for the $\bvrijhk$ bands, we use the theoretical spectral models of \citetalias{2013MNRAS.433.1745D}, \citetalias{2014MNRAS.439.3694J}, and \citetalias{2017MNRAS.466...34L} during the radioactive tail, adopting the median as the representative effective wavelength $\bar{\lambda}_x$. These values are summarized in Column~5 of Table~\ref{table:zp_and_leff}. In Column~6 we report the total-to-selective extinction ratio for $\leff_x$ ($R_{\leff_x}$), assuming the extinction curve of \citet{1999PASP..111...63F} with ${R_V=3.1}$. We also compute the $R_{\leff_x}$ values for the photospheric and transition phase ($R_x^{\mathrm{pt}}$). Those values are listed in Column~7 of Table~\ref{table:zp_and_leff}.

To obtain monochromatic fluxes corrected for reddening ($\feffx^{\mathrm{corr}}$) we use
\begin{equation}
\feffx^{\mathrm{corr}}=\feffx\cdot 10^{0.4 (\EGBV+\EhBV)R_{\leff_x}}.
\end{equation}

Using the set of ($\leff_x$, $\feffx^{\mathrm{corr}}$) values, we construct the pSED, from which the quasi-bolometric flux is computed with the trapezoidal rule, i.e.,
\begin{equation}\label{eq:flux_qbol}
\fqbol=\frac{1}{2}\sum_{i=1}^6 (\bar{\lambda}_{x_{i+1}}-\bar{\lambda}_{x_i})(\bar{f}_{x_{i+1}}^{\mathrm{corr}}+\bar{f}_{x_i}^{\mathrm{corr}}),
\end{equation}
where ${x_j=B,V,R,I,J,H,K}$ for ${j=1,\ldots,7}$. The error on $\fqbol$ due to error in photometry, $\sigma_{m_x}$, is given by
\begin{align}
\sigma_{\fqbol}=&\frac{\ln(10)}{5}\left([(\leff_{x_1}-\leff_{x_2})\bar{f}_{x_1}^{\mathrm{corr}}\sigma_{m_{x_1}}]^2 \right. \nonumber \\
&+\sum_{i=2}^6[(\bar{\lambda}_{x_{i+1}}-\bar{\lambda}_{x_{i-1}})\bar{f}_{x_i}^{\mathrm{corr}}\sigma_{m_{x_i}}]^2 \nonumber \\
&\left. +[(\leff_{x_7}-\leff_{x_6})\bar{f}_{x_7}^{\mathrm{corr}}\sigma_{m_{x_7}}]^2\right)^{1/2}.
\end{align}

To estimate the error on $\fqbol$ due to uncertainties on $\EGBV$ and $\EhBV$, we define the pSED total-to-selective extinction ratio as
\begin{equation}
R_{\mathrm{p}}=-\frac{2.5}{\EGBV+\EhBV}\log\left(\frac{\fqbol_{\mathrm{unc}}}{\fqbol}\right).
\end{equation}
Here $\fqbol_{\mathrm{unc}}$ is the quasi-bolometric flux uncorrected for $\EGBV$ and $\EhBV$, i.e., replacing $\feffx^{\mathrm{corr}}$ in equation~(\ref{eq:flux_qbol}) by $\feffx$. With this, the error on $\fqbol$ due to the uncertainties on $\EGBV$ and $\EhBV$, in magnitude scale, is
\begin{equation}
\sigma(2.5\log(\fqbol))=R_{\mathrm{p}}\sqrt{\sigma_{\EGBV}^2+\sigma_{\EhBV}^2}.
\end{equation}

\section{Model selection}\label{sec:model_selection}

Let a set of $N$ data ${X_i={\{x_i,y_i,\sigma_{y_i}\}}}$ (${i=1,\ldots,N}$) and a set of $M$ models ${\{f_{j,\theta}\}}$ (${j=1,\ldots,M}$) with parameters $\theta$. To select the model that best represents the data, we use the Bayesian information criterion \citep[BIC,][]{1978AnSta...6..461S}, performing the following procedure. First, we compute the parameters $\theta$ of each model, maximizing its log-likelihood
\begin{equation}\label{eq:lnL}
\ln\mathcal{L}(f_{j,\theta}|X_i)=-\frac{1}{2}\sum_i^N{\left[\ln(\sigma_i^2)+\frac{(y_i-f_{j,\theta}(x_i))^2}{\sigma_i^2}\right]},
\end{equation}
where ${\sigma_i^2=\sigma_{y_i}^2+\sigma_0^2}$, being $\sigma_0$ the error not accounted for the errors in $y_i$. Next, for each model we compute
\begin{equation}\label{eq:BIC}
\mathrm{BIC}_j=-2\ln\mathcal{L}_j^{\mathrm{max}}+k\ln{N}
\end{equation}
\citep{1978AnSta...6..461S}, where $\ln\mathcal{L}_j^{\mathrm{max}}$ is the maximum log-likelihood and $k$ is the number of free parameters. Then, we evaluate the Bayesian weights
\begin{equation}
p_j=e^{-0.5(\mathrm{BIC}_j-\mathrm{BIC}_l)}\left(\sum_{m=1}^M e^{-0.5(\mathrm{BIC}_m-\mathrm{BIC}_l)}\right)^{-1}
\end{equation}
\citep{Burnham_Anderson2002},
where $l$ is the index of the model with the lowest BIC value. As reference, if ${p_l/p_j>13.0}$ \citep[e.g.][]{2007MNRAS.377L..74L}, then there is a strong evidence in favour of the $l$th model over the $j$th one. Finally, among all models with ${p_l/p_j\leq13.0}$, by the principle of parsimony, we select the one with less parameters.

In the case of least-square regressions (e.g. when $\sigma_{y_i}$ values are not available), we replace equation~(\ref{eq:BIC}) by
\begin{equation}
\mathrm{BIC}_j=\ln(s_j^2)+k/N\cdot\ln{N}
\end{equation}
\citep[e.g.][]{2019MNRAS.483.5459R}, where $s_j^2$ is the average of the squared residuals around the $j$th model.

\section{$^{56}$Ni mass equation}\label{sec:MNi_equation}

Let a normal SN~II with a set of $N$ measurements of ${\{m_{x,i}\}}$ magnitudes ($K$-corrected, see Appendix~\ref{sec:K-correction}) at times $t_i$, and a set of parameters ${\{t_0,\mu,\EGBV,\EhBV,\zsnhel\}}$. The $^{56}$Ni mass (in $\msun$) can be computed using equations~(\ref{eq:Lbol}) and (\ref{eq:Qbol_vs_MNi}), which can be written as
\begin{equation}\label{eq:logMNi_simple}
\log\mni=\langle A_i -D_i\rangle+B.
\end{equation}
Here the angle brackets denote the value that maximizes the log-likelihood of a constant-only model (equation~\ref{eq:lnL}) where, using ${\bc_x=\zp_x^\mathrm{BC}+\beta_x\Delta t/100\,\mathrm{d}}$,
\begin{equation}
A_i=-\frac{m_{x,i}}{2.5}+\frac{(0.39-\beta_x/2.5)t_i}{(1+\zsnhel)100\,\mathrm{d}},
\end{equation}
\begin{equation}
B=\frac{\mu-\zp_x^\mathrm{BC}+A_x^{\mathrm{tot}} }{2.5}-\frac{(0.39-\beta_x/2.5)t_0}{(1+\zsnhel)100\,\mathrm{d}}-3.076,
\end{equation}
where ${A_x^{\mathrm{tot}}=(\EGBV+\EhBV) R_{\leff_x}}$. 
To evaluate whether it is necessary to include the $D_i$ correction in equation~(\ref{eq:logMNi_simple}), we have to check the constancy of the $\log\mni$ estimates with respect to $t_i$. For this, we use the model selection procedure given in Appendix~\ref{sec:model_selection} using the ${\{t_i,A_i,\sigma_{A_i}\}}$ data along with a zero-order and a linear polynomial. If data can be represented by a zero-order polynomial, then ${D_i=0}$ in equation~(\ref{eq:logMNi_simple}). In this case, the $\log\mni$ error is given by
\begin{equation}\label{eq:err_logMNi_simple}
\sigma_{\log\mni}=\sqrt{\sigma_{\langle A_i\rangle}^2+\sigma_B^2}.
\end{equation}
Here
\begin{equation}
\sigma_B^2=\frac{\sigma_{\mu}^2+\sigma_{\zp_x^\mathrm{BC}}^2+\sigma_{A_x^{\mathrm{tot}}}^2}{6.25}+\left[\frac{(0.39-\beta_x/2.5)\sigma_{t_0}}{(1+\zsnhel)100\,\mathrm{d}}\right]^2,
\end{equation}
where $\sigma$ denotes the error in the subscripted quantity, and ${\sigma_{\langle A_i\rangle}=\ssd_{A,k}/\sqrt{N}}$, being
\begin{equation}\label{eq:ssd_eq}
\ssd_{X,k}=\left[\frac{1}{N-k}\sum_i^N\left(X_i-\langle X_i\rangle\right)^2\right]^{1/2}
\end{equation}
the sample standard deviation, and $k$ the number of free parameters.

If data are not well represented by a zero-order polynomial, then there is a dependence of the $\log\mni$ estimates on $t_i$, so we have to include the $D_i$ term. If the time derivative of $A$ is negative as in the case of SN~2014G (left-hand side of Fig.~\ref{fig:2014G}), then we use the $D_i(T_0)$ term given by equations~(\ref{eq:Dfdep}) and (\ref{eq:fdep}). The value of $T_0$ is computed through a log-likelihood maximization (equation~\ref{eq:lnL}). Once the value of $T_0$ is determined, we compute $\log\mni$ using equation~(\ref{eq:logMNi_simple}) and ${D_i=D_i(T_0)}$. The error in $\log\mni$ is given by equation~(\ref{eq:err_logMNi_simple}) replacing $\sigma_{\langle A_i\rangle}$ by $\ssd_{A-D,k}/\sqrt{N}$ if the $\log\mni$ estimates are independent on $t_i$ (using the procedure given in Appendix~\ref{sec:model_selection}), otherwise we replace $\sigma_{\langle A_i\rangle}$ by $\ssd_{A-D,k´}$. If the time derivative of $A$ is positive as in the case of SN~2005cs (Fig.~\ref{fig:2005cs}), then we assume the result obtained with ${D_i=0}$, and ${\sigma_{\langle A_i\rangle}=\ssd_{A,k}}$ in equation~(\ref{eq:err_logMNi_simple}).

\section{Log-normal mean}\label{sec:lognormal_mean}

Let the quantity $x$ (e.g. $\mni$, $D$, $S_x$) and its log-transformation $y=\log x$. If $y$ has a normal PDF with mean $\bar{y}$ and standard deviation $\sigma_y$, then $x$ has a log-normal distribution. The mean of the latter PDF ($\bar{x}$) is not $10^{\bar{y}}$ (which indeed is the median) but
\begin{equation}\label{eq:lognormal_mean}
\bar{x} = 10^{\bar{y}+0.5\ln(10)\sigma_y^2}
\end{equation}
\citep[e.g.][]{Angus1994}. The standard deviation of the $x$ PDF is
\begin{equation}
\sigma_{x}=|\bar{x}| \sqrt{10^{\ln(10)\sigma_y^2}-1}.
\end{equation}
On the other hand, if $x$ has a normal PDF then $\bar{y}$ is not $\log\bar{x}$ but
\begin{equation}\label{eq:lognormal_mean2}
\bar{y}=\log\bar{x}-0.5\log(1+(\sigma_x/\bar{x})^2),
\end{equation}
while the standard deviation of the $y$ PDF is
\begin{equation}
\sigma_y=\sqrt{\ln(1+(\sigma_x/\bar{x})^2)}/\ln(10).
\end{equation}

\section{C3 method}\label{sec:C3}

The C3 method \citep{2014AJ....148..107R,2019MNRAS.483.5459R} relies on the assumption that, during the plateau phase, all normal SNe~II have similar linear $\vi$ versus $\bv$ C3s. For an SN   with a measurement of $\vi$ and $\bv$ (corrected for $\EGBV$ and $K$-correction, e.g. \citealt{2019MNRAS.483.5459R}) at time $t_i$ (during the plateau phase), the host galaxy color excess is given by
\begin{equation}
E^{\mathrm{h}}_{\bv,i}=A_i+\zp_\mathrm{C3}.
\end{equation}
Here $\zp_\mathrm{C3}$ is the zero-point for the C3 method, and
\begin{equation}
A_i=[(\vi)_i-m_\mathrm{C3}\cdot(\bv)_i]/(R_V-R_I-m_\mathrm{C3})
\end{equation}
\citep[e.g.][]{2019MNRAS.483.5459R}, where ${R_V=3.1}$, ${R_I=1.72}$, and ${m_\mathrm{C3}=0.45\pm0.07}$ is the $\vi$ versus $\bv$ C3 slope \citep{2019MNRAS.483.5459R}. For an SN with a set of $N$ measurement of $\{(\bv)_i,(\vi)_i\}$, the $\EhBV$ value is given by
\begin{equation}\label{eq:C3_eq}
\EhBV= \langle A_i \rangle+\zp_\mathrm{C3},
\end{equation}
where the angle brackets denote the value that maximizes the log-likelihood of a constant-only model (equation~\ref{eq:lnL}). 

The $\zp_\mathrm{C3}$ value can be computed from a set of reddening-free SNe. For each of those SNe, following equation~(\ref{eq:C3_eq}), we have $\zp_\mathrm{C3}^\mathrm{SN}=-\langle A_i \rangle^\mathrm{SN}$. Using this set of $\zp_\mathrm{C3}^\mathrm{SN}$ values, we calculate the mean and the $\ssd$ value, which we adopt as $\zp_\mathrm{C3}$ and its error, respectively. To determine $\zp_\mathrm{C3}$ we use SNe~2005ay, 2008gz, 2013ej, and 2014cx, which are found to have low reddening by other methods (see Table~\ref{table:Eh_values}), and SN~2003bn \citep{2016AJ....151...33G}, which is also affected by low reddening \citep[e.g.][]{2010ApJ...715..833O}. Using those SNe, we obtain ${\zp_\mathrm{C3}=-0.116\pm0.024}$\,mag. This value is equivalent to the $\zp_\mathrm{C3}$ inferred by \citet{2019MNRAS.483.5459R} from SNe~2003bn and 2013ej, but with a lower error as we use more SNe to determine the $\zp_\mathrm{C3}$ value.

The expression for the $\EhBV$ error is
\begin{equation}
\sigma_{\EhBV} = \sqrt{\ssd_{A,k}^2+\sigma_{\zp_\mathrm{C3}}^2+\sigma_m^2},
\end{equation}
where $\ssd_{A,k}$ is given by equation~(\ref{eq:ssd_eq}), and
\begin{equation}
\sigma_m=\left|\frac{\EhBV-\langle(\bv)_i\rangle+\overline{\langle(\bv)_i^\mathrm{ZP}\rangle}}{R_V-R_I-m_\mathrm{C3}}\right|\sigma_{m_\mathrm{C3}}
\end{equation}
is the error induced by the $m_\mathrm{C3}$ uncertainty. In the latter expression, $\overline{\langle(\bv)_i^\mathrm{ZP}\rangle}=1.065$ is the mean $\langle(\bv)_i^\mathrm{ZP}\rangle$ value of the five SNe we use to calibrate the C3 method. 

\begin{figure}
\includegraphics[width=1.0\columnwidth]{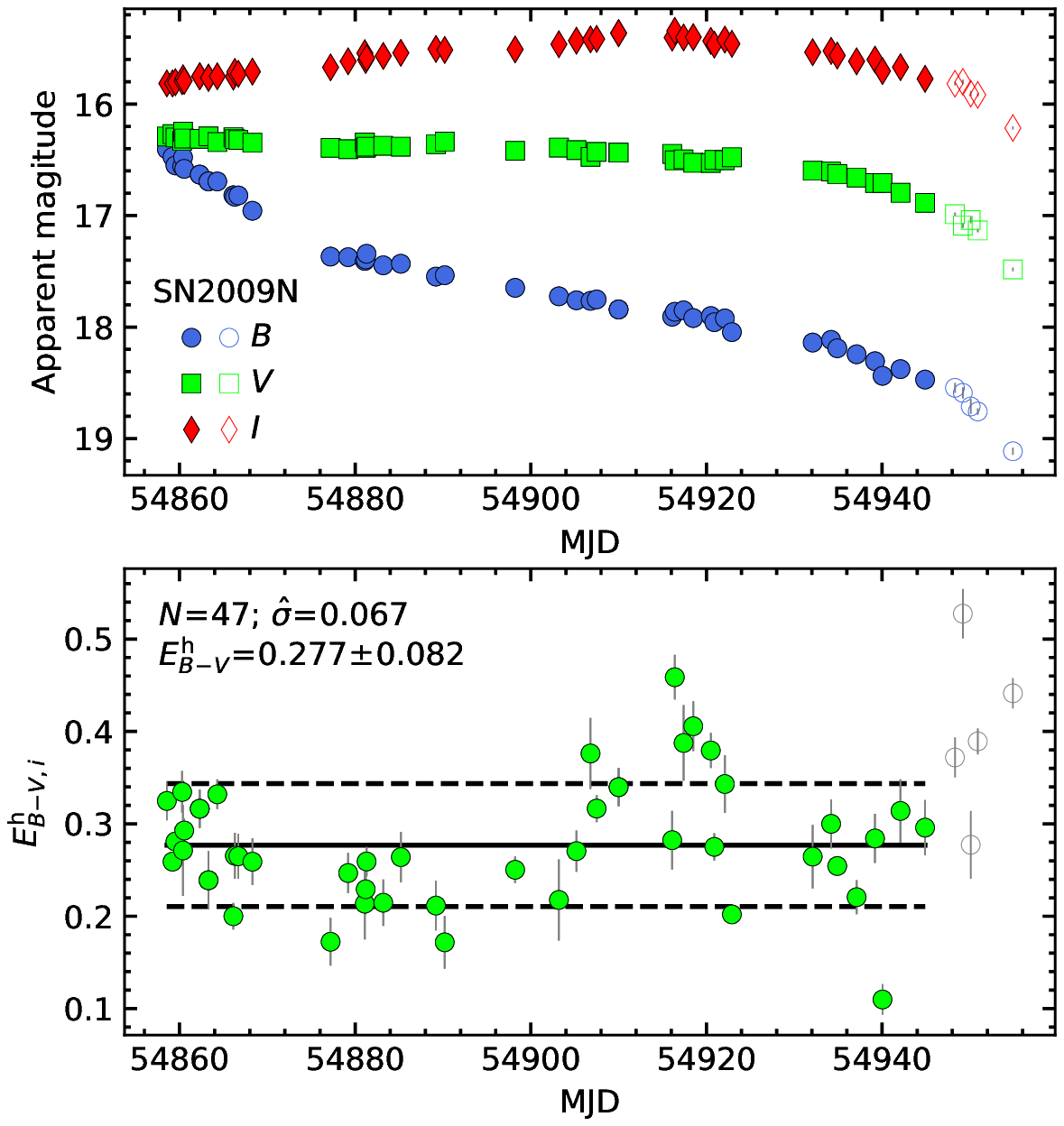}
\caption{Top panel: $BV\!I$ light curves of SN~2009N during the plateau phase. Bottom panel: $E^{\mathrm{h}}_{\bv,i}$ estimates obtained with the C3 method. Filled symbols are the data used to compute $\EhBV$. The solid line corresponds to the $\EhBV$ value that maximizes the likelihood, and dashed lines are the $\pm1\,\ssd$ limits around $\EhBV$. Error bars are $1\,\sigma$ errors due to uncertainties on photometry.}
\label{fig:2009N_c3}
\end{figure}

For example, Fig.~\ref{fig:2009N_c3} shows the $E^{\mathrm{h}}_{\bv,i}$ estimates for SN~2009N obtained with the C3 method. For that SN we compute $\EhBV=0.277\pm0.082$\,mag.

\section{Spectroscopic reddening}\label{sec:spec_EhBV}

Let a set of $N$ SN spectra $\{f_i\}$ corrected for $\EGBV$ and $\zsnhel$, where each spectrum has a dispersion $\Delta_i$ (in \AA/pixel) and $n_i$ flux measurements at wavelengths $\lambda_{i,k}$ (${k=1,\ldots,n_i}$), and a set of $M$ reddening-free spectra $\{F_j\}$. For each $\{f_i,F_j\}$ combination, the $\EhBV$ value can be inferred minimizing
\begin{equation}\label{eq:spec_EhBV}
\ssd_{i,j}^2=\frac{\Delta_i^{-1}}{n_i\!-\!2}\sum_k^{n_i}{\left[ \log{\left[\frac{f_i(\lambda_{i,k})}{F_j(\lambda_{i,k})}\right]}-\!A+\frac{\EhBV R_{\lambda_{i,k}}}{2.5} \right]^2}.
\end{equation}
Here $A$ is a constant accounting for differences in SN distances and photosphere radius sizes, and $R_{\lambda}$ is the extinction curve for the host-galaxy dust along the line of sight. The best estimation for $\EhBV$ is that provided by the $\{f_i,F_j\}$ combination with the lowest $\ssd_{i,j}$ value.

\begin{figure}
\includegraphics[width=1.0\columnwidth]{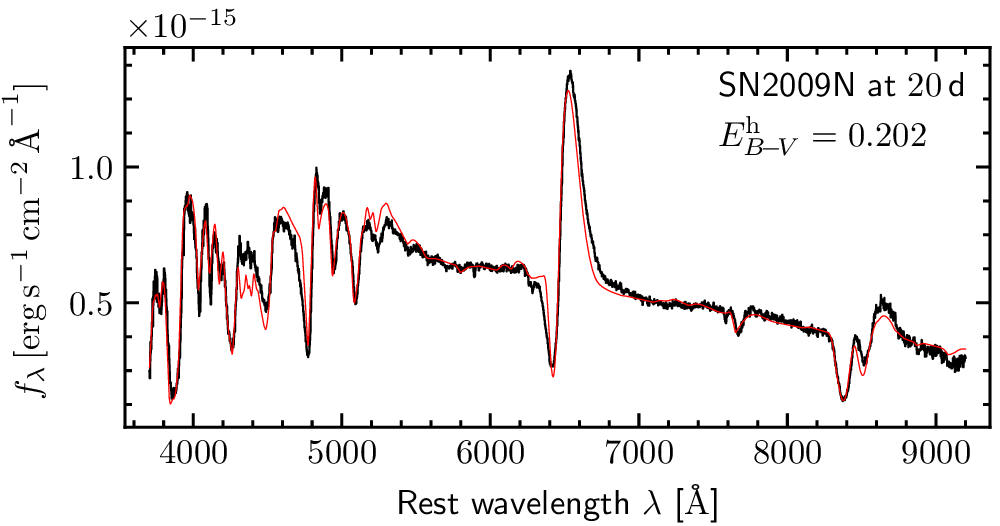}
\caption{Spectrum of SN~2009N (corrected for $\EGBV$ and $\zsnhel$, black line) and model (reddened for $\EhBV=0.202$\,mag, red line) that minimizes equation~(\ref{eq:spec_EhBV}).}
\label{fig:2001X_EhBV_spec}
\end{figure}

In this work we use the \citetalias{2013MNRAS.433.1745D} and \citetalias{2017MNRAS.466...34L} spectral models as $\{F_j\}$, and adopt the extinction curve of \citet{1999PASP..111...63F} with ${R_V=3.1}$. For an input SN we select spectra: (1) earlier that 40\,d since the explosion, (2) with a minimum wavelength coverage of 360{\textendash}700\,nm, and (3) with differences between synthetic and observed colours lower than 0.1\,mag. These criteria are such that the selected spectra are suitable to measure $\EhBV$ with an error of 0.05\,mag \citep[e.g.][]{2010ApJ...715..833O}. To be conservative, we adopt an error of 0.1\,mag. For example, Fig.~\ref{fig:2001X_EhBV_spec} shows the $\{f_i,F_j\}$ combination for SN~2009N that minimizes equation~(\ref{eq:spec_EhBV}). For that SN we compute $\EhBV=0.202$\,mag.

\section{Magnitude transformations}\label{sec:mag_transf}

We present magnitude transformations between Sloan and Johnson--Kron--Cousins filters, and between $r_\ztf$ and $R$. All magnitudes are corrected for $\EGBV$, $\EhBV$, and $K$-correction, while reported uncertainties are $1\,\ssd$ errors.

\begin{figure}
\includegraphics[width=1.0\columnwidth]{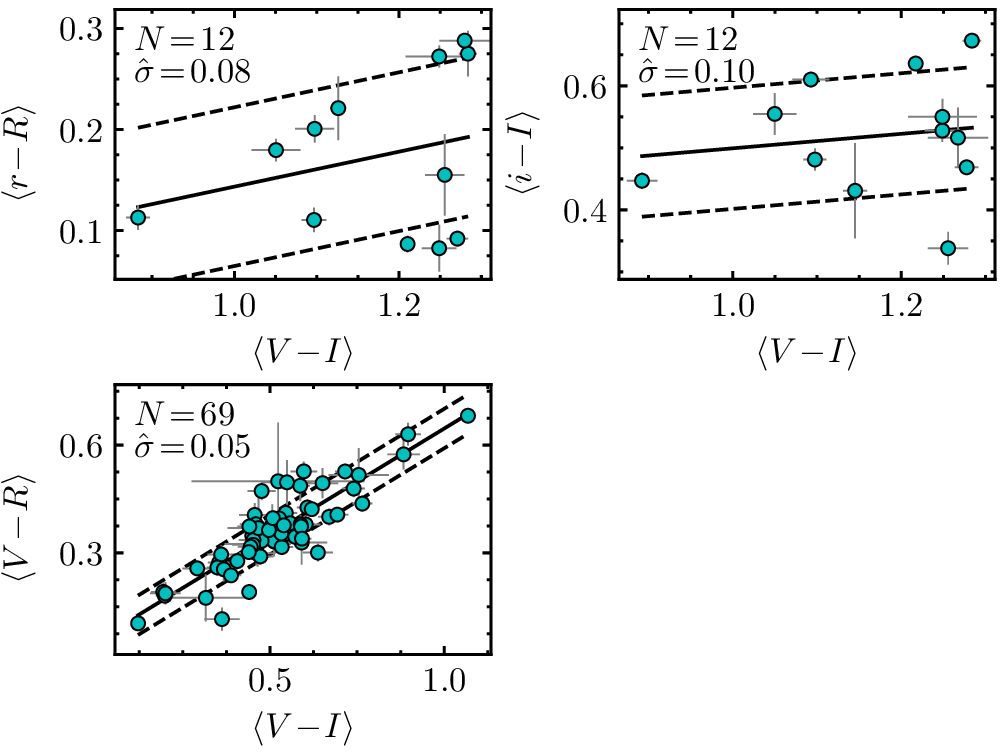}
\caption{Upper-left (upper-right) panel: mean $r\!-\!R$ ($i\!-\!I$) against mean $\vi$ for 12 normal SNe~II during the radioactive tail. Bottom panel: mean $V\!-\!R$ versus mean $\vi$ for 69 normal SNe~II during the photospheric phase. Solid and dashed lines are straight line fits and $\pm1\,\ssd$ limits, respectively. Error bars are $1\,\sigma$ errors.}
\label{fig:conv_rt}
\end{figure}

Using 12~SNe in out set having $V\!RI$ and $ri$ photometry in the radioactive tail (see Table~\ref{table:SN_sample}), we find that $r\!-\!R$ and $i\!-\!I$ have a small linear dependence on $\vi$, i.e.,
\begin{equation}\label{eq:conv_rt}
c_y=a+b(\vi),
\end{equation}
where $c_y=r\!-\!R, i\!-\!I$. Fig.~\ref{fig:conv_rt} shows the mean $r\!-\!R$ ($\langle r\!-\!R\rangle $) and $i\!-\!I$ ($\langle i\!-\!I\rangle $) values of each SN as a function of the mean $\vi$ ($\langle \vi \rangle$). The average of equation~(\ref{eq:conv_rt}) is $\langle c_y \rangle =a+b\langle \vi \rangle $, so we can compute $a$ and $b$ fitting a straight line to the data plotted in the figure. We obtain
\begin{equation}
r-R=0.17(\vi)-0.03\pm0.08
\end{equation}
and
\begin{equation}
i-I=0.12(\vi)+0.38\pm0.10.
\end{equation}
Using the same technique for $V\!-\!R$ versus $\vi$ during the photospheric phase (69~SNe, Fig.~\ref{fig:conv_rt}), we obtain
\begin{equation}
V\!-\!R=0.59(\vi)+0.05\pm0.05.
\end{equation}

\begin{figure}
\includegraphics[width=1.0\columnwidth]{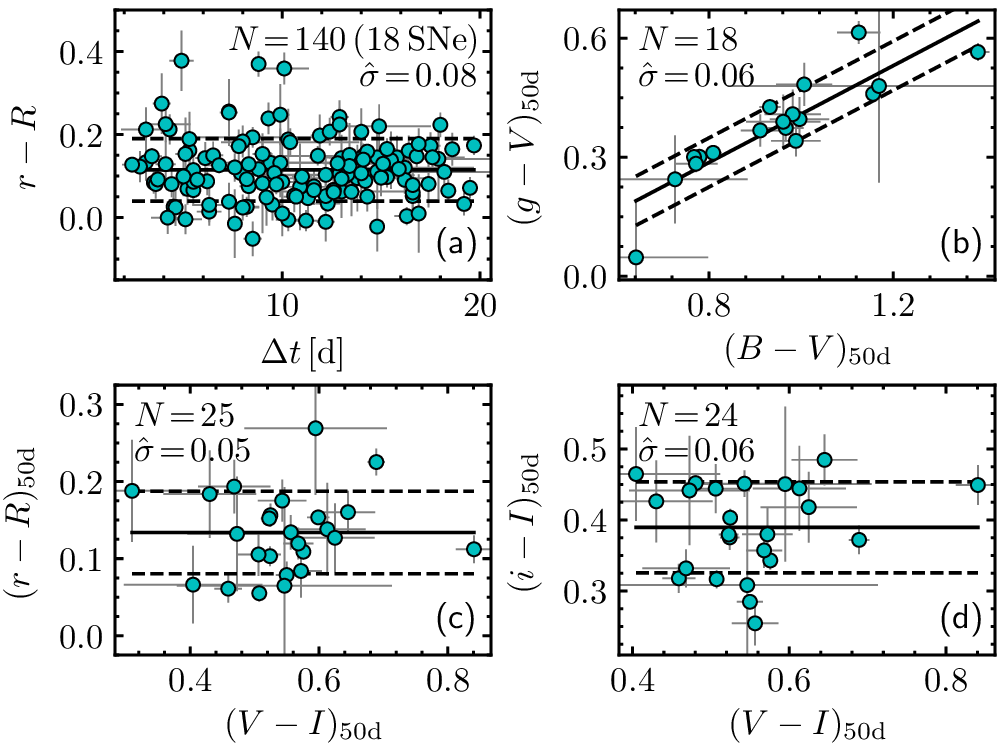}
\caption{Panel (a): $r\!-\!R$ versus time since explosion for ${\Delta t<20}$\,d. Panel (b): $g\!-\!V$ against $B\!-\!V$ at $\Delta t=50$\,d. Panels (c) and (d): $r\!-\!R$ and $i\!-\!I$ versus $\vi$ at $\Delta t=50$\,d, respectively. Solid and dashed lines are the best fits and the $\pm1\,\ssd$ limits, respectively. Error bars are $1\,\sigma$ errors.}
\label{fig:conv_max_50d}
\end{figure}

To convert $r$ to $R$ magnitudes near the maximum light, we use 18~SNe with $r\!R$ photometry at $\Delta t<20$\,d (see Table~\ref{table:SN_sample}). We find that $r\!-\!R$ is consistent with a constant value of $r\!-\!R=0.12\pm0.08$\,mag (see Fig.~\ref{fig:conv_max_50d}(a)).

To compute conversions between magnitudes at ${\Delta t=50}$\,d we use 18, 25, and 24~SNe for $g/V$, $r/R$, and $i/I$ transformations, respectively. We find $(g\!-\!V)_{50\mathrm{d}}=-0.20+0.61(B\!-\!V)_{50\mathrm{d}}\pm0.06$ (see Fig~\ref{fig:conv_max_50d}(b)), while the ${(r\!-\!R)_{50\mathrm{d}}}$ and ${(i\!-\!I)_{50\mathrm{d}}}$ values are consistent with constant values of $0.13\pm0.05$ and $0.39\pm0.06$, respectively (see Fig~\ref{fig:conv_max_50d}(c)-(d)).

The public ZTF photometry is not in the SDSS photometric system but in the native ZTF one, so we cannot just adopt $r_\ztf=r$. To assess how similar the $r_\ztf$ photometry is to the $r$- and $R$-band photometry, we compute synthetic magnitudes (Appendix~\ref{sec:syn_mag}) using SN spectra. From the spectra of the SNe in our set, we select those having differences between synthetic and observed $V\!-\!r$, $V\!-\!R$, $V\!-\!i$, and/or $\vi$ colours lower than 0.1\,mag.

\begin{figure}
\includegraphics[width=1.0\columnwidth]{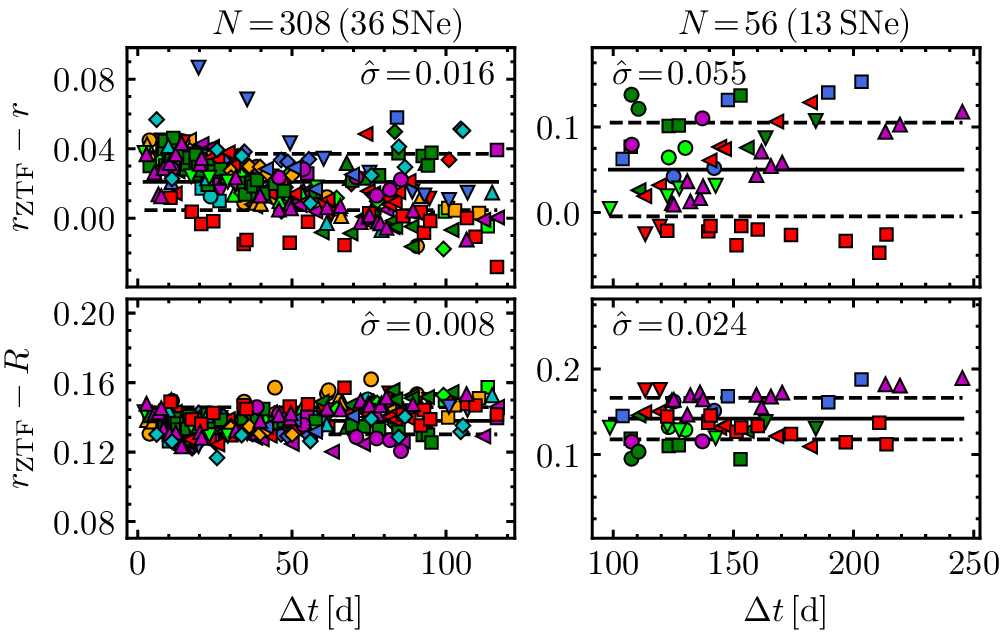}
\caption{$r_\ztf\!-\!r$ (top panels) and $r_\ztf\!-\!R$ (bottom panels) values computed from spectra of normal SNe~II in the photospheric and transition phase (left-hand panels) and in the radioactive tail (right-hand panels), against the time since explosion. Solid and dashed lines are mean values and $\pm1\,\ssd$ limits, respectively.}
\label{fig:conv_ztf}
\end{figure}

The left-hand panels of Fig.~\ref{fig:conv_ztf} show the $r_\ztf\!-\!r$ (top panel) and $r_\ztf\!-\!R$ (bottom panel) values, computed with 308 spectra of 36~SNe during the photospheric and transition phase, against the time since explosion. We see that the $r_\ztf\!-\!R$ values have a lower dependence on time and a lower scatter than the $r_\ztf\!-\!r$ estimates (the $y$-axis scale is the same in the two panels). Therefore, the $r_\ztf$ light curves during the photospheric and transition phase are more similar to the $R$-band light curves than the $r$-band ones. Given the low $\ssd$ value (0.008\,mag) around the mean ${r_\ztf\!-\!R}$ (0.138\,mag), we consider the time dependence to be negligible and adopt ${r_\ztf\!-\!R=0.138\pm0.008}$\,mag to transform $r_\ztf$ into $R$ magnitudes. We find similar results for the radioactive tail (right-hand panels, using 56 spectra of 13~SNe), for which we adopt $r_\ztf\!-\!R=0.142\pm0.024$\,mag.

In the literature we find only one normal SN~II having $r_\ztf$ and $R$ photometry (SN~2018aoq; \citealt{2019MNRAS.487.3001T}). For that SN we measure a mean ${r_\ztf\!-\!R}$ of $0.17\pm0.04$\,mag (16 points in the photospheric phase), which is consistent within its error with our result from synthetic magnitudes.

\section{$K$-corrections}\label{sec:K-correction}

We express the $x$-band $K$-correction ($K_{x,j}$) such that the apparent magnitude ($m_{x,j}$) corrected for $K$-correction is ${m_{x,j}^K=m_{x,j}-K_{x,j}}$, where
\begin{equation}
K_{x,j}=-2.5\log(1+\zsnhel)+k_{x,j}^s,
\end{equation}
being
\begin{equation}
k_{x,j}^s=2.5\log\left(\frac{d\lambda S_{x,\lambda} \lambda f_{j,\lambda}}{d\lambda' S_{x,\lambda} \lambda' f_{j,\lambda'}}\right)
\end{equation}
and $\lambda'=\lambda/(1+\zsnhel)$. To compute $k_x^s$ we use the spectral models of \citetalias{2013MNRAS.433.1745D}, \citetalias{2014MNRAS.439.3694J}, and \citetalias{2017MNRAS.466...34L}, and $\zsnhel$ values uniformly distributed between 0.001 and 0.043. Expressing $k_x^s$ as polynomials on $\Delta t$, we obtain
\begin{equation}
k_{x,j}^s=\left(\sum_{i=0} k_i\cdot\left[\frac{\Delta t_j}{100\,\mathrm{d}}\right]^i\pm\ssd\right)\zsnhel.
\end{equation}
Table~\ref{table:k_corr} lists the parameters for the photospheric and radioactive tail phase. At ${\Delta t=200}$\,d (the midpoint of the time range we use to estimate $\mni$) $K_x$ values are ${<0.03}$\,mag for $\zsnhel=0.009$ (the median redshift in our set), while for ${\zsnhel>0.02}$ (15 out of 110~SNe in our sample) $K_x$ values start to be comparable to the typical photometry errors.

\begin{table}
\caption{$K$-correction parameters.}
\label{table:k_corr}
\begin{tabular}{lccccccc}
\hline
$x$       & $k_0^*$ &$k_1^*$& $k_2^*$ &$\ssd^*$& $k_0^\dagger$ & $k_1^\dagger$ &$\ssd^\dagger$\\
\hline
 $B$      & $-1.50$ & 19.8  & $-10.8$ & 1.32 & --    & --      & --   \\
 $g$      & $-1.23$ & 16.1  & $-10.2$ & 0.90 & --    & --      & --   \\
 $V$      & $-1.46$ & 8.85  & $-5.27$ & 0.66 & 8.10  & $-1.69$ & 2.97 \\
 $r$      & $-0.05$ & 6.56  & $-4.12$ & 0.72 & 4.12  & $0.88$  & 2.21 \\
 $r_\ztf$ & $-0.70$ & 4.21  & $-2.54$ & 0.24 & 0.60  & $0.77$  & 1.22 \\
 $R$      & $-0.42$ & 3.56  & $-2.09$ & 0.21 & 1.48  & $0.57$  & 0.92 \\
 $i$      & $-1.05$ & 2.33  & $-0.87$ & 0.37 & 1.05  & $0.26$  & 1.00 \\
 $I$      & $-1.00$ & 3.60  & $-1.50$ & 0.24 & 6.31  & $-1.69$ & 1.07 \\
\hline
\multicolumn{8}{p{.95\columnwidth}}{\textit{Note.} Parameters for the photospheric ($*$) and radioactive tail ($\dagger$) phase.}\\
\end{tabular}
\end{table}

\section{Indirect Steepness value}\label{sec:Sx_from_Sb}

In order to calculate $S_x$ indirectly from steepnesses in bands other than $x$, we compute steepness transformations.

\begin{figure}
\includegraphics[width=1.0\columnwidth]{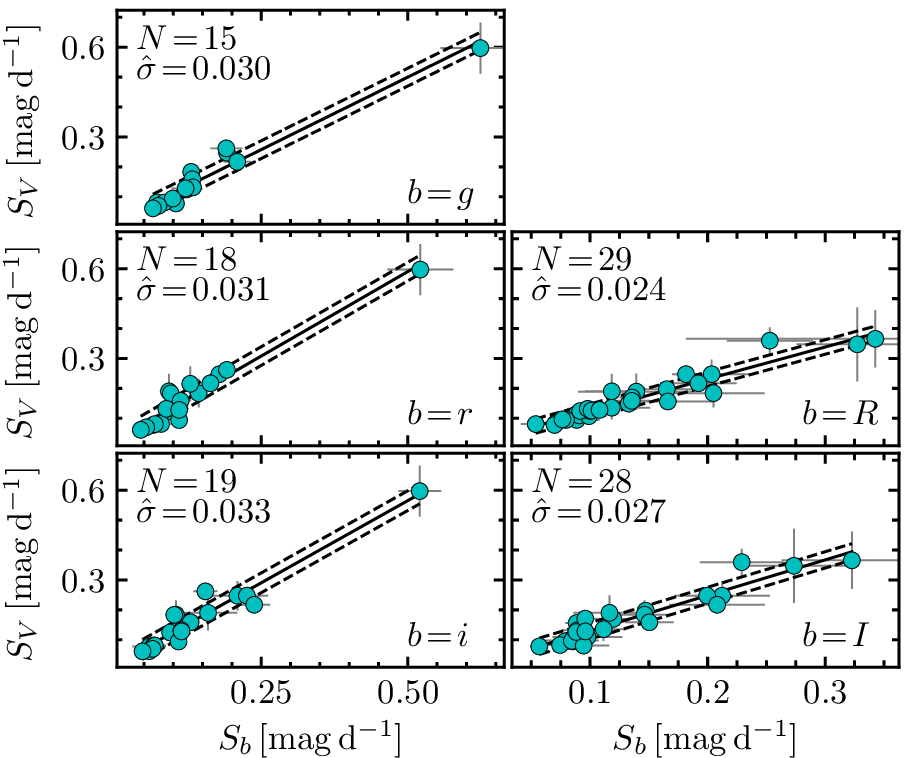}
\caption{$V$-band steepness against the steepness for the $gr\!RiI$ bands. Solid lines are straight-line fits to the data, while dashed ones are $\pm1\,\ssd$ limits around the fits. Error bars are $1\,\sigma$ errors.}
\label{fig:S_vs_Sr_Si}
\end{figure}

Fig.~\ref{fig:S_vs_Sr_Si} shows $S_V$ against the steepness for $gr\!RiI$ bands. Using the model selection procedure (Appendix~\ref{sec:model_selection}), we find that the five correlations plotted in the figure are well represented by straight lines. In general, for the 30 possible combinations with the $gV\!r\!RiI$ bands, we find that correlations between the $x$- and $b$-band steepness are given by
\begin{equation}\label{eq:SV_vs_Sx}
S_x=c_{x,b}+d_{x,b} S_b.
\end{equation}
Table~\ref{table:Sy_vs_Sx_pars} reports the $c_{x,b}$ and $d_{x,b}$ parameters, their bootstrap errors, and the $\ssd_{x,b}$ values.

For an SN with a set of $n$ steepness estimates, $\{S_{b}\}_{b\neq x}$, we can compute a set of $n$ $S_x$ values, $\{S_x\}$, using equation~(\ref{eq:SV_vs_Sx}). Our best indirect measurement of $S_x$ will be the weighted mean of the $\{S_x\}$ values, $S_x^*$, given by
\begin{equation}
S_x^*= \frac{1}{\sum_{b\neq x} w_{x,b}}\sum_{b\neq x} (c_{x,b}+d_{x,b}S_b)w_{x,b}, 
\end{equation}
where $w_{x,b}=[d_{x,b}^2 \sigma_{S_b}^2+\ssd_{x,b}^2]^{-1}$. The error on $S_x^*$ is simply
\begin{equation}
\sigma_{S_x^*}=\sqrt{\frac{1}{\sum_{b\neq x} w_{x,b}}}.
\end{equation}

\section{Reference list for the data}\label{sec:all_refs}

Here the references for Tables~\ref{table:SN_sample}, \ref{table:mu_values}, \ref{table:Eh_values}, and \ref{table:t0_values}.

(1) \citet{1982AA...116...35B}; (2) \citet{1982PASP...94..578B}; (3) \citet{1983PZ.....22...39T}; (4) \citet{1989AJ.....97..186P}; (5) \citet{1990AJ....100..782R}; (6) \citet{1991AA...247..410B}; (7) \citet{1993MNRAS.265..471T}; (8) \citet{1993AJ....105.2236S}; (9) \citet{1994AA...285..147B}; (10) \citet{1995AA...293..723C}; (11) \citet{1995AJ....110.2868B}; (12) \citet{2016AJ....151...33G}; (13) \citet{1996AJ....111.1286C}; (14) \citet{2004MNRAS.347...74P}; (15) \citet{2013AA...555A.142I}; (16) \citet{2001MNRAS.322..361B}; (17) \citet{2003MNRAS.338..939E}; (18) \citet{2009AJ....137...34K}; (19) \citet{2014MNRAS.442..844F}; (20) \citet{1999IAUC.7319....2R}; (21) \citet{2009AA...500.1013P}; (22) \citet{2002AJ....124.2490L}; (23) \citet{2006MNRAS.368.1169P}; (24) \citet{2007PZ.....27....5T}; (25) \citet{2014MNRAS.439.2873S}; (26) \citet{2005MNRAS.359..906H}; (27) \citet{2014MNRAS.445..554F}; (28) \citet{2008BASI...36...79G}; (29) \citet{2006AJ....131.2245Z}; (30) \citet{2006MNRAS.369.1780V}; (31) \citet{2009ApJ...695..619V}; (32) \citet{2011ApJ...732..109M}; (33) \citet{2014ApJ...786...67A}; (34) \citet{2006MNRAS.372.1315S}; (35) \citet{2007MNRAS.381..280M}; (36) \citet{2010MNRAS.404..981M}; (37) \citet{2012ApJ...756L..30A}; (38) \citet{2006AA...460..769T}; (39) \citet{2008ApJ...675..644D}; (40) \citet{2009MNRAS.394.2266P}; (41) \citet{2017ApJS..233....6H}; (42) \citet{2009AJ....137.4517B}; (43) \citet{2011ApJ...731...47A}; (44) \citet{2010ApJ...715..541A}; (45) \citet{2011MNRAS.417..261I}; (46) \citet{2019MNRAS.490.2799D}; (47) \citet{2012AJ....143...19V}; (48) \citet{2013AJ....146...24V}; (49) \citet{2013msao.confE.176P}; (50) \citet{2011MNRAS.414..167R}; (51) \citet{2011ApJ...736...76R}; (52) \citet{2014MNRAS.438..368T}; (53) \citet{2014PZ.....34....2T}; (54) \citet{2012MNRAS.422.1122I}; (55) \citet{2011ApJ...742....6E}; (56) \citet{2015MNRAS.450.3137T}; (57) \citet{2011MNRAS.417.1417F}; (58) \citet{2016ApJ...820...33R}; (59) \citet{2013MNRAS.434.1636T}; (60) \citet{2013ApJ...764L..13B}; (61) \citet{2013MNRAS.433.1871B}; (62) \citet{2013NewA...20...30M}; (63) \citet{2014ApJ...787..139D}; (64) \citet{2015MNRAS.448.2312B}; (65) \citet{2018MNRAS.475.1937T}; (66) \citet{2015MNRAS.450.2373B}; (67) \citet{2014ApJ...797....5Z}; (68) \citet{2016MNRAS.459.3939V}; (69) \citet{2015MNRAS.448.2608V}; (70) \citet{2014JAVSO..42..333R}; (71) \citet{2015ApJ...806..160B}; (72) \citet{2015ApJ...807...59H}; (73) \citet{2016ApJ...822....6D}; (74) \citet{2016MNRAS.461.2003Y}; (75) \citet{2018MNRAS.476.1497B}; (76) \citet{2016MNRAS.455.2712B}; (77) \citet{2016MNRAS.462..137T}; (78) \citet{2016ApJ...832..139H}; (79) \citet{2021MNRAS.tmp..825D}; (80) \citet{2018MNRAS.480.2475S}; (81) \citet{2014AcA....64..197W}; (82) \citet{2019MNRAS.490.1605D}; (83) \citet{2018MNRAS.479.2421D}; (84) \citet{2019MNRAS.485.5120B}; (85) \citet{2018MNRAS.475.3959H}; (86) \citet{2019ApJ...873L...3B}; (87) \citet{2020MNRAS.497..361M}; (88) \citet{2018ApJ...859...78N}; (89) \citet{2018ApJ...861...63H}; (90) \citet{2019ApJ...882...68S}; (91) \citet{2018ApJ...853...62T}; (92) \citet{2019ApJ...881...22A}; (93) \citet{2021ApJ...907...52T}; (94) \citet{2018ApJ...864L..20R}; (95) \citet{2018AstL...44..315T}; (96) \citet{2019ApJ...873..127T}; (97) \citet{2019ApJ...875..136V}; (98) \citet{2019ApJ...876...19S}; (99) \citet{2019MNRAS.487..832B}; (100) \citet{2019ApJ...885...43A}; (101) \citet{2020ApJ...906...56D}; (102) \citet{2021MNRAS.501.1059R}; (103) \citet{2018AJ....156..105A}; (104) \citet{2020MNRAS.496.3402D}; (105) \citet{2016ApJ...826...56R}; (106) \citet{2017AJ....154...51M}; (107) \citet{2019ApJ...882...34F}; (108) \citet{2006ApJS..165..108S}; (109) \citet{2016ApJ...828L...5C}; (110) \citet{2016MNRAS.457.1419M}; (111) \citet{2017ApJ...847...88Z}; (112) \citet{2017ApJ...836...74J}; (113) \citet{2007ApJ...661..815R}; (114) \citet{2020AstBu..75..384T}; (115) \citet{2001PhDT.......173H}; (116) \citet{2014AJ....148..107R}; (117) \citet{2010ApJ...715..833O}; (118) \citet{1998ApJ...498L.129T}; (119) \citet{2006MNRAS.370.1752P}; (120) \citet{2007ApJ...666.1093Q}; (121) \citet{2007ApJ...661.1013L}; (122) \citet{2017NatPh..13..510Y}; (123) \citet{2016PASA...33...55C}; (124) \citet{1982ApJ...257L..63T}; (125) \citet{2017ApJ...850...89G}; (126) \citet{2008AA...488..383H}; (127) \citet{2006MNRAS.369.1303H}; (128) \citet{2017PASP..129e4201S}; (129) \citet{2019TNSTR1124....1L}; (130) \citet{2016TNSTR.234....1I}; (131) \citet{2016TNSTR.673....1D}; (132) \citet{2017TNSTR.548....1W}.

\section{Tables}\label{sec:tables}

\begin{table*}
\scriptsize
\caption{SN sample.}
\label{table:SN_sample}
\begin{tabular}{l l c c c c c l}
\hline
SN     & Host galaxy & $\EGBV$ & $c\zsnhel$     &$\mu$ & $\EhBV$ & $t_0$ & References\\
       &             & (mag)   & (km\,s$^{-1}$) & (mag)& (mag)   & (MJD) &           \\
\hline
 1980K & NGC 6946 & 0.291 & $40$ & $29.44\pm0.09$ & $-0.106\pm0.084$ & $44528.2\pm5.9$ & 1, 2, 3 \\
 1986I & M99 & 0.034 & $2407$ & $30.71\pm0.40$ & $0.212\pm0.142$ & $46556.4\pm1.9$ & 4 \\
 1988A & M58 & 0.035 & $1517$ & $31.17\pm0.35$ & $0.090\pm0.111$ & $47176.2\pm0.3$ & 5, 6, 7 \\
 1990E & NGC 1035 & 0.022 & $1362$ & $30.83\pm0.26$ & $0.598\pm0.072$ & $47934.4\pm1.4$ & 8, 9 \\
 1990K & NGC 150 & 0.012 & $1584$ & $31.57\pm0.24$ & $0.227\pm0.034$ & $48013.8\pm4.2$ & 10 \\
 1991G & NGC 4088 & 0.017 & $757$ & $30.76\pm0.17$ & $0.025\pm0.071$ & $48281.5\pm5.3$ & 11 \\
 1991al & LEDA 140858 & 0.044 & $4572$ & $32.80\pm0.12$ & $0.067\pm0.026$ & $48446.7\pm3.7$ & 12 \\
 1992H & NGC 5377 & 0.015 & $1793$ & $32.07\pm0.17$ & $0.167\pm0.123$ & $48656.4\pm4.5$ & 13 \\
 1992ba & NGC 2082 & 0.050 & $1135$ & $30.78\pm0.29$ & $0.096\pm0.029$ & $48886.1\pm2.8$ & 12 \\
 1994N & UGC 5695 & 0.032 & $2940$ & $33.24\pm0.20$ & $0.045\pm0.036$ & $49453.9\pm4.5$ & 14 \\
 1995ad$^{\dagger}$ & NGC 2139 & 0.029 & $1837$ & $32.24\pm0.18$ & $0.090\pm0.111$ & $49972.0\pm4.2$ & 15 \\
 1996W & NGC 4027 & 0.036 & $1617$ & $31.86\pm0.19$ & $0.260\pm0.054$ & $50180.2\pm2.5$ & 15 \\
 1997D & NGC 1536 & 0.017 & $1217$ & $30.93\pm0.25$ & $0.090\pm0.111$ & $50361.0\pm15.0$ & 16 \\
 1999ca & NGC 3120 & 0.094 & $2791$ & $32.82\pm0.11$ & $0.076\pm0.033$ & $51273.4\pm2.9$ & 12 \\
 1999em$^{\dagger}$ & NGC 1637 & 0.035 & $800$ & $30.31\pm0.09$ & $0.082\pm0.034$ & $51474.5\pm2.0$ & 12, 17, 18, 19 \\
 1999ga & NGC 2442 & 0.173 & $1466$ & $31.51\pm0.05$ & $0.511\pm0.084$ & $51419.5\pm20.0$ & 20, 21 \\
 1999gi & NGC 3184 & 0.014 & $503$ & $30.40\pm0.18$ & $0.250\pm0.033$ & $51518.6\pm1.8$ & 19, 22 \\
 2001X & NGC 5921 & 0.034 & $1480$ & $31.52\pm0.21$ & $0.063\pm0.031$ & $51963.6\pm2.6$ & 19 \\
 2001dc & NGC 5777 & 0.009 & $2145$ & $32.69\pm0.16$ & $0.694\pm0.076$ & $52046.0\pm5.0$ & 14 \\
 2002gw & NGC 922 & 0.016 & $3143$ & $32.98\pm0.22$ & $0.138\pm0.036$ & $52554.8\pm3.0$ & 12 \\
 2002hh & NGC 6946 & 1.065 & $110$ & $29.44\pm0.09$ & $1.545\pm0.182$ & $52575.6\pm2.5$ & 23, 24 \\
 2002hx & PGC 23727 & 0.045 & $9299$ & $35.53\pm0.08$ & $0.140\pm0.043$ & $52580.3\pm3.7$ & 12 \\
 2003B & NGC 1097 & 0.023 & $1141$ & $30.62\pm0.25$ & $0.023\pm0.033$ & $52622.2\pm4.2$ & 12 \\
 2003T & UGC 4864 & 0.027 & $8368$ & $35.33\pm0.10$ & $0.154\pm0.040$ & $52652.6\pm4.6$ & 12 \\
 2003Z & NGC 2742 & 0.033 & $1289$ & $31.62\pm0.23$ & $0.109\pm0.050$ & $52665.1\pm2.4$ & 19, 25 \\
 2003fb & UGC 11522 & 0.155 & $5081$ & $34.05\pm0.13$ & $0.371\pm0.046$ & $52779.0\pm4.4$ & 12 \\
 2003gd & M74 & 0.060 & $657$ & $29.95\pm0.08$ & $0.144\pm0.040$ & $52716.5\pm21.0$ & 12, 19, 26 \\
 2003hd & ESO 543-G17 & 0.011 & $12031$ & $35.97\pm0.08$ & $0.124\pm0.042$ & $52857.2\pm2.8$ & 12 \\
 2003hk & NGC 1085 & 0.032 & $6880$ & $34.77\pm0.12$ & $0.142\pm0.054$ & $52868.3\pm2.7$ & 12, 27 \\
 2003hn$^{\dagger}$ & NGC 1448 & 0.012 & $1305$ & $31.31\pm0.04$ & $0.169\pm0.040$ & $52865.6\pm4.0$ & 12, 18 \\
 2003ho & ESO 235-G58 & 0.033 & $4091$ & $33.70\pm0.18$ & $0.640\pm0.050$ & $52848.5\pm3.3$ & 12 \\
 2003iq & NGC 772 & 0.062 & $2331$ & $32.41\pm0.16$ & $0.122\pm0.038$ & $52919.4\pm0.9$ & 12, 19 \\
 2004A & NGC 6207 & 0.013 & $852$ & $30.87\pm0.26$ & $0.177\pm0.043$ & $53012.5\pm1.7$ & 28 \\
 2004dj$^{\dagger}$ & NGC 2403 & 0.034 & $221$ & $27.46\pm0.11$ & $0.094\pm0.035$ & $53180.6\pm15.6$ & 29, 30, 31, 32 \\
 2004eg & UGC 3053 & 0.390 & $2414$ & $32.46\pm0.24$ & $0.160\pm0.150$ & $53169.5\pm30.0$ & 25 \\
 2004ej & NGC 3095 & 0.060 & $2723$ & $32.76\pm0.13$ & $0.145\pm0.116$ & $53231.5\pm4.2$ & 33 \\
 2004et$^{\dagger}$ & NGC 6946 & 0.293 & $40$ & $29.44\pm0.09$ & $0.073\pm0.043$ & $53270.5\pm0.3$ & 19, 34, 35, 36 \\
 2004fx & MCG -02-14-3 & 0.088 & $2673$ & $32.73\pm0.19$ & $0.090\pm0.115$ & $53305.1\pm1.4$ & 33 \\
 2005af & NGC 4945 & 0.159 & $563$ & $27.75\pm0.12$ & $0.090\pm0.117$ & $53320.8\pm17.0$ & 33 \\
 2005au & NGC 5056 & 0.010 & $5592$ & $34.43\pm0.12$ & $0.117\pm0.073$ & $53441.4\pm6.1$ & 37 \\
 2005ay & NGC 3938 & 0.018 & $850$ & $30.68\pm0.21$ & $0.035\pm0.037$ & $53450.7\pm1.8$ & 19, 38 \\
 2005cs$^{\dagger*}$ & M51a & 0.032 & $463$ & $29.67\pm0.07$ & $0.124\pm0.037$ & $53548.4\pm0.3$ & 19, 38, 39, 40 \\
 2005dx & ESO 550-G2 & 0.021 & $8012$ & $35.09\pm0.09$ & $0.090\pm0.142$ & $53615.7\pm4.6$ & 33 \\
 2006my & NGC 4651 & 0.012 & $788$ & $31.40\pm0.10$ & $-0.062\pm0.152$ & $53942.5\pm20.0$ & 36 \\
 2006ov$^{\ddagger}$ & M61 & 0.019 & $1566$ & $31.40\pm0.63$ & $0.275\pm0.105$ & $53973.5\pm6.0$ & 25, 41 \\
 2007aa & NGC 4030 & 0.023 & $1574$ & $31.99\pm0.27$ & $0.046\pm0.100$ & $54131.9\pm4.1$ & 33, 41, 42 \\
 2007hv & UGC 2858 & 0.418 & $5054$ & $34.05\pm0.16$ & $0.000\pm0.100$ & $54346.6\pm2.8$ & 41 \\
 2007it & NGC 5530 & 0.099 & $1193$ & $30.56\pm0.24$ & $0.109\pm0.111$ & $54348.0\pm0.6$ & 33, 43 \\
 2007od & UGC 12846 & 0.031 & $1734$ & $31.63\pm0.25$ & $0.134\pm0.070$ & $54388.2\pm4.1$ & 41, 44, 45, 46 \\
 2008K & ESO 504-G4 & 0.033 & $7964$ & $35.29\pm0.10$ & $0.090\pm0.111$ & $54466.8\pm4.8$ & 33 \\
 2008M & ESO 121-G26 & 0.039 & $2192$ & $32.62\pm0.20$ & $0.090\pm0.112$ & $54475.3\pm3.3$ & 33 \\
 2008aw & NGC 4939 & 0.035 & $3110$ & $33.19\pm0.17$ & $0.226\pm0.043$ & $54519.3\pm3.9$ & 33 \\
 2008bk$^{\dagger}$ & NGC 7793 & 0.017 & $230$ & $27.66\pm0.08$ & $0.090\pm0.111$ & $54547.1\pm2.6$ & 47, 48, 49 \\
 2008gz & NGC 3672 & 0.036 & $1862$ & $32.22\pm0.15$ & $-0.031\pm0.040$ & $54693.5\pm5.0$ & 50 \\
 2008in$^{\dagger\ddagger*}$ & M61 & 0.019 & $1566$ & $31.40\pm0.63$ & $0.037\pm0.044$ & $54824.2\pm0.1$ & 41, 46, 51 \\
 2009N$^{\ddagger*}$ & NGC 4487 & 0.018 & $905$ & $30.90\pm0.38$ & $0.265\pm0.045$ & $54850.1\pm3.4$ & 41, 46, 52 \\
 2009at & NGC 5301 & 0.015 & $1503$ & $31.82\pm0.22$ & $0.551\pm0.079$ & $54899.1\pm2.0$ & 46 \\
 2009ay & NGC 6479 & 0.034 & $6650$ & $34.74\pm0.13$ & $0.252\pm0.087$ & $54898.3\pm4.4$ & 41, 46, 53 \\
 2009bw$^{\dagger}$ & UGC 2890 & 0.197 & $1155$ & $31.07\pm0.26$ & $0.160\pm0.049$ & $54916.5\pm3.0$ & 54 \\
 2009dd & NGC 4088 & 0.017 & $757$ & $30.76\pm0.17$ & $0.285\pm0.057$ & $54916.0\pm4.2$ & 15, 41 \\
 2009hd & M66 & 0.029 & $727$ & $30.15\pm0.07$ & $1.206\pm0.056$ & $55001.6\pm3.9$ & 55 \\
 2009ib$^{*}$ & NGC 1559 & 0.026 & $1304$ & $31.41\pm0.05$ & $0.147\pm0.038$ & $55040.3\pm4.0$ & 56 \\
 2009md$^{\dagger}$ & NGC 3389 & 0.023 & $1308$ & $32.06\pm0.12$ & $0.165\pm0.048$ & $55161.9\pm7.9$ & 57 \\
 2010aj & MCG -01-32-35 & 0.029 & $6497$ & $34.78\pm0.12$ & $0.082\pm0.048$ & $55261.8\pm3.5$ & 15, 41 \\
 PTF10gva & NSA 49030 & 0.026 & $8253$ & $35.33\pm0.10$ & $0.111\pm0.111$ & $55320.3\pm0.9$ & 58 \\
 2011fd & NGC 2273B & 0.063 & $2101$ & $32.41\pm0.22$ & $0.090\pm0.111$ & $55777.9\pm4.5$ & 46 \\
 PTF11go & MCG +09-19-105 & 0.010 & $8037$ & $35.22\pm0.10$ & $0.160\pm0.150$ & $55570.9\pm1.4$ & 58 \\
 PTF11htj & SDSS J211603.12+123124.4 & 0.060 & $5096$ & $34.04\pm0.15$ & $0.160\pm0.150$ & $55751.9\pm1.5$ & 58 \\
 PTF11izt & anonymous & 0.047 & $5996$ & $34.52\pm0.11$ & $0.160\pm0.150$ & $55766.8\pm1.3$ & 58 \\
 2012A$^{*}$ & NGC 3239 & 0.027 & $753$ & $30.66\pm0.24$ & $0.040\pm0.049$ & $55929.3\pm2.6$ & 46, 59 \\
 2012aw$^{*}$ & M95 & 0.024 & $778$ & $29.93\pm0.05$ & $0.115\pm0.041$ & $56002.0\pm0.5$ & 46, 60, 61, 62, 63 \\
 2012br & SDSS J122417.04+185529.4 & 0.031 & $6805$ & $34.90\pm0.12$ & $0.199\pm0.260$ & $56000.9\pm0.2$ & 58 \\
 2012cd & CGCG 271-34 & 0.025 & $3525$ & $33.54\pm0.19$ & $0.123\pm0.134$ & $56019.4\pm1.9$ & 58 \\
 2012ec$^{\dagger\ddagger*}$ & NGC 1084 & 0.023 & $1407$ & $31.07\pm0.25$ & $0.102\pm0.048$ & $56144.5\pm3.4$ & 46, 64 \\
 PTF12grj & WISEA J012039.07+044621.5 & 0.024 & $10193$ & $35.62\pm0.09$ & $0.160\pm0.150$ & $56123.4\pm0.6$ & 58 \\
 PTF12hsx & SDSS J005503.33+421954.0 & 0.079 & $5696$ & $34.34\pm0.14$ & $0.167\pm0.221$ & $56113.0\pm0.5$ & 58 \\
 2013K & ESO 9-G10 & 0.122 & $2418$ & $32.48\pm0.21$ & $0.512\pm0.066$ & $56294.5\pm4.2$ & 65 \\
 2013ab$^{\ddagger*}$ & NGC 5669 & 0.023 & $1368$ & $31.38\pm0.20$ & $0.028\pm0.040$ & $56339.6\pm0.6$ & 46, 66 \\
 2013am$^{*}$ & M65 & 0.021 & $1114$ & $30.36\pm0.29$ & $0.536\pm0.069$ & $56371.8\pm0.4$ & 46, 65, 67 \\
 2013bu & NGC 7331 & 0.078 & $440$ & $30.84\pm0.10$ & $0.523\pm0.062$ & $56396.6\pm2.0$ & 46, 68 \\
\end{tabular}
\end{table*}
\begin{table*}
\scriptsize
\contcaption{}
\begin{tabular}{l l c c c c c l}
\hline
SN     & Host galaxy & $\EGBV$ & $c\zsnhel$     &$\mu$ & $\EhBV$ & $t_0$ & References\\
       &             & (mag)   & (km\,s$^{-1}$) & (mag)& (mag)   & (MJD) &           \\
\hline
 2013by & ESO 138-G10 & 0.188 & $1144$ & $30.46\pm0.29$ & $0.196\pm0.100$ & $56401.6\pm3.6$ & 69 \\
 2013ej$^{\dagger\ddagger*}$ & M74 & 0.060 & $657$ & $29.95\pm0.08$ & $0.044\pm0.040$ & $56496.8\pm0.2$ & 46, 70, 71, 72, 73, 74 \\
 2013fs$^{*}$ & NGC 7610 & 0.035 & $3554$ & $33.32\pm0.18$ & $0.027\pm0.037$ & $56570.8\pm0.5$ & 68, 75 \\
 2013hj & MCG -02-24-3 & 0.045 & $2072$ & $32.51\pm0.13$ & $0.047\pm0.039$ & $56636.8\pm0.9$ & 76 \\
 iPTF13dkz & SDSS J013611.64+333703.6 & 0.039 & $4797$ & $33.85\pm0.17$ & $0.144\pm0.180$ & $56547.9\pm0.3$ & 58 \\
 LSQ13dpa & LCSB S1492O & 0.032 & $7045$ & $35.00\pm0.11$ & $0.160\pm0.150$ & $56642.2\pm2.0$ & 68 \\
 2014G$^{\dagger\ddagger*}$ & NGC 3448 & 0.010 & $1160$ & $31.96\pm0.14$ & $0.268\pm0.046$ & $56669.3\pm0.8$ & 46, 76, 77 \\
 2014cx$^{\ddagger*}$ & NGC 337 & 0.096 & $1646$ & $31.51\pm0.22$ & $-0.021\pm0.045$ & $56901.9\pm0.3$ & 78 \\
 2014cy$^{*}$ & NGC 7742 & 0.048 & $1663$ & $31.54\pm0.23$ & $0.090\pm0.045$ & $56899.6\pm0.7$ & 46, 68, 79 \\
 2014dw & NGC 3568 & 0.092 & $2444$ & $32.39\pm0.13$ & $0.182\pm0.127$ & $56957.5\pm10.0$ & 68 \\
 ASASSN-14dq$^{\ddagger}$ & UGC 11860 & 0.060 & $3125$ & $32.97\pm0.21$ & $0.108\pm0.038$ & $56841.0\pm5.5$ & 80 \\
 ASASSN-14ha & NGC 1566 & 0.008 & $1504$ & $30.86\pm0.15$ & $0.090\pm0.111$ & $56909.5\pm0.6$ & 68 \\
 OGLE14-18 & ESO 87-G3 & 0.049 & $8082$ & $35.18\pm0.10$ & $0.144\pm0.137$ & $56701.7\pm0.9$ & 81 \\
 2015V & UGC 11000 & 0.031 & $1369$ & $31.63\pm0.22$ & $0.031\pm0.060$ & $57112.3\pm4.3$ & 46 \\
 2015W & UGC 3617 & 0.118 & $3984$ & $33.70\pm0.19$ & $0.157\pm0.063$ & $57015.0\pm6.5$ & 46, 68 \\
 2015an & IC 2367 & 0.009 & $2448$ & $32.79\pm0.13$ & $0.114\pm0.111$ & $57268.0\pm1.6$ & 82 \\
 2015ba & IC 1029 & 0.015 & $2383$ & $32.80\pm0.16$ & $0.416\pm0.047$ & $57347.5\pm4.9$ & 83 \\
 2015cz & NGC 582 & 0.045 & $4352$ & $33.68\pm0.16$ & $0.472\pm0.065$ & $57290.4\pm7.5$ & 79 \\
 ASASSN-15oz & LEDA 4614833 & 0.078 & $2078$ & $31.90\pm0.26$ & $0.230\pm0.056$ & $57259.1\pm1.9$ & 84 \\
 2016X$^{\ddagger*}$ & UGC 8041 & 0.019 & $1321$ & $31.32\pm0.34$ & $0.079\pm0.049$ & $57405.9\pm0.3$ & 46, 85, 86 \\
 2016aqf & NGC 2101 & 0.047 & $1204$ & $31.01\pm0.25$ & $0.180\pm0.100$ & $57442.6\pm0.3$ & 87 \\
 2016bkv$^{*}$ & NGC 3184 & 0.014 & $592$ & $30.40\pm0.18$ & $0.036\pm0.078$ & $57467.5\pm1.2$ & 88, 89 \\
 2016gfy & NGC 2276 & 0.086 & $2416$ & $32.64\pm0.22$ & $0.163\pm0.045$ & $57641.3\pm2.6$ & 90 \\
 2016ija & NGC 1532 & 0.013 & $1040$ & $30.55\pm0.17$ & $1.950\pm0.150$ & $57712.1\pm1.0$ & 91 \\
 2017it & anonymous & 0.029 & $12891$ & $36.21\pm0.07$ & $-0.087\pm0.042$ & $57746.2\pm0.7$ & 92 \\
 2017ahn & NGC 3318 & 0.067 & $2775$ & $32.84\pm0.10$ & $0.233\pm0.148$ & $57791.8\pm0.5$ & 93 \\
 2017eaw$^{\dagger\ddagger*}$ & NGC 6946 & 0.293 & $40$ & $29.44\pm0.09$ & $0.059\pm0.037$ & $57886.2\pm0.6$ & 94, 95, 96, 97, 98, 99 \\
 2017gmr$^{\dagger\ddagger*}$ & NGC 988 & 0.023 & $1510$ & $31.08\pm0.19$ & $0.332\pm0.045$ & $57999.2\pm0.6$ & 100 \\
 2018cuf & IC 5092 & 0.028 & $3248$ & $33.10\pm0.19$ & $0.221\pm0.100$ & $58291.8\pm0.3$ & 101 \\
 2018hwm & IC 2327 & 0.022 & $2684$ & $33.08\pm0.19$ & $0.150\pm0.069$ & $58424.8\pm0.9$ & 102 \\
\hline
\multicolumn{8}{m{0.98\linewidth}}{\textit{Notes}. Column~1: SN names. Column~2: SN host galaxy names, Column~3: Galactic colour excesses with random errors of 16~per~cent. Column~4: heliocentric SN redshifts. Column~5 and 6: host galaxy distance moduli and colour excesses, respectively. Column~7: explosion epochs. Column~8: references for the photometry (codes are in Appendix~\ref{sec:all_refs}).}\\
\multicolumn{8}{l}{$^\dagger$Selected to compute BCs.}\\
\multicolumn{8}{l}{$^\ddagger$With $V\!RI$ and $ri$ photometry in the radioactive tail.}\\
\multicolumn{8}{l}{$^*$Used to compute $r\!-\!R$ for $\Delta t<20$\,d.}
\end{tabular}
\end{table*}

\begin{table*}
\caption{Host galaxy distance moduli. The full table is available online as supplementary data.}
\label{table:mu_values}
\begin{tabular}{l l c c c c c c}
\hline
Host galaxy & SN & $\mu_{\mathrm{CPL}}$ & $\mu_{\mathrm{TRGB}}$ & $\mu_{\mathrm{TF}}$ & $\mu_{\mathrm{HLL}}$ & $\mu_{\mathrm{SVF}}$ & $\mu$ \\
\hline
NGC 6946                 & 1980K       & --                  & $29.44\pm0.09^{103}$& --                  & --                  & --                  & $29.44\pm0.09$      \\
                         & 2002hh      & --                  & --                  & --                  & --                  & --                  & $29.44\pm0.09$      \\
                         & 2004et      & --                  & --                  & --                  & --                  & --                  & $29.44\pm0.09$      \\
\hline
\multicolumn{8}{p{0.89\linewidth}}{\textit{Notes}. Column~1: SN host galaxy names. Column~2: SN names, Columns~3 and 4: Cepheids and TRGB distances from the literature, respectively. Column~5: Tully-Fisher distances from EDD. Column~6: distances compute with the Hubble-Lema\^{i}tre law. Column~6: distances inferred from smoothed velocity fields \citep{2020AJ....159...67K}. Column~8: adopted distance moduli. References for the data (superscript numbers) are in Appendix~\ref{sec:all_refs}.}
\end{tabular}
\end{table*}

\begin{table*}
\caption{Host galaxy colour excesses. The full table is available online as supplementary data.}
\label{table:Eh_values}
\begin{tabular}{l c c c c c}
\hline
SN       & $E_{\bv}^\mathrm{h,C3}$  & $E_{\bv}^{\mathrm{h},\vi}$   & $E_{\bv}^\mathrm{h,spec}$   & $E_{\bv}^\mathrm{h,NaID}$  & $\EhBV$\\
\hline
1980K       & $-0.106\pm0.084$         & --                       & --                       & --                       &     $0.106\pm0.084$      \\
1986I       & --                       & --                       & --                       & $\phs0.122\pm0.090^{4}$  & $\phs0.212\pm0.142$      \\
1988A       & --                       & --                       & --                       & $\phs0.000\pm0.014^{115}$& $\phs0.090\pm0.111$      \\
\hline
\multicolumn{6}{p{.78\textwidth}}{\textit{Notes}. Column~1: SN names. Column~2: $\EhBV$ values computed with the C3 method. Column~3: $\EhBV$ values measured with the $\vi$ colour method. Column~4: $\EhBV$ computed with the spectrum-fitting technique. Column~5: $\EhBV$ values estimated from the pEW of the host galaxy \ion{Na}{I}\,D line. Column~6: adopted host galaxy colour excesses. References for the data (superscript numbers) are in Appendix~\ref{sec:all_refs}.}\\
\end{tabular}
\end{table*}

\begin{table*}
\caption{SN explosion epochs. The full table is available online as supplementary data.}
\label{table:t0_values}
\begin{tabular}{l c c c l l}
\hline
SN     & $t_\mathrm{ln}$ (MJD) & $t_\mathrm{fd}$ (MJD) & $t_0$ (MJD) & Phase source & Spectral source\\
\hline 
 1980K       & $44494.0$ & $44540.0$ & $44528.2\pm5.9$    & 124, IAUC 3532       & WISeREP \\
 1986I       & --        & $46558.0$ & $46556.4\pm1.9$    & IAUC 4219            & 4       \\
 1988A       & $47175.7$ & $47176.8$ & $47176.2\pm0.3$    & IAUC 4533, IAUC 4540 & 125     \\
\hline
\multicolumn{6}{p{0.68\linewidth}}{\textit{Notes}. Column~1: SN names. Column~2 and 3: SN last non-detection and first detection epochs, respectively. Column~4: explosion epochs estimated with the \texttt{SNII\_ETOS} code ($\pm1\,\ssd$ error), unless otherwise noted. Column~5: references for $t_\mathrm{ln}$ and  $t_\mathrm{fd}$. Column~6: data source of spectroscopy used to compute $t_0$ with the \texttt{SNII\_ETOS} code. Reference codes are in Appendix~\ref{sec:all_refs}.}
\end{tabular}
\end{table*}

\begin{table*}
\scriptsize
\caption{$^{56}$Ni masses and light-curve parameters.}
\label{table:Ni_masses}
\begin{tabular}{lcccccccc}
\hline
SN       & $x$   & Range & $N$ & $T_0$  & $\log{(\mni[\msun])}$ & $\mni$   &$M_V^{50\mathrm{d}}$& $\rho$\\
         &       & (d)   &     & (d)    &(dex)                  &($\msun$) & (mag)              &        \\
\hline
1980K       & $V$ & 100--285 &   9 & --  & $-1.462\pm0.147$ & $0.03655\pm0.01273$ & $-17.348\pm0.379$ & $-0.655$ \\
1986I       & $I$ & 171--213 &   4 & --  & $-1.302\pm0.198$ & $0.05535\pm0.02661$ & $-16.911\pm0.590$ & $-0.900$ \\
1988A       & $R$ & 185--189 &   3 & --  & $-1.109\pm0.185$ & $0.08519\pm0.03800$ & $-16.488\pm0.488$ & $-0.930$ \\
1990E       & $I$ & 141--300 &   7 & --  & $-1.383\pm0.123$ & $0.04309\pm0.01245$ & $-16.702\pm0.341$ & $-0.899$ \\
1990K       & $R$ & 100--223 &   7 & 222 & $-1.470\pm0.118$ & $0.03516\pm0.00973$ & $-16.705\pm0.275$ & $-0.810$ \\
1991G       & $I$ & 135--303 &   7 & --  & $-1.778\pm0.099$ & $0.01711\pm0.00395$ & $-15.365\pm0.277$ & $-0.798$ \\
1991al      & $I$ & 101--129 &   4 & --  & $-1.629\pm0.068$ & $0.02379\pm0.00375$ & $-15.974\pm0.159$ & $-0.670$ \\
1992H       & $V$ & 140--292 &   7 & --  & $-0.788\pm0.178$ & $0.17721\pm0.07579$ & $-17.670\pm0.413$ & $-0.899$ \\
1992ba      & $I$ & 150--197 &   3 & --  & $-1.744\pm0.126$ & $0.01881\pm0.00557$ & $-15.733\pm0.305$ & $-0.925$ \\
1994N       & $R$ & 260--290 &   3 & --  & $-2.283\pm0.106$ & $0.00537\pm0.00133$ & $-14.955\pm0.233$ & $-0.801$ \\
1995ad      & $I$ & 104--229 &   7 & 258 & $-1.230\pm0.115$ & $0.06099\pm0.01644$ & $-17.086\pm0.387$ & $-0.858$ \\
1996W       & $I$ & 250--309 &   3 & --  & $-0.952\pm0.097$ & $0.11451\pm0.02590$ & $-17.469\pm0.256$ & $-0.823$ \\
1997D       & $I$ & 125--176 &   6 & --  & $-2.064\pm0.144$ & $0.00912\pm0.00311$ & $<-14.436\pm0.423$ & $-0.824$ \\
1999ca      & $I$ & 203--223 &   7 & --  & $-1.846\pm0.070$ & $0.01444\pm0.00234$ & $-16.804\pm0.167$ & $-0.644$ \\
1999em      & $I$ & 132--311 &  11 & --  & $-1.296\pm0.062$ & $0.05110\pm0.00733$ & $-16.692\pm0.138$ & $-0.661$ \\
1999ga      & $I$ & 121--196 &   3 & --  & $-1.446\pm0.120$ & $0.03720\pm0.01048$ & $<-16.693\pm0.285$ & $-0.497$ \\
1999gi      & $I$ & 133--177 &   7 & 186 & $-1.333\pm0.087$ & $0.04739\pm0.00959$ & $-16.250\pm0.207$ & $-0.845$ \\
2001X       & $I$ & 123--144 &   4 & --  & $-1.395\pm0.097$ & $0.04129\pm0.00934$ & $-16.348\pm0.231$ & $-0.877$ \\
2001dc      & $I$ & 119--153 &   7 & --  & $-2.119\pm0.110$ & $0.00785\pm0.00202$ & $-15.059\pm0.283$ & $-0.709$ \\
2002gw      & $I$ & 137--153 &   3 & --  & $-1.631\pm0.103$ & $0.02406\pm0.00579$ & $-16.017\pm0.246$ & $-0.869$ \\
2002hh      & $I$ & 167--286 &  14 & --  & $-1.082\pm0.069$ & $0.08385\pm0.01341$ & $-16.882\pm0.285$ & $-0.611$ \\
2002hx      & $I$ &  96--160 &   7 & --  & $-1.186\pm0.064$ & $0.06587\pm0.00976$ & $-16.671\pm0.165$ & $-0.612$ \\
2003B       & $I$ & 223--282 &   3 & --  & $-2.223\pm0.115$ & $0.00620\pm0.00167$ & $-14.768\pm0.271$ & $-0.876$ \\
2003T       & $V$ & 121--137 &   3 & --  & $-1.344\pm0.101$ & $0.04653\pm0.01097$ & $-16.674\pm0.164$ & $-0.595$ \\
2003Z       & $I$ & 151--208 &   3 & --  & $-2.262\pm0.116$ & $0.00567\pm0.00154$ & $-14.610\pm0.277$ & $-0.821$ \\
2003fb      & $V$ & 116--147 &   5 & --  & $-1.482\pm0.111$ & $0.03406\pm0.00885$ & $-16.016\pm0.206$ & $-0.725$ \\
2003gd      & $I$ & 128--301 &  17 & 303 & $-1.694\pm0.099$ & $0.02076\pm0.00480$ & $<-16.405\pm0.149$ & $-0.407$ \\
2003hd      & $I$ & 104--138 &   4 & --  & $-1.362\pm0.063$ & $0.04391\pm0.00640$ & $-17.056\pm0.156$ & $-0.629$ \\
2003hk      & $R$ & 117--145 &   3 & --  & $-1.572\pm0.105$ & $0.02759\pm0.00677$ & $-17.101\pm0.211$ & $-0.635$ \\
2003hn      & $I$ & 121--151 &   8 & 187 & $-1.412\pm0.056$ & $0.03905\pm0.00506$ & $-16.819\pm0.136$ & $-0.514$ \\
2003ho      & $R$ & 102--118 &   3 & --  & $-1.601\pm0.105$ & $0.02580\pm0.00633$ & $-16.554\pm0.267$ & $-0.717$ \\
2003iq      & $I$ & 120--126 &   3 & --  & $-1.318\pm0.082$ & $0.04895\pm0.00933$ & $-16.763\pm0.200$ & $-0.817$ \\
2004A       & $I$ & 122--241 &  21 & --  & $-1.604\pm0.116$ & $0.02579\pm0.00701$ & $-15.898\pm0.291$ & $-0.913$ \\
2004dj      & $I$ & 131--284 &  17 & --  & $-1.902\pm0.130$ & $0.01311\pm0.00401$ & $-15.846\pm0.178$ & $-0.320$ \\
2004eg      & $I$ & 150--203 &   4 & --  & $-2.126\pm0.188$ & $0.00822\pm0.00373$ & $<-15.077\pm0.551$ & $-0.744$ \\
2004ej      & $V$ & 120--194 &  14 & --  & $-1.793\pm0.166$ & $0.01733\pm0.00687$ & $-16.494\pm0.380$ & $-0.881$ \\
2004et      & $I$ & 135--319 &  23 & 372 & $-1.037\pm0.071$ & $0.09307\pm0.01532$ & $-17.645\pm0.213$ & $-0.759$ \\
2004fx      & $V$ & 109--161 &  24 & --  & $-1.802\pm0.173$ & $0.01708\pm0.00708$ & $-15.720\pm0.400$ & $-0.906$ \\
2005af      & $V$ & 123--162 &  20 & 234 & $-1.477\pm0.179$ & $0.03630\pm0.01562$ & $<-15.212\pm0.383$ & $-0.836$ \\
2005au      & $I$ & 102--137 &   5 & --  & $-1.195\pm0.085$ & $0.06506\pm0.01286$ & $-17.291\pm0.273$ & $-0.718$ \\
2005ay      & $R$ & 117--236 &   5 & --  & $-1.778\pm0.109$ & $0.01721\pm0.00439$ & $-15.512\pm0.239$ & $-0.828$ \\
2005cs      & $I$ & 150--311 &  18 & --  & $-2.241\pm0.076$ & $0.00583\pm0.00103$ & $-15.371\pm0.133$ & $-0.474$ \\
2005dx      & $V$ & 103--120 &   3 & --  & $-2.064\pm0.194$ & $0.00954\pm0.00448$ & $-15.583\pm0.445$ & $-0.878$ \\
2006my      & $I$ & 177--261 &   8 & --  & $-1.674\pm0.141$ & $0.02233\pm0.00745$ & $<-15.361\pm0.473$ & $-0.768$ \\
2006ov      & $I$ & 145--231 &   9 & --  & $-2.051\pm0.267$ & $0.01074\pm0.00728$ & $<-16.568\pm0.707$ & $-0.961$ \\
2007aa      & $i$ & 106--121 &   7 & --  & $-1.522\pm0.143$ & $0.03174\pm0.01074$ & $-16.556\pm0.407$ & $-0.907$ \\
2007hv      & $i$ & 113--131 &   6 & --  & $-1.326\pm0.125$ & $0.04920\pm0.01446$ & $-16.698\pm0.400$ & $-0.887$ \\
2007it      & $I$ & 137--312 &   7 & --  & $-0.987\pm0.130$ & $0.10776\pm0.03299$ & $-17.400\pm0.417$ & $-0.899$ \\
2007od      & $I$ & 239--316 &   3 & --  & $>-2.482\pm0.148$ & $>0.00349\pm0.00123$ & $-17.438\pm0.332$ & $-0.713$ \\
2008K       & $V$ & 122--182 &   4 & --  & $-1.611\pm0.180$ & $0.02669\pm0.01155$ & $-16.806\pm0.378$ & $-0.713$ \\
2008M       & $V$ &  96--120 &   5 & --  & $-1.605\pm0.172$ & $0.02686\pm0.01107$ & $-16.504\pm0.396$ & $-0.900$ \\
2008aw      & $V$ & 103--133 &   5 & 117 & $-1.074\pm0.114$ & $0.08729\pm0.02331$ & $-17.228\pm0.230$ & $-0.700$ \\
2008bk      & $I$ & 175--292 &   5 & --  & $-2.064\pm0.094$ & $0.00883\pm0.00193$ & $-14.992\pm0.366$ & $-0.806$ \\
2008gz      & $I$ & 127--273 &  26 & --  & $-1.253\pm0.081$ & $0.05683\pm0.01069$ & $<-16.366\pm0.194$ & $-0.784$ \\
2008in      & $I$ & 113--152 &   9 & --  & $-1.640\pm0.258$ & $0.02733\pm0.01778$ & $-16.134\pm0.644$ & $-0.979$ \\
2009N       & $I$ & 109--191 &  23 & --  & $-1.891\pm0.161$ & $0.01377\pm0.00528$ & $-15.357\pm0.404$ & $-0.951$ \\
2009at      & $I$ & 115--142 &   3 & --  & $-1.750\pm0.115$ & $0.01842\pm0.00496$ & $-16.153\pm0.332$ & $-0.841$ \\
2009ay      & $R$ & 100--178 &   8 & --  & $-0.932\pm0.116$ & $0.12120\pm0.03296$ & $-17.720\pm0.309$ & $-0.790$ \\
2009bw      & $I$ & 138--235 &  10 & 236 & $-1.737\pm0.120$ & $0.01904\pm0.00536$ & $-16.393\pm0.316$ & $-0.897$ \\
2009dd      & $R$ & 139--298 &   6 & --  & $-1.474\pm0.112$ & $0.03471\pm0.00910$ & $-16.385\pm0.259$ & $-0.716$ \\
2009hd      & $I$ & 158--271 &   5 & 220 & $-1.959\pm0.066$ & $0.01112\pm0.00170$ & $-16.632\pm0.188$ & $-0.677$ \\
2009ib      & $R$ & 149--262 &   4 & --  & $-1.356\pm0.075$ & $0.04472\pm0.00778$ & $-15.790\pm0.131$ & $-0.524$ \\
2009md      & $I$ & 123--180 &   4 & --  & $-2.097\pm0.078$ & $0.00813\pm0.00147$ & $-15.365\pm0.195$ & $-0.689$ \\
2010aj      & $I$ &  95--107 &   4 & --  & $-2.088\pm0.075$ & $0.00829\pm0.00144$ & $-16.764\pm0.203$ & $-0.690$ \\
PTF10gva    & $r$ & 175--213 &   3 & --  & $-1.111\pm0.138$ & $0.08146\pm0.02655$ & --              & --       \\
2011fd      & $I$ & 126--260 &  12 & 292 & $-1.498\pm0.125$ & $0.03311\pm0.00973$ & --              & --       \\
PTF11go     & $R$ &  97--160 &  15 & --  & $-1.593\pm0.160$ & $0.02732\pm0.01042$ & --              & --       \\
PTF11htj    & $r$ & 133--173 &   6 & --  & $-1.309\pm0.175$ & $0.05324\pm0.02236$ & --              & --       \\
PTF11izt    & $R$ & 111--182 &  49 & --  & $-1.651\pm0.159$ & $0.02388\pm0.00905$ & --              & --       \\
2012A       & $I$ & 125--310 &   9 & --  & $-1.766\pm0.111$ & $0.01771\pm0.00460$ & $-16.454\pm0.284$ & $-0.888$ \\
2012aw      & $I$ & 221--309 &  22 & 537 & $-1.269\pm0.055$ & $0.05426\pm0.00690$ & $-16.847\pm0.135$ & $-0.602$ \\
2012br      & $R$ &  99--263 &  37 & --  & $-1.219\pm0.254$ & $0.07166\pm0.04576$ & --              & --       \\
2012cd      & $r$ & 108--216 &   6 & 228 & $-1.038\pm0.168$ & $0.09874\pm0.03967$ & --              & --       \\
2012ec      & $I$ & 111--163 &   8 & --  & $-1.545\pm0.115$ & $0.02953\pm0.00796$ & $-16.336\pm0.290$ & $-0.891$ \\
PTF12grj    & $r$ & 107--139 &   6 & --  & $-1.565\pm0.169$ & $0.02937\pm0.01187$ & --              & --       \\
PTF12hsx    & $R$ & 135--223 & 575 & --  & $-1.090\pm0.225$ & $0.09296\pm0.05158$ & --              & --       \\
2013K       & $R$ & 211--303 &   5 & --  & $-1.593\pm0.123$ & $0.02657\pm0.00768$ & $-16.711\pm0.297$ & $-0.851$ \\
2013ab      & $I$ & 107--218 &  19 & 312 & $-1.482\pm0.095$ & $0.03376\pm0.00747$ & $-16.199\pm0.234$ & $-0.867$ \\
2013am      & $I$ & 230--314 &   5 & --  & $-1.695\pm0.163$ & $0.02166\pm0.00842$ & $-15.470\pm0.358$ & $-0.743$ \\
2013bu      & $I$ & 130--281 &   9 & --  & $-1.888\pm0.093$ & $0.01324\pm0.00287$ & $-16.104\pm0.217$ & $-0.604$ \\
\end{tabular}
\end{table*}
\begin{table*}
\contcaption{}
\scriptsize
\begin{tabular}{lcccccccc}
\hline
SN       & $x$   & Range & $N$ & $T_0$  & $\log{(\mni[\msun])}$ & $\mni$   &$M_V^{50\mathrm{d}}$& $\rho$\\
         &       & (d)   &     & (d)    &(dex)                  &($\msun$) & (mag)              &       \\
\hline
2013by      & $i$ & 104--153 &  10 & 176 & $-1.512\pm0.151$ & $0.03268\pm0.01171$ & $-17.397\pm0.436$ & $-0.902$ \\
2013ej      & $I$ & 107--226 & 107 & 172 & $-1.551\pm0.060$ & $0.02839\pm0.00394$ & $-16.950\pm0.149$ & $-0.675$ \\
2013fs      & $r$ & 102--107 &   4 & --  & $-1.323\pm0.106$ & $0.04897\pm0.01213$ & $-16.739\pm0.213$ & $-0.762$ \\
2013hj      & $I$ & 117--168 &  19 & --  & $-1.092\pm0.073$ & $0.08206\pm0.01389$ & $-17.564\pm0.178$ & $-0.767$ \\
iPTF13dkz   & $r$ & 152--174 &   5 & --  & $-1.173\pm0.205$ & $0.07506\pm0.03750$ & --              & --       \\
LSQ13dpa    & $i$ & 144--152 &   3 & --  & $-1.101\pm0.135$ & $0.08317\pm0.02649$ & $-17.181\pm0.470$ & $-0.913$ \\
2014G       & $I$ &  95--165 &  30 & 127 & $-1.146\pm0.077$ & $0.07258\pm0.01297$ & $-17.203\pm0.199$ & $-0.793$ \\
2014cx      & $I$ & 120--150 &  10 & --  & $-1.251\pm0.103$ & $0.05771\pm0.01388$ & $-16.668\pm0.263$ & $-0.884$ \\
2014cy      & $R$ & 128--164 &   4 & 113 & $-1.820\pm0.117$ & $0.01569\pm0.00431$ & $-15.584\pm0.269$ & $-0.861$ \\
2014dw      & $r$ &  97--143 &  18 & 85  & $-1.544\pm0.158$ & $0.03053\pm0.01149$ & $-16.238\pm0.487$ & $-0.734$ \\
ASASSN-14dq & $I$ & 108--204 &  10 & 266 & $-1.308\pm0.101$ & $0.05055\pm0.01192$ & $-16.967\pm0.249$ & $-0.826$ \\
ASASSN-14ha & $i$ & 140--181 &  18 & --  & $-2.030\pm0.120$ & $0.00970\pm0.00273$ & $-15.963\pm0.369$ & $-0.859$ \\
OGLE14-18   & $I$ & 182--235 &   6 & --  & $-1.367\pm0.123$ & $0.04471\pm0.01292$ & --              & --       \\
2015V       & $I$ & 124--217 &  33 & --  & $-1.643\pm0.107$ & $0.02345\pm0.00587$ & $-15.751\pm0.287$ & $-0.871$ \\
2015W       & $i$ & 253--289 &   4 & --  & $-1.372\pm0.110$ & $0.04385\pm0.01129$ & $-17.129\pm0.314$ & $-0.714$ \\
2015an      & $i$ & 148--218 &  13 & --  & $-1.575\pm0.113$ & $0.02752\pm0.00728$ & $-17.184\pm0.363$ & $-0.873$ \\
2015ba      & $i$ & 166--259 &   3 & --  & $-1.840\pm0.114$ & $0.01496\pm0.00400$ & $-17.020\pm0.221$ & $-0.613$ \\
2015cz      & $I$ & 125--143 &   4 & --  & $-1.344\pm0.093$ & $0.04634\pm0.01004$ & $-17.158\pm0.272$ & $-0.750$ \\
ASASSN-15oz & $i$ & 185--261 &  18 & 126 & $-0.875\pm0.124$ & $0.13890\pm0.04048$ & $-17.682\pm0.316$ & $-0.887$ \\
2016X       & $I$ &  98--196 &  37 & 205 & $-1.349\pm0.147$ & $0.04741\pm0.01652$ & $-16.856\pm0.371$ & $-0.937$ \\
2016aqf     & $i$ & 154--307 &  11 & 332 & $-1.670\pm0.137$ & $0.02247\pm0.00727$ & $-16.024\pm0.394$ & $-0.902$ \\
2016bkv     & $I$ & 202--307 &  29 & --  & $-1.817\pm0.106$ & $0.01570\pm0.00389$ & $-14.537\pm0.300$ & $-0.800$ \\
2016gfy     & $I$ & 122--238 &  11 & 307 & $-1.122\pm0.103$ & $0.07766\pm0.01868$ & $-17.145\pm0.263$ & $-0.883$ \\
2016ija     & $i$ &  96--105 &   6 & --  & $-1.582\pm0.144$ & $0.02766\pm0.00943$ & $-16.691\pm0.490$ & $-0.916$ \\
2017it      & $I$ & 118--135 &  18 & --  & $-1.257\pm0.060$ & $0.05587\pm0.00776$ & $-16.817\pm0.149$ & $-0.626$ \\
2017ahn     & $i$ & 101--142 &   5 & 98  & $-1.316\pm0.132$ & $0.05059\pm0.01574$ & $-16.751\pm0.464$ & $-0.914$ \\
2017eaw     & $I$ & 128--313 & 136 & --  & $-1.087\pm0.069$ & $0.08289\pm0.01325$ & $-17.265\pm0.203$ & $-0.752$ \\
2017gmr     & $I$ & 103--146 &   6 & --  & $-0.857\pm0.093$ & $0.14222\pm0.03081$ & $-17.655\pm0.235$ & $-0.852$ \\
2018cuf     & $i$ & 123--273 &  18 & --  & $-1.417\pm0.121$ & $0.03980\pm0.01131$ & $-16.898\pm0.359$ & $-0.877$ \\
2018hwm     & $r$ & 171--191 &   3 & --  & $-2.205\pm0.124$ & $0.00650\pm0.00189$ & $-15.041\pm0.284$ & $-0.822$ \\
\hline
\multicolumn{9}{p{0.78\linewidth}}{\textit{Notes}. Column~1: SN names. Column~2, 3, and 4: bands, phase ranges, and number of points used to compute $\log\mni$, respectively. Column~5: $\gamma$-ray escape times. Column~6 and 7: derived $\log\mni$ and $\mni$ values. Column~8: Absolute $V$-band magnitudes at 50\,d since explosion. Column~9: correlation coefficients between $\log\mni$ and $M_V^{50\mathrm{d}}$.}
\end{tabular}
\end{table*}

\begin{table*}
\caption{$M_R^{\mathrm{max}}$ value for the LOSS set and our SN sample.}
\label{table:MRmax}
\begin{tabular}{lclclclclc}
\hline
SN$^\dagger$ & $M_R^{\mathrm{max}}$ & SN$^\ddagger$ & $M_R^{\mathrm{max}}$ & SN$^\ddagger$ & $M_R^{\mathrm{max}}$ & SN$^\ddagger$ & $M_R^{\mathrm{max}}$ & SN$^\ddagger$ & $M_R^{\mathrm{max}}$\\
 &  (mag)   &              & (mag)    &               &  (mag)   &               &   (mag)  &               &  (mag)  \\
\hline
1999an & $-16.931$ & 1980K       & $-18.764$  & 2003hk      & $-18.303$  & 2009ay      & $-18.552$  & LSQ13dpa    & $-17.694$  \\
1999bg & $-16.473$ & 1986I       & $-16.730$  & 2003hn      & $-17.602$  & 2009bw      & $-17.309$  & 2014G       & $-18.411$  \\
1999br & $-15.526$ & 1988A       & $-16.639$  & 2003ho      & $-17.415$  & 2009dd      & $-16.993$  & 2014cx      & $-16.957$  \\
1999em & $-17.192$ & 1990E       & $-17.401$  & 2003iq      & $-17.347$  & 2009hd      & $-17.327$  & 2014cy      & $-16.122$  \\
1999gi & $-16.728$ & 1990K       & $-17.784$  & 2004A       & $-16.159$  & 2009ib      & $-16.266$  & 2014dw      & $-17.312$  \\
2000dc & $-17.921$ & 1991G       & $-15.729$  & 2004dj      & $-16.232$  & 2009md      & $-15.779$  & ASASSN-14dq & $-17.653$  \\
2000el & $-16.312$ & 1991al      & $-16.836$  & 2004eg      & $-15.693$  & 2010aj      & $-17.836$  & ASASSN-14ha & $-16.379$  \\
2001bq & $-17.577$ & 1992H       & $-18.067$  & 2004ej      & $-17.026$  & PTF10gva    & $-18.794$  & OGLE14-18   & $-17.103$  \\
2001dc & $-15.283$ & 1992ba      & $-16.218$  & 2004et      & $-17.974$  & 2011fd      & $-16.998$  & 2015V       & $-15.970$  \\
2001fz & $-15.865$ & 1994N       & $-15.371$  & 2004fx      & $-15.984$  & PTF11go     & $-16.743$  & 2015W       & $-17.893$  \\
2002bx & $-16.901$ & 1995ad      & $-17.782$  & 2005af      & $-17.413$  & PTF11htj    & $-16.946$  & 2015an      & $-17.697$  \\
2002ce & $-15.319$ & 1996W       & $-17.798$  & 2005au      & $-17.974$  & PTF11izt    & $-16.348$  & 2015ba      & $-17.750$  \\
2002dq & $-16.232$ & 1997D       & $-14.986$  & 2005ay      & $-15.912$  & 2012A       & $-17.176$  & 2015cz      & $-17.627$  \\
2002ds & $-17.197$ & 1999ca      & $-18.086$  & 2005cs      & $-15.700$  & 2012aw      & $-17.181$  & ASASSN-15oz & $-18.478$  \\
2002gd & $-15.939$ & 1999em      & $-17.192$  & 2005dx      & $-16.381$  & 2012br      & $-17.764$  & 2016X       & $-17.588$  \\
2002hh & $-17.107$ & 1999ga      & $-17.320$  & 2006my      & $-16.896$  & 2012cd      & $-18.680$  & 2016aqf     & $-16.136$  \\
2003Z  & $-14.958$ & 1999gi      & $-16.728$  & 2006ov      & $-16.195$  & 2012ec      & $-16.695$  & 2016bkv     & $-15.741$  \\
2003ao & $-16.339$ & 2001X       & $-16.726$  & 2007aa      & $-16.703$  & PTF12grj    & $-16.963$  & 2016gfy     & $-17.599$  \\
2003hl & $-17.364$ & 2001dc      & $-15.283$  & 2007hv      & $-17.158$  & PTF12hsx    & $-17.331$  & 2016ija     & $-17.238$  \\
2003iq & $-17.347$ & 2002gw      & $-16.310$  & 2007it      & $-18.234$  & 2013K       & $-16.430$  & 2017it      & $-17.178$  \\
2004et & $-17.974$ & 2002hh      & $-17.107$  & 2008K       & $-17.869$  & 2013ab      & $-16.804$  & 2017ahn     & $-18.286$  \\
2004fc & $-16.642$ & 2002hx      & $-17.596$  & 2008M       & $-17.236$  & 2013am      & $-15.958$  & 2017eaw     & $-17.828$  \\
2004fx & $-15.984$ & 2003B       & $-15.248$  & 2008aw      & $-18.181$  & 2013bu      & $-16.947$  & 2017gmr     & $-18.005$  \\
2005ad & $-15.882$ & 2003T       & $-17.131$  & 2008bk      & $-15.245$  & 2013by      & $-18.411$  & 2018cuf     & $-17.224$  \\
2005ay & $-15.912$ & 2003Z       & $-14.958$  & 2008gz      & $-17.679$  & 2013ej      & $-17.900$  & 2018hwm     & $-15.191$  \\
2006be & $-16.972$ & 2003fb      & $-16.582$  & 2008in      & $-16.596$  & 2013fs      & $-17.623$  & --          & -- \\
2006bp & $-17.622$ & 2003gd      & $-16.859$  & 2009N       & $-15.609$  & 2013hj      & $-18.270$  & --          & -- \\
2006ca & $-17.202$ & 2003hd      & $-17.635$  & 2009at      & $-17.283$  & iPTF13dkz   & $-16.470$  & --          & -- \\
\hline
\multicolumn{10}{p{0.95\linewidth}}{$^{\dagger}$LOSS set. Original $M_R^{\mathrm{max}}$ values (listed in \citealt{2011MNRAS.412.1441L}) were recalibrated using $\EGBV$ values from \citet{2011ApJ...737..103S}, $\mu$ estimates listed in Column~8 of Table~\ref{table:mu_values}, and $\EhBV$ values reported in Column~6 of Table~\ref{table:Eh_values}.}\\
\multicolumn{10}{l}{$^{\ddagger}$Our SN sample.}\\
\end{tabular}
\end{table*}

\begin{table*}
\caption{Steepness parameters. The full table is available online as supplementary data.}
\label{table:S}
\begin{tabular}{lc@{\,\,\,}c@{\,\,\,}c@{\,\,\,}c@{\,\,\,}c@{\,\,\,}cc@{\,\,\,}c@{\,\,\,}c@{\,\,\,}c@{\,\,\,}c@{\,\,\,}c}
\hline
SN       & $S_g$  & $S_V$ & $S_r$ & $S_R$  & $S_i$ & $S_I$& $S_g^*$  & $S_V^*$ & $S_r^*$ & $S_R^*$  & $S_i^*$ & $S_I^*$\\
\hline
1980K  & --           & $0.159(39)$  & --           & --           & --           & --           & $0.149(49)$  & --           & $0.118(42)$  & $0.134(40)$  & $0.131(44)$  & $0.125(37)$  \\
1991G  & --           & $0.120(46)$  & --           & $0.089(18)$  & --           & $0.069(20)$  & $0.109(14)$  & $0.104(23)$  & $0.080(14)$  & $0.083(26)$  & $0.086(24)$  & $0.090(22)$  \\
1992ba & --           & $0.098(23)$  & --           & --           & --           & $0.082(25)$  & $0.102(21)$  & $0.108(40)$  & $0.076(19)$  & $0.084(23)$  & $0.081(26)$  & $0.079(28)$  \\
\hline
\multicolumn{13}{l}{\textit{Notes.} Steepness values are in mag\,d$^{-1}$ units. Numbers in parentheses are $1\,\sigma$ errors in units of the last significant digits.}\\
\multicolumn{13}{l}{$^*$Values estimated from the steepnesses in other bands (see Section~\ref{sec:MNi_vs_S}).}
\end{tabular}
\end{table*}

\begin{table*}
\caption{$S_b$ to $S_x$ transformation parameters.}
\label{table:Sy_vs_Sx_pars}
\begin{tabular}{cccccccccccc}
\hline
$x$ & $b$ & $c_{x,b}$     & $d_{x,b}$ & $\ssd_{x,b}$        & $N$ & $x$ & $b$ & $c_{x,b}$     & $d_{x,b}$ & $\ssd_{x,b}$        & $N$  \\
    &     &(mag\,d$^{-1}$)&           &(mag\,d$^{-1}$)&     &     &     &(mag\,d$^{-1}$)&           &(mag\,d$^{-1}$)&      \\
\hline
$g$ & $V$ & $-0.008\pm0.027$ & $0.986\pm0.208$ & 0.030 & 15 & $V$ & $g$ &  $0.015\pm0.033$ & $0.969\pm0.272$ & 0.030 & 15 \\
$g$ & $r$ &  $0.004\pm0.015$ & $1.161\pm0.145$ & 0.021 & 15 & $V$ & $r$ &  $0.026\pm0.018$ & $1.129\pm0.157$ & 0.031 & 18 \\
$g$ & $R$ &  $0.042\pm0.006$ & $0.817\pm0.056$ & 0.012 &  8 & $V$ & $R$ &  $0.017\pm0.008$ & $1.072\pm0.075$ & 0.024 & 29 \\
$g$ & $i$ & $-0.000\pm0.025$ & $1.118\pm0.227$ & 0.030 & 15 & $V$ & $i$ &  $0.016\pm0.010$ & $1.099\pm0.045$ & 0.033 & 19 \\
$g$ & $I$ &  $0.054\pm0.016$ & $0.677\pm0.089$ & 0.020 &  8 & $V$ & $I$ &  $0.011\pm0.011$ & $1.187\pm0.093$ & 0.027 & 28 \\
$r$ & $g$ & $-0.001\pm0.008$ & $0.842\pm0.017$ & 0.018 & 15 & $R$ & $g$ & $-0.041\pm0.010$ & $1.146\pm0.063$ & 0.014 &  8 \\
$r$ & $V$ & $-0.014\pm0.023$ & $0.832\pm0.162$ & 0.027 & 18 & $R$ & $V$ & $-0.002\pm0.010$ & $0.853\pm0.077$ & 0.022 & 29 \\
$r$ & $R$ &  $0.026\pm0.011$ & $0.686\pm0.114$ & 0.015 & 10 & $R$ & $r$ & $-0.015\pm0.020$ & $1.256\pm0.198$ & 0.021 & 10 \\
$r$ & $i$ & $-0.001\pm0.020$ & $0.939\pm0.184$ & 0.031 & 17 & $R$ & $i$ &  $0.039\pm0.024$ & $0.680\pm0.123$ & 0.037 & 11 \\
$r$ & $I$ &  $0.022\pm0.011$ & $0.717\pm0.081$ & 0.015 & 10 & $R$ & $I$ &  $0.002\pm0.009$ & $1.053\pm0.067$ & 0.025 & 28 \\
$i$ & $g$ &  $0.007\pm0.024$ & $0.853\pm0.201$ & 0.027 & 15 & $I$ & $g$ & $-0.040\pm0.035$ & $1.189\pm0.199$ & 0.027 &  8 \\
$i$ & $V$ & $-0.004\pm0.009$ & $0.848\pm0.051$ & 0.029 & 19 & $I$ & $V$ &  $0.005\pm0.010$ & $0.755\pm0.072$ & 0.022 & 28 \\
$i$ & $r$ &  $0.011\pm0.010$ & $0.986\pm0.029$ & 0.032 & 17 & $I$ & $r$ & $-0.009\pm0.017$ & $1.203\pm0.145$ & 0.020 & 10 \\
$i$ & $R$ &  $0.022\pm0.029$ & $0.872\pm0.291$ & 0.042 & 11 & $I$ & $R$ &  $0.012\pm0.008$ & $0.855\pm0.055$ & 0.022 & 28 \\
$i$ & $I$ &  $0.000\pm0.016$ & $1.037\pm0.109$ & 0.028 & 11 & $I$ & $i$ &  $0.024\pm0.015$ & $0.792\pm0.084$ & 0.024 & 11 \\
\hline
\multicolumn{12}{l}{\textit{Notes.} $S_x=c_{x,b} + d_{x,b} S_b$.}
\end{tabular}
\end{table*}

\begin{table*}
\caption{Absolute magnitudes at 50\,d since explosion. The full table is available online as supplementary data.}
\label{table:Mx50d}
\begin{tabular}{lccccc}
\hline
SN  & $M_g^{50\mathrm{d}}$ & $M_r^{50\mathrm{d}}$ & $M_R^{50\mathrm{d}}$ & $M_i^{50\mathrm{d}}$ & $M_I^{50\mathrm{d}}$  \\
\hline
1980K       & $-17.099\pm0.429^*$  & $>-18.255\pm0.270^*$ & $>-18.385\pm0.253$   & $>-17.897\pm0.219^*$ & $>-18.287\pm0.194$   \\
1986I       & --                   & $-16.490\pm0.548^*$  & $-16.619\pm0.532$    & $-16.739\pm0.489^*$  & $-17.129\pm0.471$    \\
1990E       & --                   & $-17.084\pm0.324^*$  & $-17.214\pm0.314$    & $-16.915\pm0.302^*$  & $-17.304\pm0.289$    \\
\hline
\multicolumn{6}{l}{$^*$Values estimated using photometry in other bands (see Section~\ref{sec:NMS}).}
\end{tabular}
\end{table*}

\begin{table*}
\caption{SZB sample.}
\scriptsize
\label{table:ZTF_sample}
\begin{tabular}{lccccccccc}
\hline
IAU SN  & $\EGBV$ & $c\zsnhel$     & $\mu$ & $t_0$ & $M_R^{\max}$ & $M_{r_\ztf}^{50\mathrm{d}}$ &  $S_{r_\ztf}$   & $\log\mni^\mathrm{NMS}$ &  $\log\mni^\mathrm{tail}$\\
        & (mag)   & (km\,s$^{-1}$) & (mag) & (MJD) & (mag)        & (mag)                       & (mag\,d$^{-1}$) & (dex)                   & (dex)         \\
\hline
2018ccb & 0.139 & 4740  & $33.90\pm0.12$ & $58267.9\pm1.4$ & $-17.313$ & $-16.67\pm0.39$ & $0.116\pm0.009$ & $-1.505\pm0.145$ & $-1.399\pm0.162$ \\
2018fpx & 0.210 & 7500  & $35.03\pm0.09$ & $58356.6\pm4.0$ & $<-18.246$ & $-17.75\pm0.39$ & $0.059\pm0.011$ & $-1.033\pm0.153$ & $-0.971\pm0.161$ \\
2018hcp & 0.030 & 5934  & $34.59\pm0.11$ & $58377.0\pm1.5$ & $-17.381$ & $-16.94\pm0.39$ & $0.055\pm0.004$ & $-1.218\pm0.144$ & $-1.215\pm0.160$ \\
2019va  & 0.016 & 2637  & $33.03\pm0.19$ & $58497.6\pm3.1$ & $-17.488$ & $-17.20\pm0.41$ & $0.065\pm0.004$ & $-1.205\pm0.148$ & $-0.962\pm0.170$ \\
2019vb  & 0.011 & 6039  & $34.60\pm0.12$ & $58500.6\pm1.0$ & $-18.469$ & $-17.88\pm0.39$ & $0.074\pm0.011$ & $-1.069\pm0.149$ & $-0.872\pm0.160$ \\
2019clp & 0.031 & 7127  & $35.01\pm0.11$ & $58569.0\pm0.5$ & $-18.490$ & $-17.66\pm0.39$ & $0.068\pm0.009$ & $-1.103\pm0.147$ & $-1.074\pm0.158$ \\
2019cvz & 0.011 & 5578  & $34.39\pm0.13$ & $58577.6\pm1.0$ & $-17.282$ & $-17.09\pm0.39$ & $0.076\pm0.008$ & $-1.274\pm0.146$ & $-1.177\pm0.162$ \\
2019dma & 0.113 & 4880  & $33.89\pm0.15$ & $58560.2\pm6.4$ & $<-17.395$ & $-17.22\pm0.41$ & $0.143\pm0.017$ & $-1.426\pm0.150$ & $-1.585\pm0.167$ \\
2019dtt & 0.056 & 4673  & $33.86\pm0.15$ & $58584.2\pm5.3$ & $<-17.051$ & $-16.88\pm0.40$ & $0.096\pm0.012$ & $-1.397\pm0.149$ & $-1.279\pm0.165$ \\
2019etp & 0.044 & 4683  & $33.77\pm0.16$ & $58607.2\pm2.7$ & $<-17.788$ & $-17.45\pm0.40$ & $0.048\pm0.004$ & $-1.054\pm0.147$ & $-1.065\pm0.165$ \\
2019ffn & 0.102 & 7800  & $35.09\pm0.11$ & $58616.9\pm0.4$ & $-18.300$ & $-17.81\pm0.39$ & $0.071\pm0.003$ & $-1.078\pm0.143$ & $-1.039\pm0.159$ \\
2019gqk & 0.079 & 3900  & $33.35\pm0.19$ & $58625.6\pm4.9$ & $<-16.894$ & $-16.69\pm0.42$ & $0.056\pm0.004$ & $-1.288\pm0.149$ & $-1.247\pm0.171$ \\
2019hhh & 0.281 & 4500  & $33.72\pm0.17$ & $58635.4\pm4.5$ & $-17.059$ & $-16.69\pm0.42$ & $0.081\pm0.004$ & $-1.395\pm0.149$ & $-1.254\pm0.172$ \\
2019hkj & 0.071 & 5400  & $34.17\pm0.15$ & $58620.6\pm9.1$ & $<-17.051$ & $-16.87\pm0.40$ & $0.103\pm0.008$ & $-1.421\pm0.146$ & $-1.332\pm0.167$ \\
2019iex & 0.063 & 4211  & $33.54\pm0.19$ & $58658.9\pm1.5$ & $-17.503$ & $-17.21\pm0.41$ & $0.173\pm0.012$ & $-1.486\pm0.149$ & $-1.339\pm0.171$ \\
2019jyw & 0.029 & 3489  & $33.16\pm0.21$ & $58659.5\pm2.9$ & $-17.850$ & $-17.03\pm0.43$ & $0.070\pm0.004$ & $-1.269\pm0.150$ & $-1.140\pm0.174$ \\
2019lkx & 0.454 & 3000  & $32.81\pm0.23$ & $58677.0\pm1.5$ & $-17.220$ & $-16.85\pm0.47$ & $0.052\pm0.002$ & $-1.229\pm0.157$ & $-1.224\pm0.190$ \\
2019vus & 0.025 & 7500  & $35.08\pm0.09$ & $58806.6\pm1.0$ & $<-17.374$ & $-16.94\pm0.38$ & $0.057\pm0.009$ & $-1.224\pm0.148$ & $-1.174\pm0.163$ \\
2019wvz & 0.009 & 9571  & $35.61\pm0.09$ & $58832.5\pm0.9$ & $-18.031$ & $-17.75\pm0.38$ & $0.071\pm0.016$ & $-1.086\pm0.155$ & $-0.878\pm0.156$ \\
2020aem & 0.034 & 6650  & $34.74\pm0.13$ & $58860.6\pm6.0$ & $-17.602$ & $-16.87\pm0.39$ & $0.087\pm0.008$ & $-1.371\pm0.146$ & $-1.284\pm0.162$ \\
2020ckb & 0.075 & 7503  & $34.98\pm0.11$ & $58861.9\pm4.5$ & $<-17.867$ & $-17.50\pm0.39$ & $0.159\pm0.037$ & $-1.383\pm0.158$ & $-1.188\pm0.159$ \\
2020dpw & 0.320 & 1424  & $31.68\pm0.17$ & $58904.8\pm0.7$ & $<-16.945$ & $-16.75\pm0.42$ & $0.078\pm0.004$ & $-1.368\pm0.150$ & $-1.257\pm0.174$ \\
2020dvt & 0.013 & 4200  & $33.82\pm0.18$ & $58908.5\pm2.0$ & $-17.398$ & $-16.91\pm0.41$ & $0.217\pm0.023$ & $-1.628\pm0.150$ & $-1.719\pm0.168$ \\
2020ekk & 0.041 & 4273  & $33.80\pm0.16$ & $58918.7\pm0.6$ & $-17.905$ & $-17.22\pm0.40$ & $0.156\pm0.020$ & $-1.452\pm0.150$ & $-1.666\pm0.165$ \\
2020gcv & 0.031 & 8513  & $35.27\pm0.10$ & $58934.5\pm6.5$ & $<-17.629$ & $-17.32\pm0.40$ & $0.090\pm0.014$ & $-1.267\pm0.152$ & $-1.171\pm0.159$ \\
2020ifc & 0.173 & 4683  & $33.79\pm0.15$ & $58930.5\pm5.7$ & $-15.851$ & $-15.62\pm0.41$ & $0.133\pm0.139$ & $-1.698\pm0.290$ & $-1.760\pm0.167$ \\
2020nja & 0.026 & 5656  & $34.29\pm0.15$ & $59015.2\pm4.9$ & $<-16.702$ & $-16.50\pm0.40$ & $0.074\pm0.008$ & $-1.412\pm0.148$ & $-1.519\pm0.166$ \\
2020umi & 0.058 & 5610  & $34.43\pm0.15$ & $59119.5\pm1.0$ & $-17.860$ & $-17.00\pm0.40$ & $0.067\pm0.018$ & $-1.253\pm0.163$ & $-1.214\pm0.169$ \\
\hline
\multicolumn{10}{l}{\textit{Notes.} We adopt $\EhBV=0.16\pm0.15$\,mag, while to estimate $M_R^{\max}$ we use $r_\ztf\!-\!R=0.14$\,mag.}
\end{tabular}
\end{table*}


\bsp	
\label{lastpage}
\end{document}